\documentclass[12pt,twoside]{book}
\usepackage[latin1]{inputenc}
\usepackage{a4}
\usepackage{cite}
\usepackage[dvips]{graphicx}
\usepackage{amsmath}
\usepackage{amstext}
\usepackage{amsfonts}
\usepackage{fancyhdr}
\usepackage{psfrag}

\newcommand{\captionfonts}{\small}

\makeatletter  
\long\def\@makecaption#1#2{%
  \vskip\abovecaptionskip
  \sbox\@tempboxa{{\captionfonts #1: #2}}%
  \ifdim \wd\@tempboxa >\hsize
    {\captionfonts #1: #2\par}
  \else
    \hbox to\hsize{\hfil\box\@tempboxa\hfil}%
  \fi
  \vskip\belowcaptionskip}
\makeatother   

\numberwithin{equation}{section}

\newcommand{\½}[0]{\frac{1}{2}}

\newcommand{\abs}[1]{\lvert#1\rvert}
\newcommand{\norm}[1]{\lVert#1\rVert}
\newcommand{\bm}[1]{\boldsymbol{#1}}

\begin{document}
\thispagestyle{empty}
\begin{center}
\noindent\rule{\textwidth}{5pt}
\\
{\LARGE \bf Correlations in many-body systems with the
\\
\vspace{0.3cm}
Stochastic Variational Method
}
\\
\rule{\textwidth}{5pt}
\vspace{0.2cm}
\\
\begin{minipage}[c]{\textwidth}
\includegraphics[width=\textwidth]{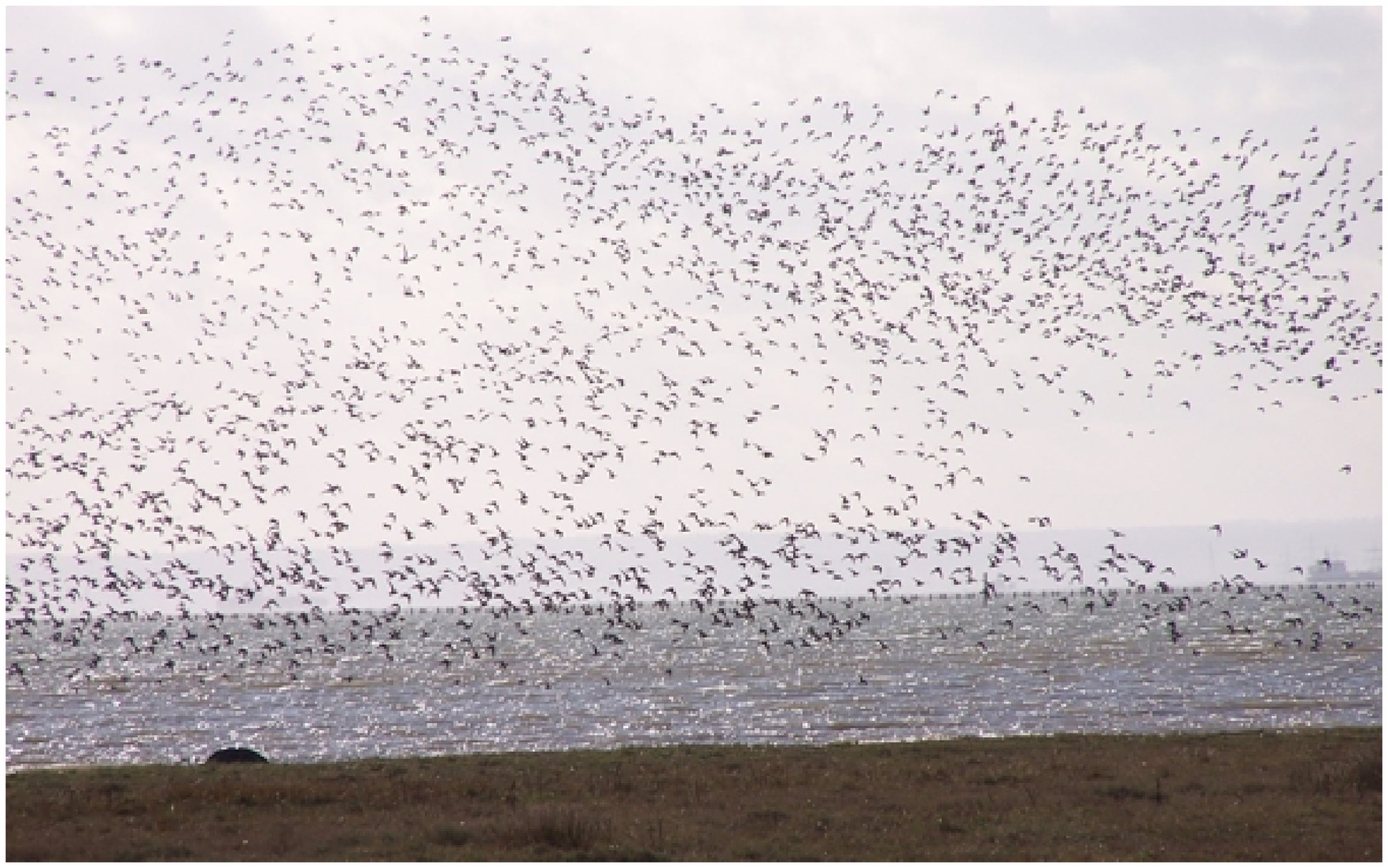}
\end{minipage}
\vspace{0.5cm}
\\
\Large
\hspace{\stretch{1}}Master Thesis\hspace{\stretch{1}}\strut
\vspace{0.5cm}
\\
Hans Henrik Sørensen
\vspace{0.8cm}
\\
\begin{minipage}[c]{4cm}
\includegraphics[width=4cm]{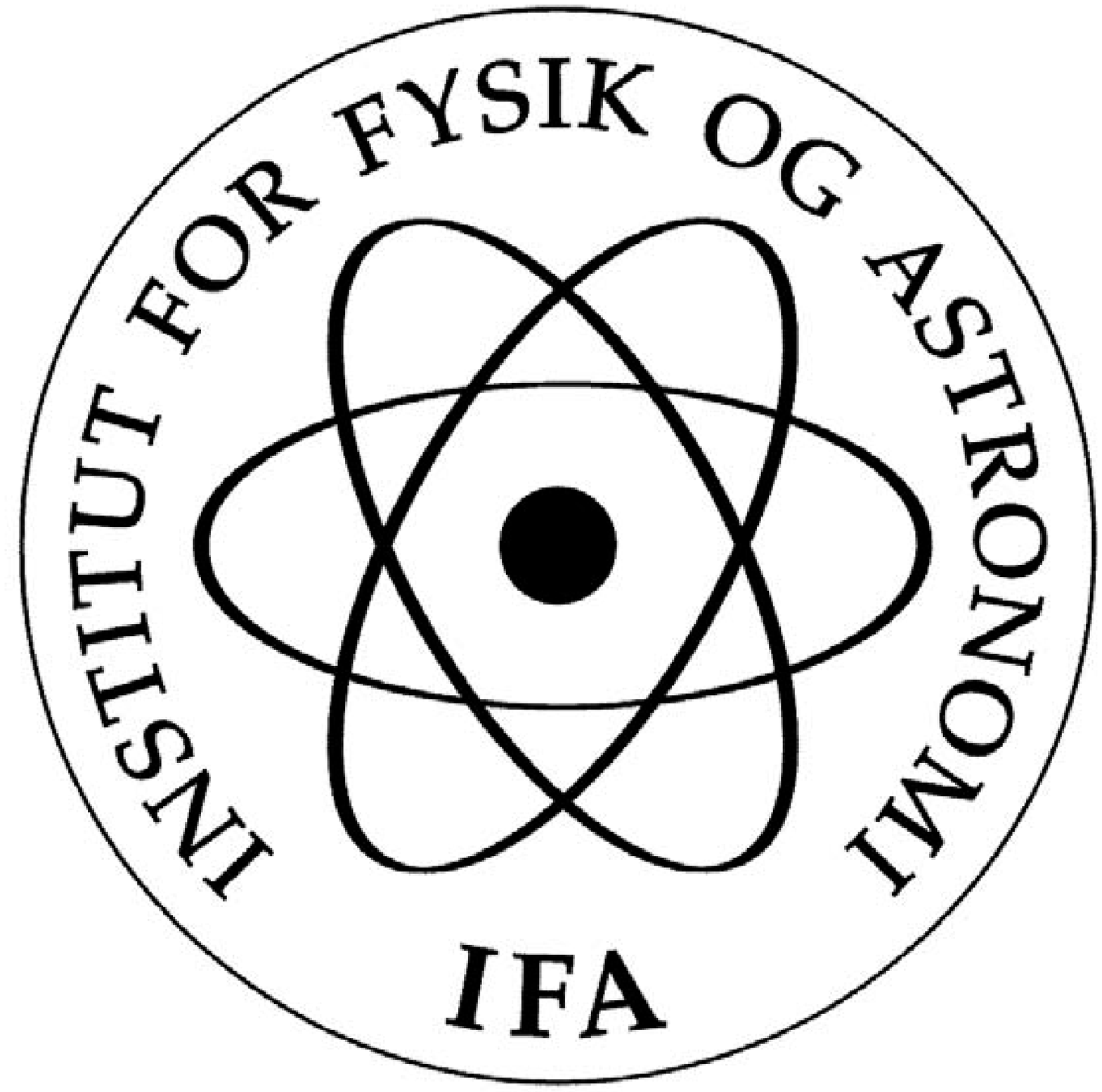}
\end{minipage}
\vspace{0.5cm}
\\
\large Institute of Physics and Astronomy
\\
University of Aarhus
\\
\vspace{0.3cm}
September 1, 2004
\end{center}

\pagenumbering{roman}
\author{Hans Henrik Sørensen}
\newpage
\tableofcontents
\newpage
\pagenumbering{arabic}
\setcounter{page}{1}

\chapter{Introduction}

Since the first realization of Bose-Einstein condensation (BEC) in trapped atomic vapors
in 1995 \cite{Anderson}, these systems have received increasing attention from both
experimental and theoretical efforts.
Most theoretical studies of these many-boson systems are based on the so-called
mean-field methods which accurately describe much of the dilute
condensate energetics.
Systems are termed as dilute when the average interparticle distance is much larger than
the range of the interaction. The main parameter characterizing the interaction in the dilute
regime is the s-wave scattering length, $a$, and the diluteness condition can be expressed
in terms of the density $n$ as $n|a|^3\ll 1$.

From the viewpoint of quantum many-body physics, the trapped atomic vapors are 
somewhat peculiar. Well
above the critical point of condensation, the gases are extremely dilute, and their description
as non-interacting bosons is very accurate. As the condensation sets in, the trapped
atoms are strongly compressed in real space. 
This makes it much more likely that the individual particles are within interaction range of
each other, and interactions suddenly become very important.
As a consequence the motion of the particles becomes correlated and both
the order of these correlations, that is, the number of particles which are simultaneously
within interaction range, and their overall influence will depend on the size of $n|a|^3$.

For the density ranges attained in BEC experiments, the diluteness condition may well
be broken, exploiting the large variation of the scattering length in the vicinity of
a Feshbach resonance \cite{Fesh}. In order to study this regime quantitatively, it is
compulsory to check the reliability of the theories adopted in the analysis. Such
work has recently been completed for the mean-field Gross-Pitaevskii (GP) theories,
typically reaching a validity estimate of $n|a|^3>10^{-3}$, \cite{Blume,DuBois,Blume2}
in combination with an instability criterion for negative scattering lengths
given by $|a|N/b_t < \sim0.67$, \cite{Sogo}.

The topic of this thesis is the description of correlations in many-boson systems
beyond the mean-field.
To achieve this, one has to consider not only BEC gas-type states
but also molecular-type states, since the instability criterion above designates the
threshold where the latter are formed.
In order to simulate a possible Feshbach resonance and break the
mean-field validity region, scattering lengths
should be allowed to cover $-\infty < a < 0$.
The main goal is then to develop the numerical tools needed to understand the nature of
interparticle relationships in BECs and estimate both the overall importance of such
correlation effects and the relative importance between the different orders.

The particular N-body technique chosen for this task is the Stochastic Variational Method (SVM).
This method provides a solid and arbitrarily improvable variational framework for the solution of
diverse bound-state problems. A special feature of the SVM is the strategy for optimization
of a variational trial function by ``controlled gambling''. This strategy has been proven to be
very effective for highly correlated nuclear few-body systems \cite{SVM}.

The computational load of the proposed numerical study with the SVM is excessive even for
the fastest modern computers, and numerical
calculations are only feasible for systems of three and four bosons.
However, since these constitute nontrivial BECs they may work as prototype systems
in the attempt to describe correlations.
In other words, the two key questions of the current study are:
\begin{itemize}
\item{To what extend does higher-order correlation effects influence the systems of three and
four trapped bosons ?}
\item{Can the main conclusions for the four-boson system be generalized to all many-boson systems ?
}
\end{itemize}
The remainder of this theoretical thesis seeks the answers to these questions.
In chapter 2 the basic theory needed in a variational treatment of an N-body system is reviewed.
The variational trial function is a crucial element of this approach. In chapter 3
it is shown how to include different levels of correlation explicitly in the functional form
of the trial funtion. The SVM is introduced in chapter 4, in combination with details of the
subsequent
application to the cases of the He atom and the N-boson systems.
Chapter 5 illustrates and discusses the numerical results, and the conclusions are collected
in chapter 6. The derivations of the matrix elements can be found in the appendices.

\section{Units and notation}

Where nothing else is indicated, the {\it Atomic Units}
\footnote{In this system of units the fundamental electron properties,
rest mass $m_e$, elementary charge $e$, Bohr radius $a_0$ and angular momentum $\hbar$
are all set equal to one atomic unit (a.u.). This makes the atomic units convenient
when describing the properties of electrons and atoms or particle systems of similar size.}
($m_e=e=a_0=\hbar=1$) are used when writing results.
Moreover, boldface is used for vectors ($\bm{a}$) and matrices ($\bm{A}$). The length of a vector is
written $|\bm{a}|$ while unit vectors have a hat ($\hat{\bm{a}}$).
The elements of vectors and matrices are always specified by subscripts ($A_{ij}$).
The elements of a set $\{\bm{A}\}$ are sometimes most conveniently denoted by
superscripts in parenthesis ($\bm{A}^{(k)}$) and sometimes by subscripts ($\bm{A}_k$).
Operators are assigned a wide hat ($\widehat{A}$).

With $\bm{x}$ being an $(N-1) \times 1$ one-column matrix of variables and
$\bm{x}^T$ it's $1 \times (N-1)$ one-row transposed matrix,
a quadratic form will be written
\begin{equation*}
\bm{x}^T \hspace{-2.0pt}A\bm{x} = \sum^{N-1}_{i=1} \sum^{N-1}_{j=1} A_{ij}\bm{x}_i \bm{x}_j
\end{equation*}
where $\bm{A}$ is a $(N-1) \times (N-1)$ symmetric matrix.

Matrix elements are written in Diracs {\it bra-ket} notation
\begin{equation*}
A_{ij} = \langle \psi_i | \widehat{A} | \psi_j \rangle
= \int \psi_i^*(\bm{\alpha}) \widehat{A} \psi_j(\bm{\alpha})d\bm{\alpha}
\end{equation*}
where $\bm{\alpha}$ denotes all the coordinates of the system
and $\psi_i$ and $\psi_j$ are square integrable functions having a
finite scalar product defined by the overlap integral
\begin{equation*}
\langle \psi_i | \psi_j \rangle
= \int \psi_i^*(\bm{\alpha}) \psi_j(\bm{\alpha})d\bm{\alpha}
\end{equation*}

Addition of angular momenta is expressed as direct products within square brackets. For example
the {\it vector-coupling} of the angular momenta $\widehat{\bm{J}}_1$ and $\widehat{\bm{J}}_2$,
each satisfying the eigenrelations $\widehat{\bm{J}}_i^2\phi_{J_iM_i}=J_i(J_i+1)\phi_{J_iM_i}$ and 
$\widehat{J}_{iz} \phi_{J_iM_i} = M_i\phi_{J_iM_i}$, is written
\begin{equation*}
[\phi_{J_1M_1} \otimes \phi_{J_2M_2}]_{JM} = \sum_{M1} \sum_{M2}
\langle J_1M_1J_2M_2|J_1J_2JM\rangle \phi_{J_1M_1} \phi_{J_2M_2}
\end{equation*}
where $\langle J_1M_1J_2M_2|J_1J_2JM\rangle$ are the Clebsch-Gordan coefficients.

Computational complexity is discussed in the {\it big oh} notation defined \cite{Papadimitriou}
\begin{equation*}
f(n)={\cal O}\big(g(n)\big)\Leftrightarrow f(n)\le c*g(n),\ {\rm for\ all}\ n \ge n_0>0
\ {\rm and} \ c>0,
\end{equation*}
which means informally that $f$ grows at the same rate as $g$ or slower.

\section{Computer programs}

In the course of this work computer programs have been developed in C++ to calculate
the numerical results (the usage information is listed in appendix \ref{prog}):

\begin{itemize}
\item{{\tt scatlen}: Calculates the scattering length for a two-body interaction of identical
bosons}
\item{{\tt bec}: Calculates the energy for a given state of an N-body system using the Stochastic
Variational Method.}
\end{itemize}

The source code for the programs {\tt bec} and {\tt scatlen} can be downloaded from my home page at:
www.phys.au.dk/$\sim$hansh. 
A few examples have been placed in footnotes throughout the thesis, indicating the explicit command
for the computation of a graph.

\chapter{Variational approach to N-body problems}\label{theory}

This chapter gives a brief description of how one solves the time
independent Schrödinger equation for N-body systems using a variational approach.
The Hamiltonian is introduced in the first section and
consists of terms for kinetic energy, two-body interaction and  possibly an external trapping field.
A section then presents the particular two-body interactions applied later in the thesis.
The most crucial points in the variational method is the introduction of a set of relative
coordinates and the construction of a flexible trial wave function from some appropriate basis
of functions. Both points are explained in subsequent sections and symmetrization is addressed.
Finally, it is shown how the variational theorem defined by the trial
function reduces to a generalized matrix eigenvalue problem and that accurate results can
be achieved with basis optimization procedures.

\section{Hamiltonian}

In the following, N-body systems of non-relativistic particles are considered, where the
$i$th particle has mass $m_i$, charge $c_i$, spin $s_i$, isospin $t_i$ and position vector
$\bm{r}_i$. The motion of the particles is given by the time-independent Schrödinger equation
\begin{equation}\label{Schr1}
\widehat{H}\Psi = E\Psi
\end{equation}
where the square integrable wave function, $\Psi(\bm{r}_1, \dots, \bm{r}_N)$, describes the state of
the system with the interpretation of $\Psi^* \Psi$ as the probability density \cite{BJ}.
Most often, this is the common starting point of both few-body and many-body treatments. However, the
magnitude of $N$ becomes significant in practice, especially when dealing with identical particles
(see below), making {\it ab initio} restrictions on $\Psi$ necessary for many-body systems.

As\-suming the particles move in an external field and that the only
particle-particle interaction is through local spin-independent two-body potentials
,$V_{ij}$, the Hamiltonian becomes
\begin{equation}\label{Hr}
\widehat{H} = \sum^N_{i=1} \Big[-\frac{\hbar^2}{2m_i}\widehat{\bm{\nabla}}_i^{2}
 + V_{ext}(\bm{r}_i)\Big] + \sum^N_{i<j}V_{ij}
\end{equation}
where $\widehat{\bm{\nabla}}_i^T= (\frac{\partial}{\partial r_{ix}},
\frac{\partial}{\partial r_{iy}}, \frac{\partial}{\partial r_{iz}})$ is the gradient
operator with respect to $\bm{r}_i$. The subsequent separation of the center-of-mass motion
from the intrinsic motion allows a translationally invariant description.
The following sections introduce the theory necessary to obtain a variational solution to
(\ref{Schr1}) for the N-body Hamiltonian (\ref{Hr}).

\subsection{Identical particles}
Many-body systems often contain a number of identical particles.
The indistinguishability of identical particles is obviously reflected in the Hamiltonian
(\ref{Hr}) by the symmetry of the operators entering. However, since it is written in
first quantization, $\widehat{H}$ does not distinguish whether identical particles are bosons or
fermions, and therefore this information should be added by hand to the wave function, $\Psi$,
in the form of a definite symmetry. For bosons the wave function is required
to be even under the interchange of any pair of particle coordinates while for fermions 
it should be odd \cite{BJ}. Achieving this in many-body problems is only feasible with
some restrictions on the form of $\Psi$ (see section \ref{sym}). Assuming the
proper symmetry is given, one may advantageously use a simpler Hamiltonian, given by
\begin{equation}\label{Hid}
\widehat{H}_{Id} =  N \Big[- \frac{\hbar^2}{2m}\widehat{\bm{\nabla}}_1^{2}
+ V_{ext}(\bm{r}_1)  + \½(N-1)V_{12} \Big]- \widehat{H}_{cm}
\end{equation}
since all terms in the sums of (\ref{Hr}) will contribute the same to the energy.
However, in the current study, the symmetric ``few-body'' Hamiltonian in (\ref{Hr})
is better suited for investigating correlations, and is therefore kept, also for identical
particles, in the following. The details of applying $\widehat{H}_{Id}$ with a symmetric
(Jastrow-type) wave function for many-boson systems can be found in ref. \cite{Fabrocini}, Chap. 2.

\section{Two-body interactions}\label{two-body}

In sufficiently dilute N-body systems only binary collisions contribute significantly to
the total energy and three- and more-body interactions can be almost completely ignored
(with one exception being three-body molecular recombination in atomic gases \cite{Esben}).
The Hamiltonian in (\ref{Hr}), introduced as the {\it ab initio} starting point of this chapter,
takes this simplification even further by assuming only spin-independent (central) two-body
interactions, $V_{ij}$. Such interactions are sufficient for the calculations of atomic
systems and gases of atoms presented later. Realistic nuclear models require spin-isospin
dependent interactions including (at least) three-body terms and are not considered here
\footnote{See ref. \cite{Fabrocini}, Chap. 10 for a treatment of the nuclear many-body problem
(Argonne/Urbana potentials).}.
Now follows a brief introduction to the two two-body potentials applied in this thesis
and of the concept of the s-wave scattering length which is essential in the description of 
BECs.

\subsection{Electrostatic interaction}
The interaction between particles carrying electric charge is the Coulomb force.
Like gravitation, this is a long range one-over-square-distance force, although many orders
of magnitude stronger. Atomic physics and solid physics, and for that matter the whole
of chemistry, can, in principle, be determined by this force combined with theories of relativity and
quantum mechanics \cite{BJ}. The corresponding interaction potential in SI units is
\begin{equation}
V(r_{12})=\frac{1}{4\pi\epsilon_0}\frac{q_1 q_2}{r_{12}}
\end{equation}
where $q_1$ and $q_2$ are the charges and $\epsilon_0$ is the permitivity of free space
$(a_0=e^2/4\pi\epsilon_0)$.
The {\it non-finite} long range character of the Coulomb potential makes solutions of the 
Schrödinger equation difficult in the case of scattering \cite{Landau}.
Moreover, it is not possible to define an s-wave scattering length (see below) for this
$1/r$-asymptotic potential.

\subsection{Atom-atom interaction}

The essential property of realistic interatomic interactions is that atoms repel at
short distances and attract when they are some distance apart. In the following
the focus will be
on the interaction of Rb from the {\it alkali atoms} group, since they in particular play a
key role in experiments on cold atomic gases (and consequently adopted as the default
particle in the numerical BEC calculations presented later).

\begin{figure}[bt]
\includegraphics[width=16cm]{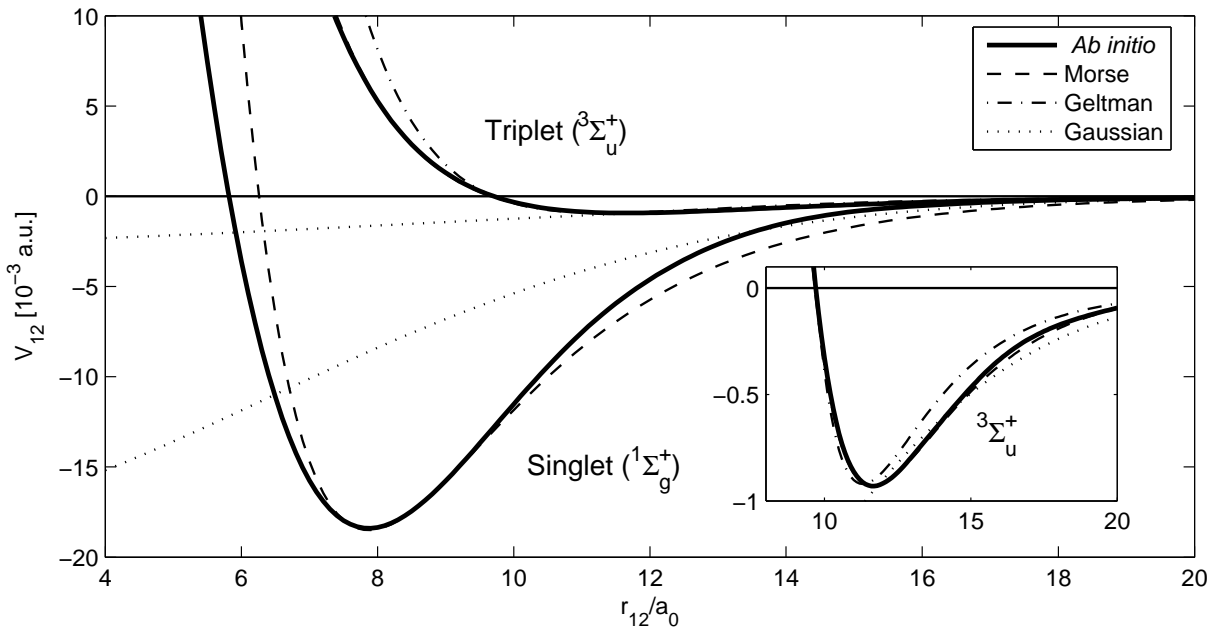}
\caption{
Rb-Rb interaction potentials (solid lines) as a function of the atomic separation.
Data is from Krauss and Stevens \cite{Krauss}.
The dashed lines corresponds to the Morse
model potential \cite{BJ}, $V(r_{12}) = D[(e^{-\alpha (r_{12}-r_0)} - 1)^2-1]$
with $r_0=7.87 a_0$, $\alpha=0.43 a_0^{-1}$ and $D=0.018$ a.u.,
for the singlet curve and $r_0=11.65 a_0$, $\alpha=0.35 a_0^{-1}$ and $D=0.00093$ a.u.,
for the triplet curve.
The dash-dotted graph represents the Geltman model for the triplet potential \cite{Geltman},
$V(r_{12}) = C_6[e^{-\alpha (r_{12}-r_c)}/r_c^6-1/r_{12}^6]$ with $C_6=4700$ a.u., $r_c=9.7 a_0$ and
$\alpha=0.9 a_0^{-1}$.
The inset shows the details of the triplet potential and the dotted lines indicate
the Gaussian models used in this work. }
\label{Rb}
\end{figure}

The ground state configuration of alkali atoms has all electrons but one occupying closed shells while
the remaining valence electron is in an s orbital of a higher shell \cite{Pethick}.
In the case of ground state collisions, the potential
energy depends solely on the internuclear separation and the orientation of
the two atoms valence electronic spins $(s_i=1/2)$ which couple into singlet $(S=0)$
or triplet $(S=1)$ configurations, where $\bm{S}= \bm{s}_1 + \bm{s}_2$, \cite{Burke}.
This is illustrated for $^{87}$Rb in figure \ref{Rb}. 
The solid curve is based on {\it ab initio} calculated data from \cite{Krauss} while the dashed
and dash-dotted curves correspond to model potentials
\footnote{The Morse model potential is an excellent approximation for the short range interaction
shape while the Geltman model potential (like the Lennard-Jones potential, \cite{BJ})
has the correct long-range behavior.} 
. For spin-polarized atoms one may assume that they interact only via
the triplet potential shown in details in the inset. 

\subsubsection{Gaussian model potential}
The key feature of the Rb-Rb potential in the current context, is that it can be assumed
to have a {\it finite range} in terms of scattering. 
This can be understood from considering the quantum mechanical
interactions of the Rb$_2$ constituents (a total of 74 identical electrons and two nuclei).
Clearly, when the nuclei are close enough (about 20Å) for the two electron clouds
to overlap the potential energy will depend greatly on the spin of the valence electrons.
This is due to the Pauli exclusion principle, since in the triplet configuration the spatial part
of the electron wave function must be anti-symmetric and so the overlap between electrons
is minimized in that case. Even at a distance there is residual overlap leading to a long range
exchange term
\footnote{This term has the form, $V_{ex}(r_{12})=\pm Ar_{12}^\alpha e^{-2\beta r_{12}}$, also found
in the Morse model potential.}.
However, when the nuclei are farther away, the energy due to the overlap of electrons
decreases exponentially and the interatomic potential is dominated by the {\it van der Waals} force
\footnote{As the electrons move, small fluctuations occur in the charge density surrounding each atom
so, in turn, one atom can polarize the other setting up an momentary dipole moment which
then attracts the first.}.
This {\it dispersion effect} can be expanded in a multipole expansion such as
\cite{Marinescu}
\begin{equation}\label{dispersion}
V_{disp}(r_{12})=-\frac{C_6}{r_{12}^6}-\frac{C_8}{r_{12}^8}-\frac{C_{10}}{r_{12}^{10}}-\cdots
\end{equation}
where the leading (van der Waals) term is $1/r_{12}^6$. With such an infinite 
tail on the potential it seems invalid to talk about a finite range and scattering length.
Fortunately it can be shown that for scattering via power-law potentials, $1/r^n$, the decrease
is fast enough to be considered of finite range provided that $n>3$, \cite{Pethick}.

In this work, the atom-atom interaction is represented by a simple Gaussian model
potential, $V(x) = V_0 e^{-x^2/b^2}$, of finite range (the dotted lines in fig. \ref{Rb}).
For weak interactions in the low-energy (i.e. ultra-cold) limit, the properties of the
two-body interaction are basically determined by the scattering length, $a$, introduced in the
next section, alone.
This means, that the exact shape of the potential is insignificant (see e.g. section \ref{Born})
and that the apparent lack of a hard core at small $r_{12}$ in the Gaussian model is acceptable.
More details of the Rb-Rb interaction can be found in \cite{Burke}.

\subsection{Scattering length for finite interactions}
\begin{figure}[bt]
\includegraphics[width=16cm]{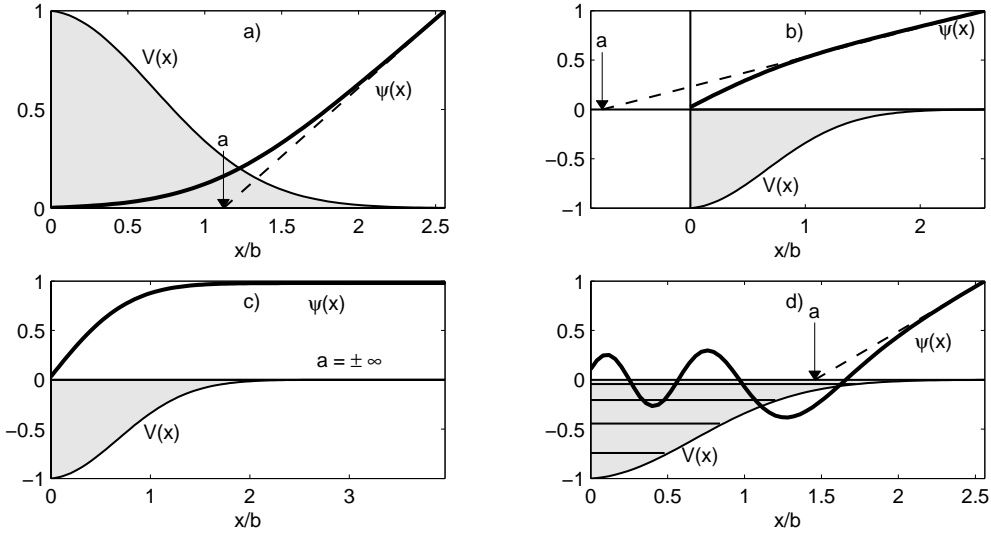}
\caption{
Scattering lengths, $a$, and wave functions, $\Psi(x)$, for Gaussian potentials
$V(x) = V_0 e^{-x^2/b^2}$, with $b=18.9$ fixed and different strengths:
a) repulsive, $V_0 = 2.11 \times 10^{-7}$; 
b) weak attractive, $V_0 = -2.11 \times 10^{-8}$;
c) more attractive at bound state threshold, $V_0 = -4.743 \times 10^{-8}$;
d) strong attractive with 4 bound states, $V_0 = -2.11 \times 10^{-6}$.
The functions have been scaled to fit $[-1;1]$. Other numbers are in atomic units.
}
\label{scatlengths}
\end{figure}

The following is a very brief account of basic scattering theory which can be found in
details in Refs. \cite{BJ,Sak}.
Consider the situation of two isolated particles with masses $m_1$ and $m_2$ that interact via
a central potential $V(r_{12})$, where $r_{12} = |\bm{r}_1 - \bm{r}_2|$
is the interparticle distance. Further assume that the interaction vanishes rapidly
(faster than $\propto r_{12}^{-3}$)
for large separations, i.e. $V(r_{12}) \rightarrow 0$ for $r_{12} \rightarrow \infty$.
As outlined in section \ref{relative}, the motion of the particles separates into the trivial
center-of-mass motion and the relative motion described by a single coordinate wave 
function, $\Psi(\bm{x})$, satisfying the Schrödinger equation
\begin{equation}\label{two-schr}
\bigg[ -\frac{\hbar^2}{2\mu}\widehat{\bm{\nabla}}_{\bm{x}}^2 + V(x)\bigg] \Psi(\bm{x}) 
=E \Psi(\bm{x}) 
\end{equation}
where $\bm{x} = \bm{r}_{12}$ and $\mu = m_1m_2/(m_1+m_2)$ is the reduced mass.
Solutions with $E<0$ correspond to bound states of the potential.
Scattering is described by the Lippmann-Schwinger solutions \cite{Sak}
with positive energy $E = \hbar^2 k^2/2\mu$, and the asymptotic form
\begin{equation}
\Psi_{\bm{k}}^{(+)}(\bm{x}) \stackrel{x \rightarrow \infty}{=}
\frac{1}{(2\pi)^{3/2}} \big[ e^{i\bm{k}\cdot\bm{x}} + f(\bm{k}^\prime,\bm{k})\frac{e^{ikx}}{x} \big]
\end{equation}
corresponding to the sum of an incoming plane wave with relative momentum $\hbar\bm{k}$ and a
scattered spherical wave (i.e. the (+) superscript) with amplitude
\begin{equation}\label{scatamp}
f(\bm{k}^\prime,\bm{k})\equiv -\frac{4\pi^2\mu}{\hbar^2}
\int d\bm{x}^\prime \frac{e^{-i\bm{k}^\prime\cdot\bm{x}^\prime}}{(2\pi)^{3/2}}V(x^\prime)
\Psi_{\bm{k}}^{(+)}(\bm{x}^\prime)
\end{equation}
For a spherically symmetric potential the scattering amplitude only depends on the angle,
$\theta$, between the relative momentum of the particles before and after the scattering, 
$f(\bm{k}^\prime,\bm{k})\equiv f(k,\theta)$.
In the low energy limit, $k\rightarrow 0$, where isotropic s-wave scattering is dominant
\footnote{A partial wave expansion of $\Psi(\bm{x})$ in Legendre polynomials, $P_l(cos \theta)$,
gives a radial Schrödinger equation where the effective potential includes a centrifugal
barrier, i.e. the term $\hbar^2 l(l+1)/(2\mu x^2)$, \cite{Landau}. Thus waves with energies much
lower than this barrier are simply reflected leaving only the s-wave $(l=0)$ contribution.
},
the scattering amplitude approaches a constant,
$f(0,0) = -a$, and the wave function reduces to
\begin{equation}\label{scatred}
\Psi_{\bm{k}}^{(+)}(\bm{x}) \stackrel{k \rightarrow 0}{=} \Psi^{(+)}(x) = 1 - \frac{a}{x}
\end{equation}
The constant $a$ is the s-wave scattering length and can thus be determined as the interception
of the asymptotic wave function and the $\hat{\bm{x}}$ axis, that is $\Psi(a)=0$ for
the zero energy solution to the Schrödinger equation.
Figure \ref{scatlengths}
\footnote{The numerical data for fig. \ref{scatlengths} is produced by calling
the program {\tt scatlen}. Graph a) is generated with the command lines:
{\tt scatlen -V0 2.11e-7 -compare}, {\tt scatlen -V0 2.11e-7 -printwave} and
{\tt scatlen -V0 2.11e-7 -printpot}.
For b), c) and d) just change {\tt 2.11e-7} to the corresponding values of $V_0$.
}
demonstrates this in the case where two identical particles are interacting via a finite Gaussian
potential, $V(x) = V_0 e^{-x^2/b^2}$. The scattering length is always positive and finite 
for repulsive interactions, $V_0 > 0$, while for attractive interactions, $V_0 < 0$, it can be both
negative and positive and becomes divergent when changing sign. This behavior (zero-energy
resonance) occurs each time the potential is just deep enough to support a new bound state.

\section{Relative coordinates}\label{relative}
The most convenient way to remove the center-of-mass motion is to express the Hamiltonian
in terms of relative coordinates that do not change when the system moves or rotates as a
whole. The obvious choice is the scalar {\it interparticle distances}
\begin{equation}\label{RelCoor}
r_{ij}=|\bm{r}_i-\bm{r}_j |, \ \ i \neq j = 1, \dots ,N
\end{equation}
where $\bm{r}_i$ is the position of the $i$th particle, giving $N(N-1)/2$ relative but dependent
coordinates. Choosing one relative coordinate, $r_{12}$,
in the two-body case is trivial. In three-body problems the truly
independent positive {\it perimetric coordinates}
\begin{equation}
u_i=\½(r_{ik}+r_{ij}-r_{jk}),\ \ i\ne j\ne k=(1,2,3),
\end{equation}
are often preferred as this simplifies integral evaluations over the coordinates 
(used in appendix \ref{basicint}).
Since there is only $3N-6$ internal space degrees-of-freedom in an $N$-body problem (for $N>2$),
the set of scalar relative coordinates include unnecessary extra coordinates when 
$N(N-1)/2>3N-6 \Leftrightarrow N>4$. This complicates the use of interparticle
coordinates in many-body systems significantly \cite{Frolov5}.

A different approach, also convenient if $N>4$, is to introduce a set of re\-la\-tive vector
coordinates $\bm{x}^T= (\bm{x}_1,\bm{x}_2,\dots,\bm{x}_{N-1})$ and the explicit
center-of-mass coordinate $\bm{x}_N$. They are related to the single-particle coordinates 
$\bm{r}^T= (\bm{r}_1,\bm{r}_2,\dots,\bm{r}_N)$ by a linear transformation
\begin{equation}\label{xtrans}
\bm{x}_i = \sum^N_{j=1} U_{ij} \bm{r}_j, \ \ i=1, \dots, N
\end{equation}
written in matrix form as $\bm{x}=\bm{U}\bm{r}$, where $\bm{U}$ is a suitable $N \times N$
transformation matrix
\footnote{
Obviously one can readily generalize the scalar definition (\ref{RelCoor}) to an equivalent vector
description with $\bm{x}^T=(\bm{r}_{12},\bm{r}_{13},..,\bm{r}_{1N},
\bm{r}_{23},..,\bm{r}_{2N},..,\bm{r}_{(N-1)N}, \bm{x}_M)$, given by an appropriate linear $M \times N$
transformation of the single-particle coordinates.}
.
A widely used choice for $\bm{x}$, which is also employed for the many-body problems considered
in this thesis, is the {\it Jacobi coordinate set} \cite{Fabrocini} defined by
\begin{equation}\label{Jacobi}
\begin{aligned}
\bm{x}_i &= \sqrt{\mu_i}(\bm{C}_i - \bm{r}_{i+1}), \ \ i = 1, \dots, N-1
\\\bm{x}_N &= \bm{C}_N,
\end{aligned}
\end{equation}
where $\bm{C}_i$ is the center-of-mass and $\mu_i = \frac{m_{i+1}m_{12\cdots i}}{m_{12\cdots i+1}}$
the reduced mass of the first $i$ particles.
The corresponding transformation matrix is
\begin{equation}\label{UJ}
\bm{U}_J =
\begin{pmatrix} 
\sqrt{\mu_1} & -\sqrt{\mu_1} & 0 & \cdots & 0 \\ 
\sqrt{\mu_2}\frac{m_1}{m_{12}} & \sqrt{\mu_2}\frac{m_2}{m_{12}} & -\sqrt{\mu_2} & \cdots & 0 \\ 
\vdots & \vdots & \vdots & \ddots & \vdots \\ 
\sqrt{\mu_{N-1}}\frac{m_1}{m_{12 \cdots N-1}} & \sqrt{\mu_{N-1}}\frac{m_2}{m_{12 \cdots N-1}} 
& \sqrt{\mu_{N-1}}\frac{m_3}{m_{12 \cdots N-1}} & \cdots & -\sqrt{\mu_{N-1}} \\ 
\frac{m_1}{m_{12 \cdots N}} & \frac{m_2}{m_{12 \cdots N}} & \frac{m_3}{m_{12 \cdots N}} & \cdots &
\frac{m_N}{m_{12 \cdots N}} \\ 
\end{pmatrix}
.
\end{equation}
where the short notation means $m_{12 \cdots i} = m_1+m_2+ \cdots m_i$,
making $m_{12 \cdots N}$ the total mass of the system. A specific set of Jacobi coordinates,
$\{\bm{x}_i\}$, are related to the interparticle distances, $r_{ij}$, and the hyperradius,
$\rho$, through the relation \cite{Ole}
\begin{equation}\label{hyperradius}
\rho^2 \equiv \frac{1}{N} \sum_{i<j}^{N} \bm{r}_{ij}^2 = \sum_{i=1}^{N} \bm{r}_i^2 - N \bm{C}_N^2
= \sum_{i=1}^{N-1} \bm{x}_i^2
\end{equation}
Moreover, since the first Jacobi coordinate, $\bm{x}_1$, of the set defined in (\ref{UJ})
is equal to
$\sqrt{\mu_1}(\bm{r}_1-\bm{r}_2) \equiv \sqrt{\mu_1}\bm{r}_{12}$, the two-body interaction,
$V_{12}$, depending on this interparticle-distance, has a simple form
$V(\bm{x}_1/\sqrt{\mu_1})$. By permutations of the particle labels
$(1,2, \cdots, N)$, and the corresponding columns in (\ref{UJ}), one can generate different
Jacobi coordinate sets $\bm{x}^{(p)}$, known as arrangements or partitions, where the
first Jacobi coordinate is $\sqrt{\mu_1}\bm{r}_{ij}$ and the form of $V_{ij}$ is simple.
This is an advantage when trying to obtain analytical expressions for the integrals
(see (\ref{Hmatrix}) and (\ref{Smatrix}) derived in the next section) that have to be evaluated
in the variational method.

\subsection{Separation of the center-of-mass}

The many-body Hamiltonian (\ref{Hr}) written in terms of relative coordinates separates
into a translationally invariant part and a part involving only the center-of-mass coordinate.
Corresponding to the change of coordinates (\ref{xtrans}) the single-particle
gradient operators
$\widehat{\bm{\nabla}}^T=(\widehat{\bm{\nabla}}_1,\widehat{\bm{\nabla}}_2,
..,\widehat{\bm{\nabla}}_N)$,
entering the kinetic energy part, are transformed by
\begin{equation}\label{ptrans}
\widehat{\bm{\nabla}} = \bm{U}^T \widehat{\bm{\nabla}}_{\bm{x}}
\end{equation}
and since the transformation matrix $\bm{U}$ can be assumed to satisfy the relations
$U_{Ni}= \frac{m_i}{m_{12 \dots N}}$ and $\sum^N_{j=1} U_{ij} = \delta_{Ni}$ \cite{SVM},
easily verified for $\bm{U}_J$, one has
\begin{align}\label{cut3}
-\sum^N_{i=1} \frac{\hbar^2}{2m_i} \widehat{\bm{\nabla}}_i^{2}&= -\frac{\hbar^2}{2}
\sum^N_{i=1} \frac{1}{m_i}
\big(\sum^N_{k=1}U_{ki}\widehat{\bm{\nabla}}_{\bm{x}_k}\big)
\big(\sum^N_{l=1}U_{li}\widehat{\bm{\nabla}}_{\bm{x}_l}\big)
\nonumber
\\&= -\frac{\hbar^2}{2} \sum^{N-1}_{k=1}\sum^{N-1}_{l=1}\sum^{N}_{i=1}
\frac{U_{ki}U_{li}}{m_i} \widehat{\bm{\nabla}}_{\bm{x}_k} \widehat{\bm{\nabla}}_{\bm{x}_l}
-\frac{\hbar^2}{2m_{12 \cdots N}}{\widehat{\bm{\nabla}}_{\bm{x}_N}}^2
\nonumber
\\&= -\frac{\hbar^2}{2}  \sum^{N-1}_{k=1}\sum^{N-1}_{l=1} \Lambda_{kl}
\widehat{\bm{\nabla}}_{\bm{x}_k} \widehat{\bm{\nabla}}_{\bm{x}_l} + \widehat{T}_{cm}
\end{align}
where $\widehat{T}_{cm} = \frac{\widehat{\bm{p}}_{cm}^2}{2m_{12 \dots N}}
= -\frac{\hbar^2}{2m_{12 \dots N}}\widehat{\bm{\nabla}}_{\bm{x}_N}^2$
is the center-of-mass kinetic energy
\footnote{Corresponding to the total momentum
$\widehat{\bm{p}}_{cm}=-\sum^N_{k=1}i\hbar\widehat{\bm{\nabla}}_k =
-i\hbar\sum^N_{j=1} \delta_{Nj} \widehat{\bm{\nabla}}_{\bm{x}_j} =
-i\hbar\widehat{\bm{\nabla}}_{\bm{x}_N}$.},
and 
\begin{equation}\label{Lambda}
\Lambda_{kl} = \sum^{N}_{i=1} \frac{U_{ki}U_{li}}{m_i}, \ \ \ k,l=1, \dots, N-1
\end{equation}
The external field potential, $V_{ext}$, separates in a similar way when applying transformation
(\ref{xtrans}), see section \ref{BEC}.
Thus the Hamiltonian (\ref{Hr}) can also be expressed in terms of the quadratic form
\begin{equation*}
\widehat{\bm{\nabla}}_{\bm{x}}^T \bm{\Lambda} \widehat{\bm{\nabla}}_{\bm{x}} =
\sum^{N-1}_{k=1}\sum^{N-1}_{l=1} \Lambda_{kl} \widehat{\bm{\nabla}}_{\bm{x}_k}
\widehat{\bm{\nabla}}_{\bm{x}_l}
\end{equation*}
as
\begin{equation}\label{Hj}
\widehat{H} = -\frac{\hbar^2}{2}\widehat{\bm{\nabla}}_{\bm{x}}^T \bm{\Lambda}
\widehat{\bm{\nabla}}_{\bm{x}}
+\sum^{N-1}_{i=1} V_{ext}(\bm{x}_i)
+ \sum^N_{i<j}V_{ij} + \widehat{H}_{cm}
\end{equation}
with $V_{ext}$ and $V_{ij}$ depending on the relative coordinates
$\{\bm{x}_1,\bm{x}_2, \dots, \bm{x}_{N-1}\}$ and
$\widehat{H}_{cm} = \widehat{T}_{cm}+ V_{ext}({\bm{R}})$. Explicit insertion of the Jacobi
transformation matrix, $\bm{U}_J$, in the expression (\ref{Lambda}) for $\bm{\Lambda}$, produces
the important result $\bm{\Lambda} \equiv \bm{I}$. This means, that in the specific case of the
Jacobi coordinates the quadratic form is dissolved and only terms of the Laplacians,
$\widehat{\bm{\nabla}}_{\bm{x}_i}^2$, have to be considered in a
calculation of the kinetic energy. 

\section{Matrix representation}
The general {\it stationary-state} solution, $\Psi$, to the Schrödinger equation (\ref{Schr1})
will be a linear superposition of the eigenfunctions $\Psi_n$ of $\widehat{H}$:
\begin{equation}\label{super}
\Psi = \sum^\infty_{n=1} a_n \Psi_n,
\end{equation}
where $\Psi_n$ satisfies
\begin{equation}\label{eigen}
\widehat{H} \Psi_n = E_n \Psi_n, \ \ \ n=1, 2, \dots .
\end{equation}
If the Hamiltonian (\ref{Hj}) is Hermitian, the eigenvalues $E_n$ are real and
the eigenfunctions, $\Psi_n$, called the energy eigenstates, form a complete and
or\-tho\-go\-nal set $\{\Psi_n\}$, \cite{BJ}.
With focus on bound-state solutions, in particular the ground state and lowest excited
states, one has to find the lowest discrete energies, $E_n$, and corresponding eigenstates, $\Psi_n$,
from eq. (\ref{eigen}). Unfortunately, except for the two-body cases, the explicit form of the
eigenstates is not known a priori, making it impossible to solve the eigenvalue problem
analytically.

The best alternative is to consider a finite set of known functions
$\{\psi_1, \psi_2, \dots, \psi_K\}$,
that are linearly independent and possibly non-orthogonal.
A general function in the space $\cal{V}_K$ spanned by this set can be written
\begin{equation}\label{func}
\Psi = \sum_{i=1}^K c_i \psi_i
\end{equation}
The state vector $\bm{c}$ uniquely defines $\Psi$ in the function space $\cal{V}_K$.
Inserting this form into the Schrödinger equation gives
\begin{equation}\label{eigen2}
\sum_{j=1}^K (\widehat{H}-\epsilon) c_j \psi_j = 0,
\end{equation}
which is the eigenvalue problem for $\widehat{H}$ inside $\cal{V}_K$.
From this restricted problem one can determine
eigenvalues $\epsilon_1,\epsilon_2,\dots,\epsilon_K$
and corresponding state vectors $\bm{c}^{(1)},\bm{c}^{(2)},\dots,\bm{c}^{(K)}$
that are approximations to the exact solution (\ref{eigen}).
As will be discussed in the next section it is actually the best solution
within $\cal{V}_K$ from a variational standpoint.

Equation (\ref{eigen2}) is conveniently expressed in a matrix representation by
multiplying from the left by $\psi_i^*$ and integrating
over all coordinates $\bm{\alpha}$ that $\psi_i$ depend on, giving
\begin{equation}
\sum_{j=1}^K (H_{ij}-\epsilon S_{ij}) c_j = 0, \ \ \ i=1,2,\dots, K.
\end{equation}
where
\begin{equation}\label{Hmatrix}
H_{ij} = \langle \psi_i | \widehat{H} | \psi_j \rangle
= \int \psi_i^*(\bm{\alpha}) \widehat{H} \psi_j(\bm{\alpha})d\bm{\alpha}
\end{equation}
are the elements of the $K \times K$ {\it Hamiltonian matrix} and
\begin{equation}\label{Smatrix}
S_{ij} = \langle \psi_i | \psi_j \rangle
= \int \psi_i^*(\bm{\alpha})  \psi_j(\bm{\alpha})d\bm{\alpha}
\end{equation}
are the elements of the $K \times K$ {\it overlap matrix}.
In this way, the problem of determining the eigenvalues of the
operator $\widehat{H}$ in (\ref{eigen2}) has been transformed
to a generalized eigenvalue problem of square matrices
\footnote{The connection between the algebra of linear operators and  square matrices is quite
fundamental, see \cite{Dahl} sec. 5.10.}
, most elegantly written
\begin{equation}\label{geneigen}
\bm{H}\bm{c} = \epsilon\bm{S}\bm{c}. 
\end{equation}
If the basis functions are chosen to be orthogonal, $\bm{S}$ becomes the identity matrix, and
equation (\ref{geneigen}) reduces to a standard eigenvalue problem.

\section{The linear variational method}
In this thesis, a variational method is used to obtain the approximate bound-state energies
and wave functions of systems described by the N-body Hamiltonian (\ref{Hr}). It is linked to the fact
that an arbitrary wave function corresponds to an energy
higher or equal to the true ground state energy.
The so-called {\it variational theorem}
states
\begin{equation}\label{vartheorem}
{\cal E} \equiv \frac{\langle \Psi | \widehat{H} | \Psi \rangle}
{\langle \Psi | \Psi \rangle} \ge E_1
\end{equation}
for any square-integrable $\Psi$.  The functional ${\cal E}$ is
the expectation value of $\widehat{H}$ and the equality holds only
if $\Psi$ is the ground state of $\widehat{H}$ with the eigenvalue $E_1$. 
A proof of this theorem is elementary and can be found in textbooks
on quantum mechanics, e.g. \cite{BJ} p. 116.

The variational theorem is the basis of the widely used
{\it Rayleigh-Ritz variational method}.
The idea behind this method is to choose a trial function as $\Psi$ that
depends on a number of variational parameters.
Evaluating the expectation value ${\cal E}$ yields a function of these parameters,
and by minimizing with respect to the parameters, one obtains the best approximation to 
$E_1$ that the explicit form of $\Psi$ allows.

The {\it linear variational method} is a variant of the Rayleigh-Ritz method in which
one works with trial functions of the form (\ref{func}), described in the previous
section. The expansion coefficients
$\bm{c}^T = \{c_1, c_2, \dots, c_K\}$ serve as the variational parameters and
minimization consists of demanding that
\begin{equation}\label{cond}
\frac{\partial {\cal E}}{\partial c_i} = 0 \ \ \ {\rm and} \ \ \ 
\frac{\partial {\cal E}}{\partial c_i^*} = 0, 
\end{equation}
for all $i=1,2,\dots, K$.
Considering first the latter condition in
(\ref{cond}) by differentiating ${\cal E}$ with respect to $c_i^*$ gives
\begin{align}
\nonumber
\frac{\partial {\cal E}}{\partial c_i^*}
&=\frac{
\langle \Psi | \Psi \rangle \frac{\partial}{\partial c_i^*}
\langle \Psi | \widehat{H} | \Psi \rangle
-\langle \Psi | \widehat{H} | \Psi \rangle \frac{\partial}{\partial c_i^*}
\langle \Psi | \Psi \rangle
}
{\langle \Psi | \Psi \rangle^2}
\\&=
\nonumber
\frac{
\frac{\partial}{\partial c_i^*}
\langle \Psi | \widehat{H} | \Psi \rangle
-{\cal E} \frac{\partial}{\partial c_i^*}
\langle \Psi | \Psi \rangle
}
{\langle \Psi | \Psi \rangle}
\end{align}
For this expression to vanish the numerator must be zero.
Introducing the expansion (\ref{func}) as $\Psi$ in the matrix element expressions, one has
\begin{equation}
\langle \Psi | \widehat{H} | \Psi \rangle=
\sum_{i=1}^K \sum_{j=1}^K c_i^* c_j \langle \psi_i | \widehat{H} | \psi_j \rangle=
\sum_{i=1}^K \sum_{j=1}^K c_i^* c_j H_{ij}
\end{equation}
and
\begin{equation}
\langle \Psi | \Psi \rangle=
\sum_{i=1}^K \sum_{j=1}^K c_i^* c_j \langle \psi_i | \psi_j \rangle=
\sum_{i=1}^K \sum_{j=1}^K c_i^* c_j S_{ij}
\end{equation}
where $H_{ij}$ and $S_{ij}$ are again the elements of the Hamiltonian and overlap
matrices, and the condition $\frac{\partial {\cal E}}{\partial c_i^*} = 0$ reduces to
\begin{equation*}
\frac{\partial}{\partial c_i^*}
\langle \Psi | \widehat{H} | \Psi \rangle
-{\cal E} \frac{\partial}{\partial c_i^*}
\langle \Psi | \Psi \rangle
=\sum_{j=1}^K (H_{ij}-{\cal E} B_{ij}) c_j = 0, \ \ \ i=1,2,\dots, K
\end{equation*}
The condition $\frac{\partial {\cal E}}{\partial c_i} = 0$ will reduce to the complex
conjugate of this equation \cite{Dahl} and hence gives no new information. Thus imposing
the conditions (\ref{cond}) on the expectation value ${\cal E}$ has produced precisely
the same matrix eigenvalue problem
\begin{equation}\label{eigen3}
\bm{H}\bm{c} = \epsilon\bm{S}\bm{c}
\end{equation}
that was derived in the previous section. This is an important
connection and means that by choosing an arbitrary basis $\{\psi_1, \psi_2, \dots, \psi_K\}$
of known functions and solving eq. (\ref{eigen3}), also called the {\it secular equation},
one will in fact get the best approximate solution inside ${\cal V}_K$.
One could say, that the minimization with respect to the parameters $\{c_1, c_2, \dots, c_K\}$
is implicit in the solution. Of course, the functions $\psi_i$ can be made
dependent on additional {\it nonlinear} variational parameters, giving further flexibility
to the trial function.

The variational theorem implies that the lowest of the eigenvalues
determined by eq. (\ref{eigen3}) will be an upper bound
to the real ground state energy $E_1$. In turn, all the eigenvalues
$\epsilon_i, i=1,2,\dots,K$,
are upper bounds to eigenstate energies of the full Hamiltonian (see \cite{SVM}, theorem 3.3).
Arranging in increasing order
the $K$ eigenvalues $\epsilon_1 \le \epsilon_2 \le \dots \le \epsilon_K$
of the truncated problem and the discrete eigenvalues $E_1 \le E_2 \le \dots$
of the full problem
(\ref{eigen}), it can be shown that
\begin{equation}\label{varthe2}
E_1 \le \epsilon_1, E_2 \le \epsilon_2, \dots, E_K \le \epsilon_K
\end{equation}
Expanding ${\cal V}_K$ by increasing the number of functions in the basis
will  bring (\ref{eigen3}) closer to the full Hilbert space
problem, and obviously improve on the approximate eigenvalues $\epsilon_i$ by lowering
them towards the exact values $E_i$
\footnote{This can be explicitly proven, see \cite{SVM}, p.27.}.

\section{Basis functions}
A crucial point when using the linear variational method is the choice of basis functions.
An expansion in the basis should give a good representation of the physical shape
of the wave function for the quantum system in question.
It is important, in general, that two basic requirements are satisfied:
\begin{itemize}
\item{The basis should form a {\it complete set} so
that the result obtained by a systematic increase of the number of basic functions
will converge to the exact eigenvalue.}
\item{Furthermore, all matrix elements should be {\it analytically calculable}, for the variational
approach to be practical.}
\end{itemize}
In addition, to solve an N-particle problem accurately and with a high convergence rate,
the trial function, $\Psi = \sum_{i=1}^K c_i \psi_i$, and hence the basis functions, $\{\psi_i\}$,
should describe the correlation between the particles well, have the proper symmetry and
encompass the appropriate degrees of freedom, e.g. orbital and spin angular momenta.
In this thesis, it is assumed that a basis function with total angular momentum $J$ and
projection $M$, can be written in the form \cite{SVM, Kukulin2}
\begin{equation}\label{BF}
\psi_k= \widehat{\cal P} \Big\{ \Phi_k(\bm{x})[ \theta_{LM_L}(\hat{\bm{x}})
\otimes \chi_{SM_S}]_{JM} \hspace{2.0pt} \eta_{TM_T} \Big\},
\end{equation}
where $\widehat{\cal P}$ is a sum of permutation operators ensuring the proper symmetry,
$\Phi_k(\bm{x})$ describes the spatial dependence,
$\theta_{LM_L}(\hat{\bm{x}})$ specifies the orbital motion with definite angular momentum $L$,
and $\chi_{SM_S}$ and $\eta_{TM_T}$ are the spin and isospin parts
\footnote{
Parts for other degrees of freedom, like color and flavor,
can be added correspondingly.}.
Since all correlations between particles in the systems treated later are
state-independent, they can be fully represented in the spatial part, $\Phi_k(\bm{x})$,
of $\psi_k$, as discussed in detail in chapter \ref{correlations}.
More elaborate descriptions of the angular momentum parts can be found in appendix \ref{AppBasis}.

\subsection{Symmetry}\label{sym}

From experiments it is known that particles of zero or integral spin, such as
the photon and $^4$He, are bosons. Particles with half-integral spin values, such as electrons and
nucleons, are fermions. Atoms constitute bosons when they contain an equal number of nucleons,
otherwise fermions. When a system consists of a number of {\it identical and
 indistinguishable particles} the wave function must have the proper symmetry with respect
to any interchange of the space and spin coordinates of the identical particles. 
The proper symmetry of the trial wave function can be achieved by operating on the basis functions
with the operator \cite{Fabrocini};
\begin{equation}\label{P}
\widehat{\cal P} =\sqrt{\frac{1}{N!}}\sum_P \alpha_P \widehat{P}
\end{equation}
using $\alpha_P=1$ for identical bosons and $\alpha_P=(-1)^p$,
where p=0,1 is the parity of the permutation, $P$, for identical fermions.  
Here, the permutation operator, $\widehat{P}$, permutes the variable indices $(1, 2, \dots, N)$
of identical particles to $(p_1, p_2, \dots, p_N)$ and the summation over $P$ includes all necessary
permutations. Thus $\widehat{\cal P}$ corresponds to a symmetrizer ($\widehat{\cal S}$) for bosons
and an antisymmetrizer ($\widehat{\cal A}$) for fermions.

Obviously, the permutation operator, $\widehat{\cal P}$, commutes with the symmetric many-body
$\widehat{H}$ in (\ref{Hr}), but, except for the case $N=2$, the
$N!$ different permutation operators do not commute among themselves. This means, that an
eigenfunction of $\widehat{H}$ is not necessarily an eigenfunction of all $\widehat{P}$. Only
a {\it totally symmetric} eigenfunction, $\Psi_S$, or a {\it totally antisymmetric}
eigenfunction, $\Psi_A$, can be common eigenfunctions of $\widehat{H}$ and all $\widehat{P}$.
The two types of wave functions, $\Psi_S$ and $\Psi_A$, are thought to be sufficient to describe
all systems of identical particles
\footnote{This is the so-called {\it symmetrization postulate}, \cite{BJ}.}.
Accordingly, if a system is composed of different kinds of identical particles,
its wave function must be separately totally symmetric (bosons) or totally antisymmetric
(fermions) with respect to permutations of each kind of identical particles, \cite{BJ}.

\section{Basis optimization}

The most {\it direct approach} to a variational solution of a quantum mechanical bound-state
problem is to solve the secular equation defined by a basis of functions,
$\{\psi_1, \psi_2, \dots, \psi_K\}$, containing no nonlinear parameters. Two main steps
are involved: the calculation of the overlap and Hamiltonian matrix elements and the
solution of the generalized eigenvalue problem. To achieve the desired accuracy one only needs
to add many (linearly independent) functions to the basis. Unfortunately, calculating all
matrix elements takes time ${\cal O}(K^2)$ on a computer and solving eigenvalue equations
is an ${\cal O}(K^3)$-procedure, \cite{Bunge}. Any variational approach is thus only feasible on
a basis of reasonable size consisting of functions that allow fast matrix element evaluation.
The direct method in particular suffers from this limitation since the convergence is
often slow increasing the demand for a large number of basic functions, \cite{Kamada}.

Another approach, designed to avoid a huge basis dimension, is {\it basis optimization}.
The idea is to only select the specific basis functions that give good results.
To this end, the shape of the basis functions is made dependent on nonlinear parameters,
which, in effect, determine how well the variational function space, ${\cal V}_K$, contain
the true eigenfunction. An optimal basis of definite size, $K$, can be established by
minimizing the variational energy function with respect to these parameters.
However, although numerous elaborate methods are available for multidimensional function
minimization (see \cite{James}), the optimization of the nonlinear parameters in a trial wave function
is by no means a trivial task. In fact, computational complexity studies show that the general
problem takes time exponential in the number of parameters
\footnote{
Vavasis \cite{Vavasis} reports the worst-case complexity of minimizing a Lipschitz
constant function, $f(x_1,x_2,..,x_d)$, in a box to be ${\cal O}((\frac{L}{2\epsilon})^d)$,
where $L$ is the Lipschitz constant.}
and it is therefore rated as intractable, i.e. not amenable to a practically efficient
solution. This means that the sheer number of nonlinear parameters quickly becomes the bottle-neck
in basis optimization.

Different strategies have been employed for basis optimization in few-body problems related to
the following scenarios:

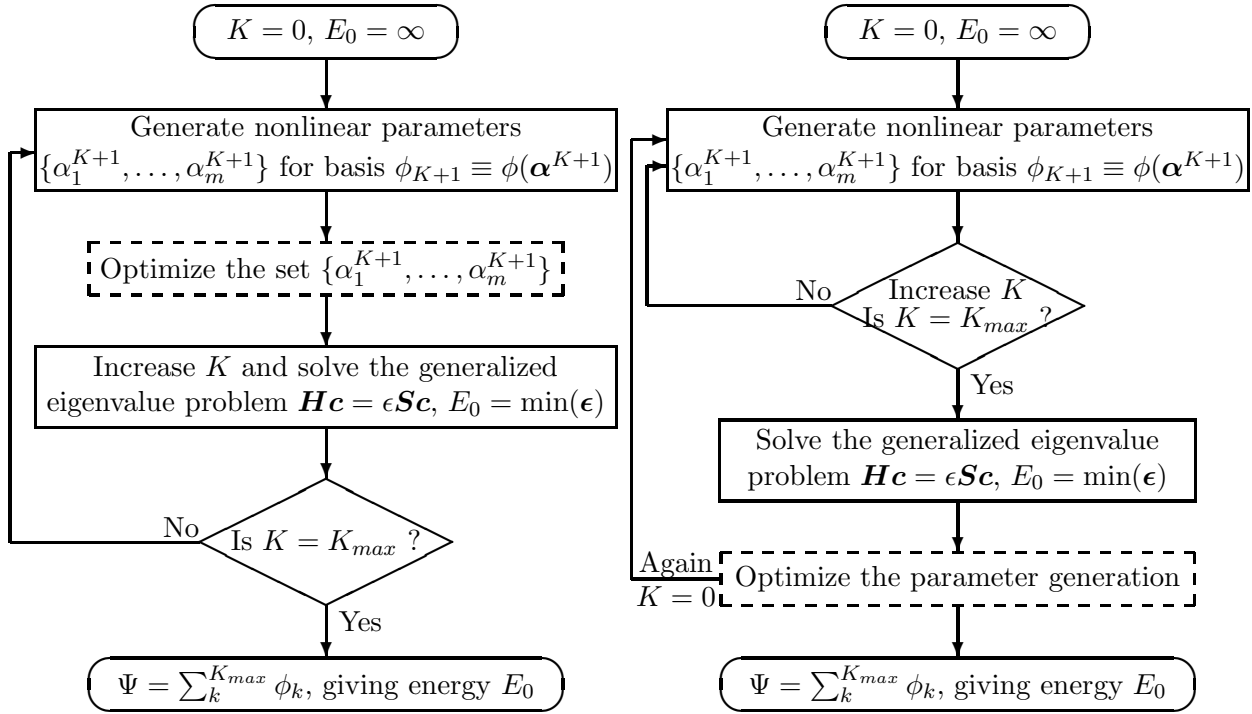
\begin{figure}[bt]
\setlength{\unitlength}{0.7cm}
\small
\begin{picture}(23.500000,13.400000)(-3.500000,-13.400000)
\thicklines
\put(2.5000,-0.5000){\oval(5.0000,1.0000)}
\put(0.0000,-1.0000){\makebox(5.0000,1.0000)[c]{\shortstack[c]{
$K=0$, $E_0=\infty$
}}}
\put(2.5000,-1.0000){\vector(0,-1){1.0000}}
\put(-3.0000,-3.5000){\framebox(11.0000,1.5000)[c]{\shortstack[c]{
Generate nonlinear parameters
\\
$\{\alpha_1^{K+1}, \dots, \alpha_m^{K+1}\}$ for basis $\phi_{K+1}\equiv\phi(\bm{\alpha}^{K+1})$
}}}
\put(2.5000,-3.5000){\vector(0,-1){1.0000}}
\put(-2.0000,-5.5000){\dashbox{0.2}(9.0000,1.0000)[c]{\shortstack[c]{
Optimize the set $\{\alpha_1^{K+1}, \dots, \alpha_m^{K+1}\}$
}}}
\put(2.5000,-5.5000){\vector(0,-1){1.0000}}
\put(-3.0000,-8.0000){\framebox(11.0000,1.5000)[c]{\shortstack[c]{
Increase $K$ and solve the generalized
\\
eigenvalue problem $\bm{H}\bm{c} = \epsilon\bm{S}\bm{c}$, $E_0 ={\rm min}(\bm{\epsilon})$
}}}
\put(2.5000,-8.0000){\vector(0,-1){1.0000}}
\put(0.1000,-10.2000){\line(2,1){2.4000}}
\put(0.1000,-10.2000){\line(2,-1){2.4000}}
\put(4.9000,-10.2000){\line(-2,-1){2.4000}}
\put(4.9000,-10.2000){\line(-2,1){2.4000}}
\put(0.1000,-11.4000){\makebox(4.8000,2.4000)[c]{\shortstack[c]{
Is $K = K_{max}$ ?
}}}
\put(0.1000,-9.7200){\makebox(0,0)[rt]{No}}
\put(2.7400,-11.8800){\makebox(0,0)[lb]{Yes}}
\put(0.1000,-10.2000){\line(-1,0){3.6000}}
\put(-3.5000,-10.2000){\line(0,1){7.4000}}
\put(-3.5000,-2.8000){\vector(1,0){0.5000}}
\put(2.5000,-11.4000){\vector(0,-1){1.0000}}
\put(2.5000,-12.9000){\oval(9.0000,1.0000)}
\put(-2.0000,-13.4000){\makebox(9.0000,1.0000)[c]{\shortstack[c]{
$\Psi=\sum_k^{K_{max}}\phi_{k}$, giving energy $E_0$
}}}
\put(14.5000,-0.5000){\oval(5.0000,1.0000)}
\put(12.0000,-1.0000){\makebox(5.0000,1.0000)[c]{\shortstack[c]{
$K=0$, $E_0=\infty$
}}}
\put(14.5000,-1.0000){\vector(0,-1){1.0000}}
\put(9.0000,-3.5000){\framebox(11.0000,1.5000)[c]{\shortstack[c]{
Generate nonlinear parameters
\\
$\{\alpha_1^{K+1}, \dots, \alpha_m^{K+1}\}$ for basis $\phi_{K+1}\equiv\phi(\bm{\alpha}^{K+1})$
}}}
\put(14.5000,-3.5000){\vector(0,-1){1.0000}}
\put(12.1000,-5.7000){\line(2,1){2.4000}}
\put(12.1000,-5.7000){\line(2,-1){2.4000}}
\put(16.9000,-5.7000){\line(-2,-1){2.4000}}
\put(16.9000,-5.7000){\line(-2,1){2.4000}}
\put(12.1000,-6.9000){\makebox(4.8000,2.4000)[c]{\shortstack[c]{
Increase $K$
\\
Is $K = K_{max}$ ?
}}}
\put(12.1000,-5.2200){\makebox(0,0)[rt]{No}}
\put(14.7400,-7.3800){\makebox(0,0)[lb]{Yes}}
\put(12.1000,-5.7000){\line(-1,0){3.5000}}
\put(8.6000,-5.7000){\line(0,1){2.6500}}
\put(8.6000,-3.0500){\vector(1,0){0.4000}}
\put(14.5000,-6.9000){\vector(0,-1){1.0000}}
\put(10.0000,-9.4000){\framebox(9.0000,1.5000)[c]{\shortstack[c]{
Solve the generalized eigenvalue
\\
problem $\bm{H}\bm{c} = \epsilon\bm{S}\bm{c}$, $E_0 ={\rm min}(\bm{\epsilon})$
}}}
\put(14.5000,-9.4000){\vector(0,-1){1.0000}}
\put(10.0000,-11.4000){\dashbox{0.2}(9.0000,1.0000)[c]{\shortstack[c]{
Optimize the parameter generation
}}}
\put(10.0000,-10.9000){\line(-1,0){1.7000}}
\put(8.3000,-10.9000){\line(0,1){8.3500}}
\put(8.3000,-2.5500){\vector(1,0){0.7000}}
\put(14.5000,-11.4000){\vector(0,-1){1.0000}}
\put(14.5000,-12.9000){\oval(9.0000,1.0000)}
\put(10.0000,-13.4000){\makebox(9.0000,1.0000)[c]{\shortstack[c]{
$\Psi=\sum_k^{K_{max}}\phi_{k}$, giving energy $E_0$
}}}
\put(4.6500,-11.4000){\makebox(9.0000,1.0000)[c]{\shortstack[c]{
Again
\\
$K=0$
}}}
\end{picture}
\caption{The control flow diagrams for two common basis optimization strategies: (left) Optimizing
while increasing one basis at a time and (right) optimizing the parameter generation or intervals.}
\label{optflow}
\end{figure}
\begin{itemize}
\item{
In few-body problems the number of nonlinear parameters needed in each basis function is
reasonably low, and one might be able to perform a {\it full simultaneous optimization} of the basis
parameters. The strategies employed can be placed into two categories: the deterministic and the
stochastic. The former is based on stepping or gradient strategies (e.g. the conjugate-gradient
method, see \cite{Kinghorn1}) and are sensitive to a given starting point, always reaching the same
minimum from the same initial condition. The solution in these cases may not be the global minimum
sought but a local minimum. Convergence depend heavily on the initial guesses for the parameters.
A stochastic minimization tends to convergence much slower but eliminates the risk of
ending up in a local minimum \cite{SVM}. Combining analytical gradient and stochastic techniques in
a mixed approach, as proposed recently in \cite{Bubin}, seems to be very economical.
}
\item{
When the above approach is too time-consuming, which is most often the case, one can use {\it grid
methods} to reduce the number of parameters to a smaller number of tempering parameters,
either by fixing parameters through a geometric progression \cite{Kameyama} 
or pseudo-randomly \cite{Alexander}. In the latter case this corresponds to optimizing
the limits of the intervals from which the pseudo-random numbers are selected \cite{Frolov1}.
The drawback is, that parameters may be assigned values disregarding whether the corresponding
basis functions contribute to the solution or not.
}
\item{
Alternatively, a {\it partial optimization} can be performed where only a few parameters are
optimized at a time and all others fixed. In particular, if only one specific basis function
is optimized, then only one
row of the (symmetric) Hamiltonian and overlap matrices are affected. Even the
consecutive solving of the full generalized eigenvalue problem can be avoided in the
optimization procedure. This is employed in the Stochastic Variational Method and, as described
in the chapter \ref{chapSVM}, takes only a fraction of the computational time of a full optimization.
}
\end{itemize}
Fig. \ref{optflow} shows two control flow structures that can be used 
with the optimization strategies discussed here. The left diagram corresponds to the case
where one basis function is optimized and added to the basis at a time and the right diagram is for
a procedure where the entire trial function is constructed and subsequently optimized. The dashed
boxes designate a (possibly complicated) optimization method in which place the SVM trial and error
technique will be considered in chapter \ref{chapSVM}.

\chapter{Correlations in many-body systems}\label{correlations}

The basic theory necessary for treating particle correlations in N-body systems with a variational
approach is presented in this chapter. The main goal is to develop several
descriptions, each representing a different level of correlation, and allow for a
direct comparison of the corresponding correlation energies. This is based on the vital
assumption, that correlations can be explicitly included in a description, by embedding them,
{\it ab initio}, in the form of the variational trial function. To this end, respective
sections treat first the uncorrelated Hartree trial function used within the
Hartree-Fock theory, then the effective interaction (mean-field) approach based on the
pseudopotential approximation and last an explicitly correlated trial function designed to
handle two-body, three-body and higher-order correlations. To begin with, however, a short
remark about correlation as a concept.

\section{Defining correlations}\label{defcor}

Since there are various definitions of the term correlation available in the physics literature,
it is appropriate to define the concept clearly before deriving a theoretical description.
In the dictionary, correlation is explained as ``a shared relationship'' or ``causal connection''.
Within the physics context of N-body systems, correlation correspondingly designates the
possibly complex interparticle relationship among the particles. However, in some textbooks,
the energy connected with such correlated behavior, $E_{corr}$, is defined as the difference
between the energy of the similar non-interacting system and the exact measured or calculated
energy \cite{Landau}. In other theoretic areas, like the atomic Hartree-Fock theory,
the correlation energy is regarded as the difference between the energy obtained with an
independent particle model based on the Hartree product wave function (see below) and the exact
energy \cite{Dahl,BJ}. Although both interpretations have valid argumentation, they are also very
distinct on the important question of what defines an uncorrelated system.

Here, and in the remainder of this thesis, the following definition is adopted:
\begin{center}
\begin{minipage}[c]{15.0cm}
\it
In an N-body system, where the interaction between the particles is
state-independent, the (inherently) correlated motion of the particles can be represented
by a wave function of the form \cite{Fabrocini}
\end{minipage}
\end{center}
\begin{equation}\label{meanphi}
\Psi(\bm{r}_1, \bm{r}_2, \dots \bm{r}_N) = F(\bm{r}_1, \bm{r}_2, \dots \bm{r}_N)
\Phi(\bm{r}_1, \bm{r}_2, \dots \bm{r}_N)
\end{equation}
\begin{center}
\begin{minipage}[c]{15.0cm}
\it
where $F$ is a correlation factor and $\Phi$ is an uncorrelated wave function corresponding
to a system of independent particles. The specific correlation energy included in such a
representation, $E_{corr}$, is defined by
\end{minipage}
\end{center}
\begin{equation}
E_{corr} = |E_{interaction}| = |<\Psi|\sum^N_{i<j}V_{ij}|\Psi>|
\end{equation}
\begin{center}
\begin{minipage}[c]{15.0cm}
\it
where $V_{ij}$ is the two-body interaction potential.
\end{minipage}
\end{center}
With this definition, the energy reference point corresponding to an uncorrelated system
is given by the energy
of the non-interacting system. Contrary to other interpretations, this allows even
a mean-field theory with $F=1$ to represent correlation energy, since any type of
interparticle interaction is tantamount to correlation effects.
The belief of the writer is, that such a standpoint is more true to the ``civil''
perception of the word correlation,
and in any case, advantageous in the current context, because the primary aim here is to compare
different levels of correlation where the widely used mean-field approach is just one candidate.

\section{Hartree-Fock mean-field description}

The {\it independent particle model}, originally formulated by Hartree in 1928 \cite{Hartree} and
generalized with symmetry by Fock and Slater \cite{Fock},
is based on the ansatz that a many-body wave function can be written as a properly
symmetrized product of orthogonal single-particle states, given by
\begin{equation}\label{hartree}
\Psi_{HF}(\bm{r}_1, \bm{r}_2, \dots \bm{r}_N)
 = {\cal \widehat{P}} \Psi_{H}(\bm{r}_1, \bm{r}_2, \dots \bm{r}_N)
 = {\cal \widehat{P}} \prod_{i=1}^N \phi_i(\bm{r}_i)
\end{equation}
where ${\cal \widehat{P}}$ is defined in (\ref{P}) as  the symmetrizer for bosons
and the antisymmetrizer for fermions.
In the Hartree-Fock method \cite{Fock} this form of wave function is applied with the
variational theorem in (\ref{vartheorem}), by demanding that the variation of the energy functional
value is zero, i.e
$\delta {\cal E}_{HF}[\Psi_{HF}] = \delta <\Psi_{HF}|\widehat{H}|\Psi_{HF}> = 0$.
To derive the theory of this method in the case where $\widehat{H}$ is
the non-relativistic N-body Hamiltonian (\ref{Hr}), one may conveniently rewrite $\widehat{H}$ as
\begin{equation}\label{HFH}
\widehat{H} = \sum^N_{i=1} \widehat{h}_i + \sum^N_{i<j}V_{ij},
\quad {\rm where} \quad
\widehat{h}_i = -\frac{1}{2m_i}\widehat{\bm{\nabla}}_i^{2} + V_{ext}(\bm{r}_i)
\end{equation}
having the first term explicitly given by a sum of N identical one-body Hamiltonians,
$\widehat{h}_i$. Taking into account that $\widehat{H}$ is invariant under the permutation
of particle coordinates, i.e. $[\widehat{H}, {\cal \widehat{P}}]=0$, and
${\cal \widehat{P}}^2 = {\cal \widehat{P}}$ by definition, the 
expectation value of $\widehat{h}_i$ is simply
\begin{equation}\label{HFh}
<\Psi_{HF}|\widehat{h}_i|\Psi_{HF}>=
<\Psi_{H}|\widehat{h}_i {\cal \widehat{P}}|\Psi_{H}>=
<\phi_i|\widehat{h}_i|\phi_i>
\equiv \int d\bm{r} \phi^*_i(\bm{r})\widehat{h}_i \phi_i(\bm{r})
\end{equation}
assuming the single-particle states are orthonormal, $<\phi_i|\phi_j>=\delta_{ij}$.
Using the same arguments the expectation value of the two-body interaction, $V_{ij}$, becomes
\begin{equation}\label{HFV}
<\Psi_{HF}|V_{ij}|\Psi_{HF}>=
<\Psi_{H}|V_{ij}{\cal \widehat{P}}|\Psi_{H}>=
<\phi_i \phi_j|V_{ij}|\phi_i \phi_j \pm \phi_j \phi_i>
\end{equation}
where $\pm$ indicates a $+$ for bosons and a $-$ for fermions.
Here a two-particle matrix element involves a double integral over the coordinates of both particles
\begin{equation}
<\phi_i \phi_j|V_{ij}|\phi_i \phi_j>
\equiv \int \int d\bm{r} d\bm{r}^\prime \phi^*_i(\bm{r})\phi^*_j(\bm{r}^\prime)
V(\bm{r}-\bm{r}^\prime) \phi_i(\bm{r})\phi_j(\bm{r}^\prime)
\end{equation}
Proceeding by taking the variation of $E_{HF}$ with respect to the single-particle states,
$\phi_i$, while imposing the orthonormality constraints on the $\phi_i$'s
by introducing (diagonal) Lagrange multipliers, ${\cal E}_i$, yields
\begin{equation}
\delta E_{HF} - \sum_{i=1}^N {\cal E}_i \delta <\phi_i|\phi_i> = 0
\end{equation}
After some algebra (see ref. \cite{BJ}) this variation leads to the $N$ Hartree-Fock
{\it integro-differential equations} for the single-particle wave functions
\begin{equation}\label{HF}
\widehat{h}_i\phi_i(\bm{r}) + V_{HF}^{D}(\bm{r}) \phi_i(\bm{r})
+ \int d\bm{r}^\prime V_{HF}^{Ex}(\bm{r},\bm{r}^\prime) \phi_i(\bm{r}^\prime) 
= {\cal E}_i \phi_i(\bm{r})
\end{equation}
where the {\it direct} potential is
\begin{equation}\label{Vd}
V_{HF}^{D}(\bm{r}) = \sum_{j=1}^N \int d\bm{r}^\prime \phi_j^*(\bm{r}^\prime) V(\bm{r}-\bm{r}^\prime)
\phi_j(\bm{r}^\prime)
\end{equation}
and the {\it exchange} potential is
\begin{equation}\label{Ex}
V_{HF}^{Ex}(\bm{r},\bm{r}^\prime) = \pm\sum_{j=1}^N \phi_j^*(\bm{r}^\prime)
V(\bm{r}-\bm{r}^\prime) \phi_j(\bm{r})
\end{equation}
with $+$ for bosons and $-$ for fermions. Adding the requirement of {\it self-consistency} between
the approximate individual single-particle states, $\phi_i(\bm{r})$, and the variational interaction
potential, $V(\bm{r}-\bm{r}^\prime)$, the equations
(\ref{HF})-(\ref{Ex}) can be solved with a simple iterative procedure \cite{BJ}.

\subsection{Correlations in the Hartree-Fock description}

The derived Hartree-Fock equations provide useful physical insight.
First of all, they show that if the N-body wave function is approximated by the single-particle
product (\ref{hartree}), the corresponding variational solution describes a model where each
particle moves in an effective potential generated by the other $N-1$ particles (i.e.
a {\it mean-field}).
A further striking feature of the integro-differential equations is that they involve the {\it joint}
probability for finding particles in states $i$ and $j$ at points $\bm{r}$ and
$\bm{r}^\prime$. This obviously imposes a relationship among the coordinates of the particles
which indicates they are to some degree correlated
\footnote{One may note that the correlations induced by the exchange term are repulsive
for fermions (on a range comparable to the size of the system)
and corresponds to the Pauli blocking effect.}.

The direct term represents the average potential due to the local
presence of the other particles. The exchange term takes into account the symmetry
effects from exchanging particles and indicates that the effective single-particle potential
is both state dependent and nonlocal. Determining one $\phi_i(\bm{r})$ requires the states for 
all other particles throughout the system as well as all other $\bm{r}^\prime$.
This means that the independent particle approximation does in fact not entirely
neglect particle-particle correlations. Rather it assumes that most of their important
effects can be taken into account with a sufficiently clever (variational) choice of the two-body
interaction potential form $V_{ij}$. As explained in the next section the optimal potential
is not the exact particle-particle interaction.

It is clear, that in the first quantized Hartree-Fock derivation given above, the only distinction
made between bosons and fermions is the definite symmetry of the wave function.
This seems only to have minor implications given by a sign in the exchange potential.
However, at the ultra-low temperature quantum level this difference in the
exchange correlations of bosons and fermions becomes very pronounced.
While bosons eagerly fall into a single quantum state to form a Bose-Einstein condensate fermions
tend to fill energy states from the lowest up, with one particle per quantum state.
To exemplify and complete the Hartree-Fock description, an expression for the
ground state energy of the N-boson system is now derived, since this is the case of interest later.
The corresponding derivation for systems of identical fermions, which are not considered
further here, is briefly addressed in appendix \ref{AppHFferm}.

\subsection{Ground state of identical bosons}
Bosons in a many-body system obey Bose-statistics with no restrictions on the
allowed quantum states. The ground state for identical bosons will then have all particles
occupying the {\it lowest orbital}, $\phi_i(\bm{r}_i) \equiv \psi_0(\bm{r}_i)$, and
the symmetric Hartree wave function
\begin{equation}
\Psi_{HF}^{(0)} = \psi_0(\bm{r}_1)\psi_0(\bm{r}_2)\dots\psi_0(\bm{r}_N)
\end{equation}
is appropriate as the starting point of the Hartree-Fock method. The spin part of the
wave function, left out here, is similarly a product of single-boson spin functions but otherwise
does not enter the calculation. Since the permutation operator, ${\cal \widehat{P}}$, is superfluous
in the ground state derivation, the exchange term in the integro-differential equations
(\ref{HF}) vanishes.
Removing the self-interaction contribution $(i=j)$ from the direct term
the ground state Hartree-Fock equations for identical bosons $(m_i\equiv m)$ is then
\begin{equation}\label{bHF}
\Big[-\frac{\hbar^2}{2m}\widehat{\bm{\nabla}}^{2} + V_{ext}(\bm{r})
 + (N-1)\int d\bm{r}^\prime \psi_0^*(\bm{r}^\prime) V(\bm{r}-\bm{r}^\prime)
\psi_0(\bm{r}^\prime) \Big] \psi_0(\bm{r}) = \mu \psi_0(\bm{r})
\end{equation}
where $m$ is the mass and $\mu$ corresponds to the chemical potential encountered in the
Bogoliubov theory. The total ground state energy of the system of N identical bosons becomes
\begin{equation}\label{bHFE}
E_{HF}^{(0)} = N \mu - \frac{N(N-1)}{2} <\psi_0 \psi_0|V_{ij}|\psi_0 \psi_0>
\end{equation}
where the second term is due to double counting (see app. \ref{AppHFferm}).
Assuming that the system is sufficiently cold and dense for the single-boson wave functions,
$\psi_0(\bm{r})$, to overlap, the Hartree wave function, $\Psi_H^{(0)}$, corresponds to a
condensed state, as explained previously. Then $E_{HF}^{(0)}$ is the Hartree-Fock approximation
to the BEC energy and apparently takes the boson-boson interactions into account.
However, another consequence of interactions is
collisional excitations, where the bosons are scattered in and out of their single-particle
states, leading to quantum depletion of the lowest state even at $T=0$. This
is entirely neglected in the derivation of (\ref{bHF}) and (\ref{bHFE}).

\section{Pseudopotential mean-field description}\label{pseudo}

A wave function in the Hartree form (\ref{hartree}) is explicitly uncorrelated. Consequently,
it is clear that the Hartree-Fock description in the previous section ignores real dynamical
correlations although incorporating exchange and mean-field effects. In the dilute limit, however,
it is possible (and in some cases imperative, see below) to introduce some extend of
(short range) correlation effects in the variational solution by adopting an effective interaction
potential instead of the exact $V_{ij}$. The details of this important refinement of the
Hartree-Fock theory are presented in the following.

As a first approximation one must assume that the system consists of {\it weakly interacting}
particles (what exactly constitutes a weak interaction will be addressed in section \ref{valpseudo}).
When this is the case, the interaction potential is so short that the single-particle wave
functions do not vary over the interaction region
\footnote{In the independent particle model weakly interacting particles are
roughly free particles given by plane waves, $\phi_i(\bm{r})=(2\pi)^{-3/2}e^{i\bm{k} \cdot \bm{r}}$.
Thus the approximation $\phi_i(\bm{r})\approx \phi_i(\bm{r}^\prime)$ amounts to
the requirement that the thermal de Broglie wavelength, $\lambda_T =(\frac{2\pi \hbar^2}{mkT})^{1/2}$,
is much larger than the range of $V(\bm{r}-\bm{r}^\prime)$.
}. Then one can rewrite, as
\begin{equation}
\int d\bm{r}^\prime \phi_j^*(\bm{r}^\prime) V(\bm{r}-\bm{r}^\prime) \phi_j(\bm{r}^\prime)
\approx |\phi_j(\bm{r})|^2 \int d\bm{r}^\prime V(\bm{r}-\bm{r}^\prime)
\end{equation}
simplifying the intergro-differential Hartree-Fock equations (\ref{HF}) to
\begin{equation}
\Big[\widehat{h}_i+V_{MF}(\bm{r}) \Big]\phi_i(\bm{r}) = {\cal E}_i \phi_i(\bm{r})
\end{equation}
where the {\it mean-field potential} is
\begin{equation}\label{MF}
V_{MF}(\bm{r}) =
\sum_{j=1}^N (1 \pm \delta_{ij}) |\phi_j(\bm{r})|^2\int d\bm{r}^\prime V(\bm{r}-\bm{r}^\prime)
\end{equation}
where $+$ is for bosons and $-$ is for fermions.
Thus, in the bosonic ground state described above, the exchange term simply doubles
the direct term. For identical fermions in equivalent single-particle states
the terms cancel instead as expected due to Pauli exclusion. 

\subsection{Effective interactions}

Intuitively the mean-field interaction, $V_{MF}$, should be mediated by the elastic collisions
in the system. As mentioned, only two-body scattering described by the two-body potential,
$V_{ij}$, are significant in a dilute system. However, there are several reasons not to use the
exact two-body interaction potential in the Hartree-Fock approach. First of all, it is quite
difficult to determine the exact potential precisely, and a small error in the shape of $V_{ij}$
may produce a large error in other results, e.g. in the scattering length, $a$.
Secondly, the exact potential is very deep and supports many bound states. Such strong interactions
cannot be treated within the weak field assumption (see below). Finally, the hard-core repulsion
at short distances makes the evaluation
of the mean-field integral (\ref{MF}) somewhat troublesome. For an extreme example, consider the
hard-sphere potential ($V$ is infinite if $(\bm{r}-\bm{r}^\prime) < r_c$,
zero if $(\bm{r}-\bm{r}^\prime) > r_c$). Obviously this potential causes $V_{MF}(\bm{r})$
to be infinite for any nonzero value of $\phi_i(\bm{r})$, regardless of the size of the hard-sphere
radius $r_c$. This is unreasonable since even in the limit of an infinitesimal radius, the
contribution would still be infinite.

The explanation for the discrepancy of the hard-sphere example is that in the independent particle
approximation, no dynamical correlations between individual particles are allowed. In reality,
there would be no particles closer to each other than the radius $r_c$ in the hard-sphere
scattering example. However, there is no way for the naive Hartree-Fock theory to account for this.
Simply neglecting the hard-sphere interaction, as there are no particles that close anyway,
is not sensible either since this would allow the single-particle wave functions of two
neighboring particles to overlap inside the hard-core radius 
\footnote{The single-particle wave functions need to bend away in the forbidden
hard-core region so that the resulting curvature of such $\phi_i$'s contribute correctly
to the kinetic energy of the system, \cite{Roberts}.}.
The solution is instead to replace the exact interaction potential by a model potential
that (1) has the same scattering properties at low energies, i.e. is the same scattering length
and (2) will work in the independent particle approximation.
To some extend, the short wave length components of the wave function that reflect the dynamic
correlations between particles are then implicitly taken into account.
This implies that the Born approximation in the case of scattering (see below) and the
Hartree-Fock method for calculating bound state energies give better results provided that
the simple effective interaction is used rather than the real one.
To exactly what extend such a mean-field approach succeeds in including correlations is 
somewhat clarified in chapter \ref{Results}.

\subsection{The pseudopotential approximation}
The model potential satisfying the two requirements stated above
with the minimal number of parameters (one!)
is the zero-range pseudopotential initially introduced by Fermi \cite{Fermi} and Huang \cite{Huang}:
\begin{equation}\label{pse}
V_{pseudo}(\bm{r}_{12}) = g \delta(\bm{r}_{12}) \frac{\partial}{\partial r_{12}}r_{12}
\end{equation}
where the coupling constant, $g=\frac{2\pi\hbar^2}{\mu}a$, is directly proportional to the s-wave
scattering length, $a$. It is valid for dilute systems (typically stated as $n|a|^3 \ll 1$,
where $n$ is the characteristic density) at low energies, although making $g$ energy or
density dependent can extend the validity region \cite{Ole, Blume2}.
The pseudopotential involves
a Dirac $\delta$-function and a regularizing operator, $\frac{\partial}{\partial r_{12}}r_{12}$,
that removes a possible divergence of the wave function at $r_{12}=0$. When the wave function
is regular at $r_{12}=0$ the regularizing operator has no effect and the pseudopotential can be
viewed as a mere contact potential, $V(\bm{r}_{12}) = g \delta(\bm{r}_{12})$. The widely used
{\it Gross-Pitaevskii equation}
\footnote{
Using the pseudopotential (\ref{pse}) with $g=\frac{4\pi\hbar^2}{m}a$ ($\mu=m/2$)
in the boson ground state Hartree-Fock mean-field equations
(\ref{bHF}) leads directly to the Gross-Pitaevskii equation:
\begin{equation}\label{GP}
\Big[-\frac{\hbar^2}{2m}\widehat{\bm{\nabla}}^{2} + V_{ext}(\bm{r})
+\frac{4\pi\hbar^2a_B}{m}(N-1) |\phi_0(\bm{r})|^2 \Big]\phi_0(\bm{r}) = \mu \phi_0(\bm{r})
\end{equation}
where $\mu$ is the chemical potential and $\phi_0$ is the (regular) single-particle wave function.
} 
corresponds to a mean-field approach for systems of bosons based on the
pseudopotential effective interaction.

In the low energy limit the pseudopotential reproduces the scattered wave function
of the exact two-body potential asymptotically and gives the correct scattering length.
However, the possible change of the wave function inside the (finite) interaction range
is effectively ignored. This is also known as the shape-independent approximation, \cite{Esry2}.
As demonstrated above with the hard-sphere potential example, it is the large repulsive core
of the exact interaction which makes $V_{MF}$ huge regardless of the other details of the potential.
Remarkably, this leads to the {\it counterintuitive conclusion} that using a realistic
two-body potential in the Hartree-Fock equations yields a much poorer result than using
a $\delta$-function potential with the same asymptotic scattering properties
\footnote{In \cite{Esry}, B. D. Esry illustrates this by comparing Hartree-Fock calculations for the
pseudopotential and for the realistic Morse potential with the exact hyperspherical result
for three atoms in a harmonic trap.}.
This means, that it is not only convenient to make the
shape-independent approximation in the Hartree-Fock approach but actually essential
in the case of hard-core potentials in order to obtain quantitatively correct results.
At the same time it is important to
stress that the pseudopotential only works within the independent particle approximation, that
is with the Hartree wave function, and should not be applied in an exact solution
(see e.g. \cite{Esry2}).

\subsection{The Born approximation}\label{Born}

The Born series is the perturbation expansion of the scattering wave function
or equivalently the scattering amplitude in powers of the interaction potential.
It is interesting in this mean-field context because the condition of the pseudopotential
to neglect the distortion of the (incoming) wave function in the region of the two-body potential,
is precisely the same requirement that makes the Born approximation scheme valid in scattering theory,
\cite{Sak}. In particular the {\it first} term of the Born series follows directly from the
assumption that the initial wave function is an undistorted plane wave, that is
\begin{equation}
\Psi_{\bm{k}}^{(+)}(\bm{x}) \approx 
\Psi_{\bm{k}}^{(+)}|_{B1} =(2\pi)^{-3/2}e^{i\bm{k}\cdot\bm{x}}
\end{equation}
Plugging this into (\ref{scatamp}) gives the first-order Born scattering amplitude 
\begin{equation}
f(\bm{k}^\prime,\bm{k})|_{B1} = -\frac{\mu}{2\pi\hbar^2}
\int d\bm{x} e^{-i(\bm{k}^\prime-\bm{k})\cdot\bm{x}}V(\bm{x})
\end{equation}
which in the low energy limit, $k \rightarrow 0$, yields the
Born approximation scattering length
\begin{equation}\label{born}
a_B \stackrel{k \rightarrow 0}{\equiv} - f(\bm{0},\bm{0})|_{B1}
= \frac{\mu}{2\pi\hbar^2} \int d\bm{x} V(\bm{x})
=\frac{2\mu}{\hbar^2} \int_0^\infty dr r^2 V(r)
\end{equation}
where the last equality holds for central potentials. In the specific case of the
zero-range pseudopotential (\ref{pse}) it follows that $a_B = a$, that is,
the first-order Born approximation to the scattering length actually coincides with the
s-wave scattering length. This fortunate property greatly simplifies the treatment of the N-body
problem when one models a sufficiently weak interaction with the pseudopotential
\footnote{In a more general treatment the Born scattering amplitude is
replaced by the two particle T -matrix element which holds regardless of the
interaction potential strength. In the low energy s-wave scattering case
the T -matrix element is proportional to $a$, giving the same results as the Born approximation
\cite{Pethick}.
}
\cite{Esry,Blume3}.

\begin{figure}[bt]
\includegraphics[width=16cm]{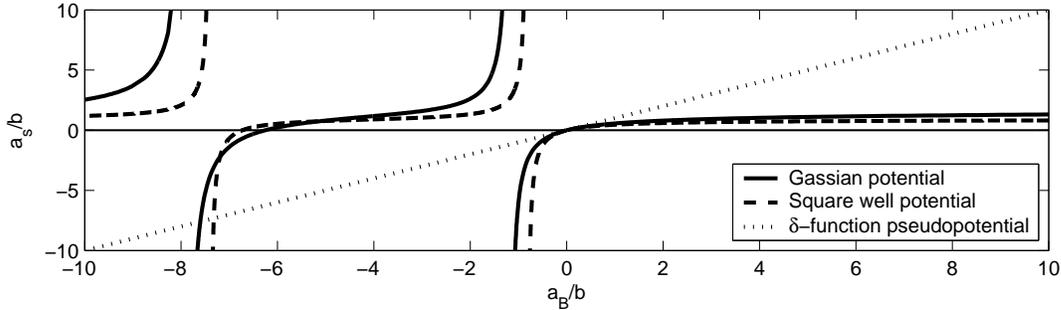}
\caption{
Scattering length, $a$, as a function of the Born approximation to the scattering
length, $a_B$, in units of $b=18.9 a_0$, for a Gaussian two-body potential 
$V(x) = V_0 e^{-x^2/b^2} \ (a_B = \sqrt{\pi}\mu b^3 V_0/ 2 \hbar^2)$
, a square well two-body potential 
$V(x) = V_0, \ x \le b,\ V(x) = 0, \ x > b \ (a_B = 2\mu b^3 V_0/ 3 \hbar^2)$
and the contact pseudopotential $V(x) = 4\pi\hbar^2 a\delta(x)/m \ (a_B = a)$.
}
\label{abscat}
\end{figure}

It is also worth noting, that the value of $a_B$ is in general very different
from the s-wave scattering length, $a$, for all potentials other than the
pseudopotential. The numerical relation between $a$ and $a_B$ is illustrated in figure \ref{abscat}
(\footnote{Data for the Gaussian potential is numerically calculated (in atomic units)
from subsequent runs: {\tt scatlen -V0 $V_0$}, where $-10^{-6} \le V_0 \le 10^{-6}$
with steps of $10^{-9}$. For the square well potential one has analytically \cite{BJ}:
$a = b(1-{\rm tan}(\gamma)/\gamma)$, where $\gamma = \sqrt{2\mu b^2|V_0|/\hbar^2}$.
}) for the cases of a Gaussian potential, a square well potential and the contact pseudopotential.
When the former two potentials are attractive $(a_B < 0)$ the emergence of bound states and
corresponding energy resonances are clearly visible. The similar overall behavior of the curves
for these cases reflects the {\it shape independence} at low scattering energies. Obviously, the Born
scattering length is a good approximation only when the interaction is very weak and $a \ll b$.

\subsection{Validity range of the pseudopotential approximation}\label{valpseudo}

Since the condition for the application of the pseudopotential and the first Born approximation
are quite similar it is interesting to consider the validity range of the latter
in more detail. One well known range where the Born Approximation describes the
scattering amplitude nicely is the high-energy regime, where the energy of the incoming particle is
much greater than the energy scale of the scattering potential, \cite{BJ}. This is
not relevant for the low energy pseudopotential approximation described here.
However, the requirement that the incoming wave function is not significantly altered
in the region of the potential, or equivalently that the second term in the Born
perturbation expansion is very small, can be related via the Lippmann-Schwinger equation
to the condition (\cite{Sak}, p. 388)
\begin{equation}\label{bvalid}
\bigg| \frac{\mu}{2\pi\hbar^2} \int d\bm{x} \frac{e^{ikx}}{x}V(\bm{x})e^{i\bm{k}\cdot\bm{x}}
\bigg| \ll 1
\end{equation}
For central potentials and at low collision energies $(k\rightarrow0, e^{ikx}\rightarrow 1)$
this gives
\begin{equation}\label{bvalid2}
\frac{2\mu}{\hbar^2} \bigg| \int dx\ xV(x) \bigg| \ll 1
\end{equation}
which is obviously satisfied for a sufficiently weak scattering potential, $V(x)$.
The $\delta$-function in the pseudopotential fulfills the condition by definition $(0\ll1)$.
In the case of a Gaussian two-body potential, $V(\bm{r})=V_0 e^{-\bm{r}^2/b^2}$, the simple
integral evaluates to
\begin{equation}\label{pseudovalid}
\mu b^2 |V_0|/\hbar^2 \ll 1
\end{equation}
which is also the expression obtained for the square-well potential.
This requirement may be compared with the condition for the potentials to develop a two-body bound
state, that is a solution to (\ref{two-schr}) with $E<0$, where $E = -\hbar^2/2\mu a^2$
(\cite{Landau}, p. 57). For the square-box potential the bound-state condition is
$\mu b^2|V_0|/\hbar^2 > \pi^2/8 \approx 1.2$ (by inserting the analytic expression for the
scattering length, see caption of fig. \ref{abscat}).
This is quite the opposite of condition (\ref{pseudovalid}).
A similar analytic expression for the Gaussian potential is not available, but in the case
where $b=11.65$ a.u., $\mu = 1.58\times10^5/2$ a.u. ($^{87}$Rb), it is possible to determine
numerically (by using the {\tt scatlen} program) that the first two-body bound state occurs at
$V_0 = -1.25\times10^{-7}$ a.u.. Plugging this into (\ref{pseudovalid}) gives
$\mu b^2|V_0|/\hbar^2 \approx 1.34$, which also clearly violates the condition.
In other words, if the potential is strong enough
to develop a bound state, the Born approximation and the pseudopotential approach
will probably give misleading results. The conclusion is then, that a weak (attractive)
interaction in the current mean-field theory, is one that is far from supporting a bound state.
This observation is also visible in the numerical results presented in chapter \ref{Results}
(see e.g. fig. \ref{N3correlation}).

\section{Explicitly correlated description}\label{explicit}

The key point of the preceding section is that the pseudopotential can be used under
certain conditions as a mean-field effective interaction and without the necessity of calculating
detailed short range correlations. The conditions where found to be satisfied at low energies
by weakly interacting dilute systems, where the particles are mostly far away from each other and
correlations in head-to-head collisions are expected to be negligible. However, the importance of
correlations must increase with the density of the system and the strength of the interaction,
and at some point the mean-field approach becomes inadequate.
Going {\it beyond mean-field theory} is only possible with the explicit
inclusion of correlation effects in the wave function, that is, an appropriate choice of
the correlation factor $F$ in (\ref{meanphi}). The following section describes a simple method
for constructing $F$, in a way very similar to the discussion of translationally invariant clusters
in coordinate space given by Bishop {\it et al} in \cite{Bishop}.

\subsection{Two-body correlations}

As the first step towards a systematic approach to the exact correlated ground state wave function
one can consider a correlation factor containing only two-body correlations, e.g.
\begin{equation}\label{JastrowF}
F_2(\bm{r}_1, \dots , \bm{r}_N) = \prod_{i<j}^N f_2(r_{ij})
\end{equation}
where $f_2$ is a properly chosen pair correlation function depending only on the interparticle
distance
\footnote{A {\it proper} correlation function has to satisfy certain requirements (e.g. approach
unity at large particle separation). See the complete list on
page 62 in \cite{Fabrocini}.}
. This is the widely used {\it Jastrow ansatz}
\cite{Jastrow}. As stated at the beginning of this section the function $f_2$
should go to unity, i.e. to the mean-field limit, at large separations manifesting the absence
of correlations when the particles are far away from each other. At short distances
the correlation function is expected to deviate from unity and writing $f_2(r_{ij})=1+c_2(r_{ij})$,
where $c_2(r_{ij})$ represents the short range deviation, yields
\begin{equation}\label{Fexp}
F_2(\bm{r}_1, \dots , \bm{r}_N) = \prod_{i<j}^N \Big(1+c_2(r_{ij})\Big)
=1+\sum_{i<j}^N c_2(r_{ij})+\frac{1}{2!}\sum_{(i<j)\neq(k<l)}^N c_2(r_{ij})c_2(r_{kl})+\cdots
\end{equation}
where the indices of the interparticle coordinates, appearing in the summed products, at all
orders, never overlap.
The second term in this expansion corresponds to the effect of pair correlation while
the third term induces separate correlations between two independent pairs of particles (clusters)
and so forth. For a sufficiently dilute system it is unlikely that two or more independent pairs
simultaneously are close in space and the expansion can be truncated after the first two terms, giving
\begin{equation}\label{F2}
F_2(\bm{r}_1, \dots , \bm{r}_N) \approx 1+\sum_{i<j}^N c_2(r_{ij})
=\sum_{i<j}^N \Big(\frac{1}{N(N-1)/2}+c_2(r_{ij}) \Big)\equiv\sum_{i<j}^N C_2(r_{ij})
\end{equation}
where $C_2(r_{ij})$ is the simple redefinition of $c_2(r_{ij})$ that absorb the factor one in the
expansion. 

\subsection{Three-body and higher-order correlations}
The most obvious improvement of the two-body correlation factor in (\ref{F2}) is to include the
next term in the expansion (\ref{Fexp}). This is easily done by replacing $C_2(r_{ij})$ with a
more general function, $C_2^{(2)}(r_{ij},r_{kl})$, depending on two interparticle distances,
and realizing that it is possible to absorb all lower-order terms within this form, thus giving
$F_{2}(\bm{r}_1, \dots , \bm{r}_N) \approx \sum_{i<j\neq k<l}^N C_2^{(2)}(r_{ij},r_{kl})$.
In most cases, however, the corresponding improvement is small and the introduction of three-body
correlations is much better (see \cite{Bishop}, table 2). In particular, extending the Jastrow
formulation to include three-body correlations, leads to
\begin{equation}\label{F3}
F_3(\bm{r}_1, \dots , \bm{r}_N) = \prod_{i<j}^N f_2(r_{ij})
\prod_{i<j<k=1}^N f_3(r_{ij},r_{ik},r_{jk})
\approx \sum_{i<j<k=1}^N C_3(r_{ij},r_{ik},r_{jk})
\end{equation}
where $f_3$ is a proper triplet correlation function and the freedom in choosing the functional form
of $C_3(r_{ij},r_{ik},r_{jk})$ has been utilized to absorb all lower-order (cluster) terms.
Following the same ideas, a strait forward generalization of the Jastrow approach (as done by Feenberg
\cite{Feenberg}) to include all higher-order correlations in the wave function gives
\begin{equation}
F_N(\bm{r}_1, \dots , \bm{r}_N) \approx \widehat{\cal S} \ C_N(r_{12},r_{13},\dots,r_{(N-1)N})
\end{equation}
where the symmetrization operator, $\widehat{\cal S}$, ensures that the correlation factor
does not influence the exchange symmetry of the wave function, and the functional form
$C_N(r_{12},r_{13},\dots,r_{(N-1)N})$ is assumed to completely describe all the correlations of
the N-body system. It should be noted, that
in order to make a calculation manageable in practice it is necessary to further expand
the unknown correlation functions, $C_n$, in a set of simple functions (for example Gaussians,
as described in detail in section \ref{basisfunc}).

\subsection{Validity range of the Jastrow-Feenberg description}

Several points concerning the validity of the correlation description above are important:
\begin{itemize}
\item{
Firstly, it is apparent that the application is limited to homogeneous and isotropic systems, 
that is, $f_1(\bm{r}_i)=1$ and $f_2(\bm{r}_i,\bm{r}_j)=f_2(r_{ij})$, etc. The translational
invariance resulting from this is essential to avoid problems with the center-of-mass motion
\cite{Bishop}.}
\item{
Secondly,
the correlations do not depend on internal quantum numbers such as spin. This is inappropriate
in cases where the interactions are state dependent like in nuclear physics. Unfortunately, including
state dependence in the Jastrow-type correlation functions, $f_i$, turns them effectively into
non-commuting operators demanding further symmetrization of the product form of $F$. This considerably
complicates the formalism and is not considered here (for details see e.g. the 
FHNC single-operator-chain (FHNC/SOC) method \cite{Fabrocini2} or CBF theory \cite{Fabrocini}).
}
\item{
Thirdly, the lack of momentum-dependency in the Jastrow-Feenberg ansatz makes it questionable
when dealing with the ground state of Fermi systems, since there is no information about
the specific location of a given particle within the Fermi sea. Such a treatment is perhaps
acceptable for ``integrated'' quantities like the energy, but it is not at all clear whether
it works for physical properties, like the specific heat, depending primarily on the
``active'' particles close to the Fermi surface \cite{Ashcroft}. To examine this question, it
is again necessary to go beyond the Jastrow-correlated wave function (see \cite{Fabrocini}, chap. 7). 
}
\item{
Finally, returning to the discussion central in the pseudopotential approximation above,
the particular choice of a wave function parametrization always corresponds to a restriction of the
full Hilbert space solution. While this restriction is quite severe in the mean-field
Hartree wave function (with $F=1$), leading to failure in combination with the exact interaction
potential, it is much less pronounced with the Jastrow-type correlation factor. Still, the 
truncated factors, $C_2(r_{ij})$ and $C_3(r_{ij},r_{ik},r_{jk})$,
are clearly not able to take hard-core repulsion into account, since this requires that all
pairs are simultaneously correlated. Only $C_N(r_{12},r_{13},\dots,r_{(N-1)N})$ has this feature and
thus seems to be valid with hard-core potentials. But whether a solution based on a realistic
or, as in this work, a Gaussian interaction, with the given inclusion of two-, three- or
N-body correlations, will reproduce reasonable results is, however, not obvious.
The most convincing argument is, that many, if not most, of the
key methods in modern quantum many-body theory are based on the Jastrow-Feenberg approach or
similar ideas, and they reach such a high level of accuracy, also when limited to pair or triplet
correlations, that it has been debated, although recently disproved (see \cite{Bishop2}), if the
ansatz could be generally exact. This obvious success story continues with the results
presented in chapter \ref{Results}.
}
\end{itemize}

\chapter{The Stochastic Variational Method}\label{chapSVM}

The SVM has been developed through the search for precise solutions of nuclear few-body problems
\cite{SVM}. In this chapter it is shown how to employ the method to atoms and
N-body systems of trapped bosons. The main aim is to develop the SVM formalism
to a point where one can
systematically include the effects of two- and higher-order correlations in a way
which is both intuitive and computationally efficient.
Subsequent sections treat the trial and error selection procedure, the explicitly
correlated basis functions and the details of how to symmetrize the trial function in practice.
The three-body system constituting the Helium atom is used to benchmark the method towards
treating the more intricate case of BECs.

\section{Stochastic trial and error procedure}\label{secTE}
\begin{figure}[htbp]
\setlength{\unitlength}{0.7cm}
\small
\begin{picture}(22.000000,28.749998)(-8.400000,-28.749998)
\thicklines
\put(2.5000,-0.5000){\oval(5.0000,1.0000)}
\put(0.0000,-1.0000){\makebox(5.0000,1.0000)[c]{\shortstack[c]{
$K=0$, $E_0=\infty$
}}}
\put(2.5000,-1.0000){\line(0,-1){0.0000}}
\put(2.5000,-1.0000){\vector(0,-1){1.0000}}
\put(-7.0000,-3.5000){\framebox(19.0000,1.5000)[c]{\shortstack[c]{
Generate nonlinear parameters $\{\alpha_1^{K+1}, \dots, \alpha_m^{K+1}\}$ for basis $\phi_{K+1}\equiv\phi(\bm{\alpha}^{K+1})$
}}}
\put(2.5000,-3.5000){\vector(0,-1){1.0000}}
\put(-0.7000,-6.1000){\line(2,1){3.2000}}
\put(-0.7000,-6.1000){\line(2,-1){3.2000}}
\put(5.7000,-6.1000){\line(-2,-1){3.2000}}
\put(5.7000,-6.1000){\line(-2,1){3.2000}}
\put(-0.7000,-7.7000){\makebox(6.4000,3.2000)[c]{\shortstack[c]{
Iterate $i$ through
\\
the $m$ parameters
}}}
\put(5.7000,-5.4600){\makebox(0,0)[lt]{Done}}
\put(2.8200,-8.3400){\makebox(0,0)[lb]{$i<m$}}
\put(2.5000,-7.7000){\vector(0,-1){1.0000}}
\put(-6.0000,-10.2000){\framebox(17.0000,1.5000)[c]{\shortstack[c]{
Stochasticly select $t$ trial values $\{\tilde{\alpha}_1, \dots, \tilde{\alpha}_{t}\}$ for the $i$'th parameter
}}}
\put(2.5000,-10.2000){\vector(0,-1){1.0000}}
\put(-6.0000,-13.2000){\framebox(17.0000,2.0000)[c]{\shortstack[c]{
Calculate the $t$ ground state energies $\{E_1, \dots, E_t\}$ that correspond
\\
to $\alpha_i^{K+1}=\tilde{\alpha}_1, \dots, \tilde{\alpha}_{t}$ in the trial wave function $\Psi=\sum_k^{K+1}\phi(\bm{\alpha}^{K+1})$
}}}
\put(2.5000,-13.2000){\vector(0,-1){1.0000}}
\put(-5.5000,-15.2000){\framebox(16.0000,1.0000)[c]{\shortstack[c]{
Locate the index $min$ of the lowest energy from $\{E_1, \dots, E_t\}$
}}}
\put(2.5000,-15.2000){\vector(0,-1){1.0000}}
\put(0.1000,-17.4000){\line(2,1){2.4000}}
\put(0.1000,-17.4000){\line(2,-1){2.4000}}
\put(4.9000,-17.4000){\line(-2,-1){2.4000}}
\put(4.9000,-17.4000){\line(-2,1){2.4000}}
\put(0.1000,-18.6000){\makebox(4.8000,2.4000)[c]{\shortstack[c]{
Is $E_{min} < E_0$ ?
}}}
\put(0.1000,-16.9200){\makebox(0,0)[rt]{No}}
\put(2.7400,-19.0800){\makebox(0,0)[lb]{Yes}}
\put(0.1000,-17.4000){\line(-1,0){6.6000}}
\put(-6.5000,-17.4000){\line(0,1){7.9000}}
\put(-6.5000,-9.5000){\vector(1,0){0.5000}}
\put(2.5000,-18.6000){\vector(0,-1){1.0000}}
\put(-3.0000,-20.6000){\framebox(11.0000,1.0000)[c]{\shortstack[c]{
Choose $\alpha_i^{K+1}=\tilde{\alpha}_{min}$ and set $E_0 = E_{min}$
}}}
\put(-3.0000,-20.1000){\line(-1,0){4.0000}}
\put(-7.0000,-20.1000){\line(0,1){14.0000}}
\put(-7.0000,-6.1000){\vector(1,0){6.3000}}
\put(5.7000,-6.1000){\line(1,0){0.5000}}
\put(6.2000,-6.1000){\vector(1,0){1.0000}}
\put(7.2000,-6.1000){\line(2,1){2.4000}}
\put(7.2000,-6.1000){\line(2,-1){2.4000}}
\put(12.0000,-6.1000){\line(-2,-1){2.4000}}
\put(12.0000,-6.1000){\line(-2,1){2.4000}}
\put(7.2000,-7.3000){\makebox(4.8000,2.4000)[c]{\shortstack[c]{
Repeat a
\\
few times...
}}}
\put(9.8400,-4.4200){\makebox(0,0)[lt]{Again}}
\put(12.0000,-5.6200){\makebox(0,0)[lt]{Done}}
\put(9.6000,-4.9000){\line(0,1){0.4000}}
\put(9.6000,-4.5000){\vector(-1,0){7.1000}}
\put(12.0000,-6.1000){\line(1,0){0.6000}}
\put(12.6000,-6.1000){\line(0,-1){16.5000}}
\put(12.6000,-22.6000){\line(-1,0){0.5000}}
\put(12.1000,-22.6000){\vector(-1,0){1.0000}}
\put(-5.9000,-23.3500){\framebox(17.0000,1.5000)[c]{\shortstack[c]{
Increase $K$ and solve the generalized eigenvalue problem $\bm{H}\bm{c} = \epsilon\bm{S}\bm{c}$
}}}
\put(2.6000,-23.3500){\vector(0,-1){1.0000}}
\put(0.2000,-25.5500){\line(2,1){2.4000}}
\put(0.2000,-25.5500){\line(2,-1){2.4000}}
\put(5.0000,-25.5500){\line(-2,-1){2.4000}}
\put(5.0000,-25.5500){\line(-2,1){2.4000}}
\put(0.2000,-26.7500){\makebox(4.8000,2.4000)[c]{\shortstack[c]{
Is $K = K_{max}$ ?
}}}
\put(0.2000,-25.0700){\makebox(0,0)[rt]{No}}
\put(2.8400,-27.2300){\makebox(0,0)[lb]{Yes}}
\put(0.2000,-25.5500){\line(-1,0){7.7000}}
\put(-7.5000,-25.5500){\line(0,1){22.8000}}
\put(-7.5000,-2.7500){\vector(1,0){0.5000}}
\put(2.6000,-26.7500){\vector(0,-1){1.0000}}
\put(2.6000,-28.2500){\oval(17.0000,1.0000)}
\put(-5.9000,-28.7500){\makebox(17.0000,1.0000)[c]{\shortstack[c]{
Optimized wave function is $\Psi=\sum_k^{K_{max}}\phi_{k}$ giving energy $E_0$
}}}
\thinlines
\put(-8.4000,-21.0500){\dashbox{0.2}(22.0000,17.0000)[c]{\shortstack[c]{
}}}
\end{picture}
\caption{The control flow diagram for the Stochastic Variational Method.}
\label{SVMflow}
\end{figure}
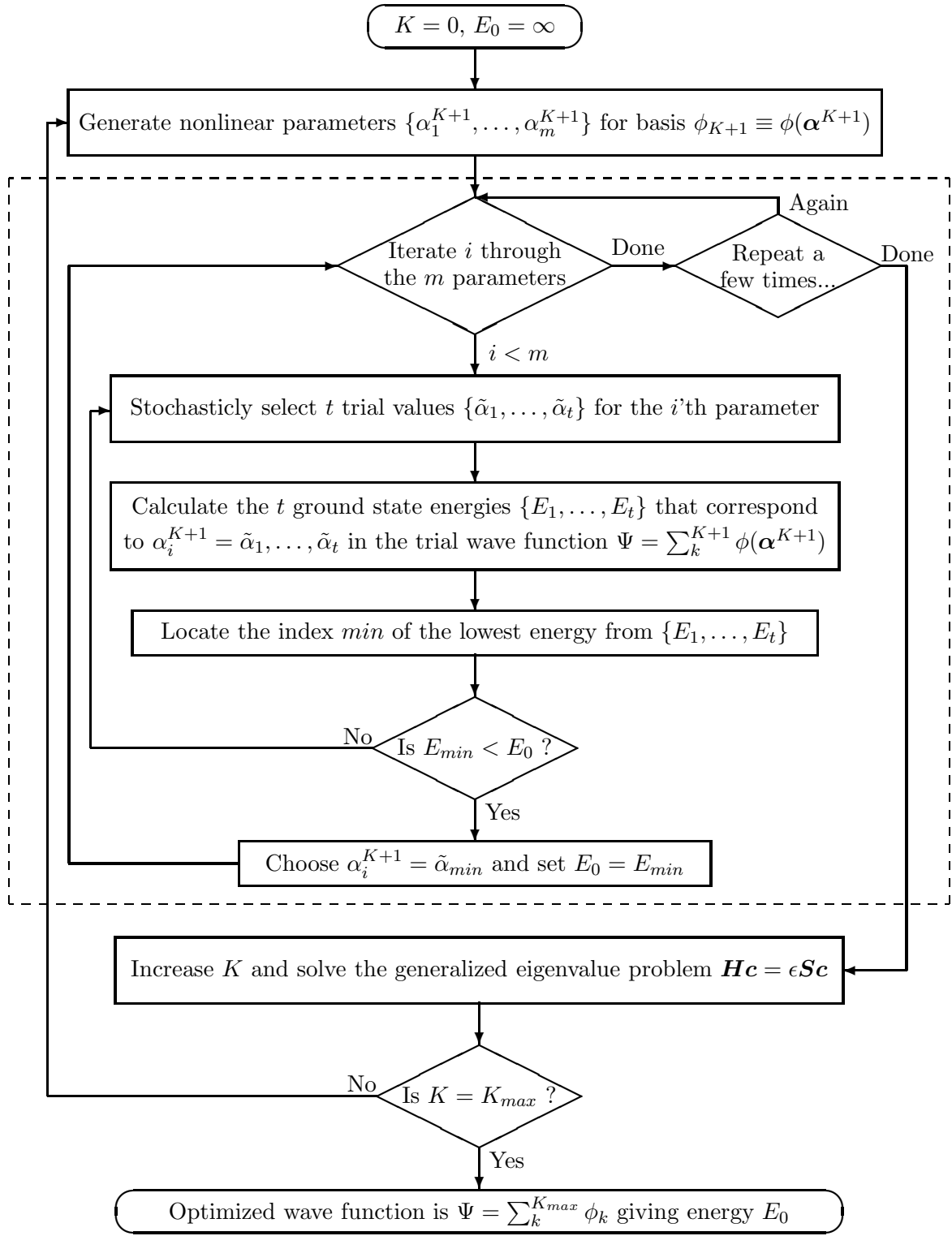

The variational foundation of the time-independent Schrödinger equation presented in chapter
\ref{theory} provides a solid and arbitrarily improvable framework for the solution of diverse
bound-state problems. A key point is that the quality of a variational calculation
crucially depends on the trial function, $\Psi = \sum_{i=1}^K c_i \psi_i$, and
consequently on the choice of the basis functions, $\{\psi_1,\dots,\psi_K\}$. Assuming
that each basis function depend on a set of nonlinear parameters,
$\psi_i\equiv \psi(\alpha^{(i)}_1, \dots, \alpha^{(i)}_p), i=1..K$, the SVM attempts to set
up the most appropriate basis by a {\it stepwise strategy}: One generates a would-be basis function by
choosing the nonlinear parameters randomly, judges it's utility by the energy gained when including it
in the basis, and either keeps or discards it. In turn, each parameter is then changed
(still randomly) in the search for additional improvement. One repeats this ``trial and error''
procedure until the basis found leads to convergence and no further energy is gained. The
control flow structure for this method is shown in figure \ref{SVMflow} (next page)
and corresponds to a detailed version of the optimization flow diagram displayed on the left of
figure \ref{optflow}.
This selection strategy has several advantages, where the most important are:
\begin{itemize}
\item{The optimization iteration (the flow within the dotted box) is clearly separated from the
computationally demanding solving of the generalized eigenvalue problem. This means that
the nonlinear parameters can be improved repeatedly without the need of diagonalizing a
$K$-dimensional matrix.}
\item{Due to the stepwise optimization procedure, a relative small number of matrix
elements have to be calculated to test a new basis function candidate and the corresponding
ground state energy is easily determined by finding the lowest root of a simple equation (see below).}
\item{The variational principle (\ref{vartheorem}) ensures that the energy of the $K$-dimensional
basis is always lower than that of an $(K-1)$-dimensional one. The procedure is therefore
guaranteed to lead to a better and better upper bound of the ground state energy.}
\end{itemize}
Even though it is rarely the case, one still has to make sure that the solution is not on the
plateau of some local minima. This is most easily done by confirming that independent calculations
starting from different first basis states (i.e. different random seeds) lead to practically
the same solution.

\subsection{Gram-Schmidt diagonalization}
\begin{figure}[bt]
\includegraphics[width=16.5cm]{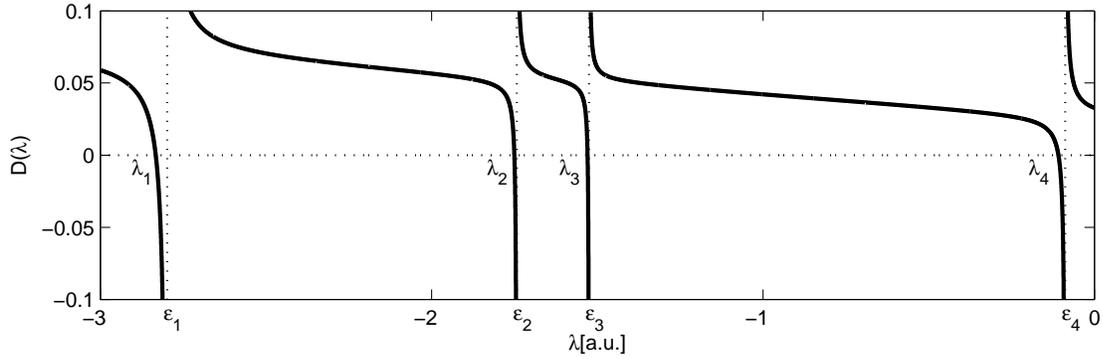}
\caption{Graphic illustration of the characteristic polynomial
$D(\lambda) = \sum_{i=1}^K \frac{\abs{h_i}^2}{(\epsilon_i-\lambda)} - \lambda + h_{K+1}$,
where the roots $\lambda_1, \dots, \lambda_{K+1}$ determine the eigenvalues of the
eigenvalue problem (\ref{deig}). The example here corresponds to the adding of a candidate
function $\psi_8$ to an existing basis $\{\psi_1, \dots, \psi_7\}$ in the calculation of
$^\infty$He (see section \ref{SecHe}). There are four additional roots ($\lambda>0$) not shown here.}
\label{GS}
\end{figure}

For the stochastic optimization to be practical, it is essential that the ground state energy
corresponding to a trial function candidate is evaluated with minimal computational effort.
Otherwise it is simply not possible to test enough candidates to cover a reasonable parameter spread.
A full diagonalization performed by solving the general eigenvalue problem is out of the question.
Fortunately this can be avoided if only one basis function, let's say $\psi_{K+1}$, is changed or
added at a time and the eigenvalue problem, $\bm{H}c = \epsilon\bm{S}c$, for the other basis
functions, $\psi_i\equiv \psi(\bm{\alpha}^{(i)}), i=1..K$, has been solved.
The idea is to evaluate the eigenvalues of the $(K+1)$-dimensional problem in a basis of
orthonormal functions. Obviously, the $K$-dimensional solution has produced eigenvalues
$\epsilon_1, \epsilon_2, \dots, \epsilon_K$ and corresponding
eigenvectors $\bm{c}^{(1)}, \bm{c}^{(2)}$, $\dots$, $\bm{c}^{(K)}$ satisfying
$\bm{c}^\dagger\bm{S}\bm{c} = 1$, and can be written in standard diagonal form
\begin{equation}
\begin{pmatrix}
\epsilon_1 & 0 & \cdots & 0 \\ 
0 & \epsilon_2 & \cdots & 0 \\ 
\vdots & \vdots & & \vdots \\ 
0 & 0 & \cdots & \epsilon_K \\ 
\end{pmatrix}
\begin{pmatrix}
d_1 \\
d_2 \\
\vdots \\
d_K \\
\end{pmatrix}
=\epsilon
\begin{pmatrix}
d_1 \\
d_2 \\
\vdots \\
d_K \\
\end{pmatrix}
\end{equation}
in a basis of orthonormal functions, $\{\phi_1, \phi_2, \dots, \phi_K\}$, where
$\phi_{i} = \sum^K_{j=1}c_j^{(i)}\psi_j, i = 1..K$.  The $(K+1)$-dimensional solution can then
be obtained by first applying Gram-Schmidt's method to construct $\phi_{K+1}$ from $\psi_{K+1}$
so that it is orthogonal to all $\phi_1, \phi_2, .. , \phi_K$, i.e. \cite{LA}
\begin{equation}\label{GramPhi}
\phi_{K+1} = \frac{\psi_{K+1}
-\sum_{i=1}^K \phi_i\langle \phi_i | \psi_{K+1} \rangle}
{\sqrt{\langle \psi_{K+1} | \psi_{K+1} \rangle
-\sum_{i=1}^K |\langle \phi_i | \psi_{K+1} \rangle|^2}}
\end{equation}
and then solving for the eigenvalues of
\begin{equation}\label{deig}
\begin{pmatrix} 
\epsilon_1 & 0 & \cdots & 0 & h_1 \\ 
0 & \epsilon_2 & \cdots & 0 & h_2 \\ 
\vdots & \vdots & & \vdots & \vdots \\ 
0 & 0 & \cdots & \epsilon_K & h_K \\
h^\ast_1 & h^\ast_2 & \cdots & h^\ast_K & h_{K+1} \\
\end{pmatrix}
\begin{pmatrix}
d_1 \\
d_2 \\
\vdots \\
d_K \\
d_{K+1} \\
\end{pmatrix}
=\lambda
\begin{pmatrix} 
d_1 \\
d_2 \\
\vdots \\
d_K \\
d_{K+1} \\
\end{pmatrix}
\end{equation}
where $h_j=\langle \phi_j | H | \phi_{K+1} \rangle$ and $h^\ast_j$ is the complex conjugate of $h_j$.
For this to work, the candidate $\psi_{K+1}$ has to be linearly independent
of the previous basis functions as required by the Gram-Schmidt orthogonalization
\footnote{In practice, this must be explicitly verified during the stochastic selection procedure.}.
The characteristic equation of (\ref{deig}) is
\begin{equation}
{\rm det}(\tilde{\bm{H}} - \lambda I) 
\equiv
\begin{vmatrix}
\epsilon_1-\lambda & 0 & \cdots & 0 & h_1 \\ 
0 & \epsilon_2-\lambda & \cdots & 0 & h_2 \\ 
\vdots & \vdots & & \vdots & \vdots \\ 
0 & 0 & \cdots & \epsilon_K-\lambda & h_K \\
h^\ast_1 & h^\ast_2 & \cdots & h^\ast_K & h_{K+1}-\lambda \\
\end{vmatrix}
=0
\end{equation}
which, assuming that all $h_i$ are nonzero, has a straightforward reduction
\begin{equation}
\begin{split}
{\rm det}(\tilde{\bm{H}} - \lambda I) 
&=
(-1)^{K+1}h^\ast_1
\begin{vmatrix} 
0 & \cdots & 0 & h_1 \\ 
\epsilon_2-\lambda & \cdots & 0 & h_2 \\ 
\vdots & & \vdots & \vdots \\ 
0 & \cdots & \epsilon_K-\lambda & h_K \\
\end{vmatrix}
+ \cdots + (h_{K+1}-\lambda)\prod_{i=1}^K(\epsilon_i-\lambda)
\\
&=\sum_{i=1}^K \Biggl(\abs{h_i}^2
\prod_{\begin{smallmatrix} j=1 \\ j\neq i\end{smallmatrix}}^K(\epsilon_j-\lambda) \Biggr)
+ (h_{K+1}-\lambda)\prod_{i=1}^K(\epsilon_i-\lambda)
\\
&=\left(\sum_{i=1}^K \frac{\abs{h_i}^2}{(\epsilon_i-\lambda)} + h_{K+1}-\lambda\right)
\prod_{j=1}^K(\epsilon_j-\lambda)
=0
\end{split}
\end{equation}
Thus, when $(\epsilon_j - \lambda) \neq 0$ ($\Leftrightarrow h_j\neq 0$), the $K+1$ eigenvalues
$\lambda_1, \dots, \lambda_{K+1}$ are obtaining by simply finding the roots of the function
\begin{equation}
D(\lambda) = \sum_{i=1}^K \frac{\abs{h_i}^2}{(\epsilon_i-\lambda)} - \lambda + h_{K+1}
\end{equation}
as graphically exemplified in figure \ref{GS}.
The variational theorem ensures that
$\epsilon_1 < \lambda_1 < \epsilon_2 < \lambda_2 < \dots \epsilon_K < \lambda_{K+1}$,
assuming both $\epsilon_1, \dots, \epsilon_K$ and $\lambda_1,\dots,\lambda_{K+1}$ are arranged
in increasing order. This is also helpful in setting up intervals for a
root-finding algoritm (see app. \ref{apRoot}).

\subsection{Refining process}
At any particular time, only one parameter out of the possible large number of nonlinear parameters
in every of
the $K$ basis functions can be considered optimal, since the optimization procedure is applied
consecutively, element by element, rather than simultaneously. It is then reasonable to
expect that at least some of the previously added basis functions are no longer optimal or
even needed. Especially when the basis functions are nonorthogonal to each other this might
be the case, since, even though none of them are really indispensable, any of them can be omitted
or changed because some others will compensate for the loss. To help shake off these flaws a
so-called {\it refining process} is introduced. After having successfully found $K_{max}$ basis
functions, one can further improve the energy without increasing the basis dimension. This is done
by iterating through the current functions still optimizing their parameters in the spirit of the
trial and error algoritm. After a few of these refining runs all of the basis functions
play an active role again, and depending on the value of $K$, this process often improve the result
considerably.

\section{Explicitly correlated basis functions}
\label{basisfunc}

From the outset, the SVM works well with any basis function
that leads to analytical closed form evaluation of the required integrals of the Hamiltonian and
overlap matrices. In addition, the stepwise optimization of the variational parameters
allows the efficient handling of a relatively large set of nonlinear parameters,
$\{ \alpha^{(i)}_1, \dots, \alpha^{(i)}_p \}$, per basis function. This is important when trying to
incorporate flexibility in the basis functions to treat complicated correlation effects.
In the following it is described how to use explicitly correlated basis functions of the
kind introduced in section \ref{explicit} with the SVM.
This treatment is applicable to both few-body and many-body systems, all though in practice,
the {\it full} correlated description is only feasible for $N<\sim5$.

As discussed in chapter \ref{correlations}, the correlation description adopted in this thesis
is based on the Jastrow-type trial function form, $\Psi = F(r_{12},r_{13},\dots,r_{(N-1)N})
\Phi(\bm{r}_1, \bm{r}_2, \dots \bm{r}_N)$, where $F$ is the correlation factor
and $\Phi$ is the mean-field model state. Moreover, the factor $F$ was approximated by
symmetrized correlation functions, i.e. $F\approx\widehat{\cal S}C_n(r_{12},r_{13},\dots,r_{(n-1)n})$
in the case where up to n-body correlations are considered. To apply this with the SVM, it
is necessary to expand the $C_n$'s in a mathematically complete set of
functions where each term is simple enough to give analytical expressions for the matrix elements.
Both Varga \cite{SVM} and Wilson \cite{Wilson} argue that the only set of functions
which meets such requirements for $N$-body systems is the so-called {\it contracted Gaussian basis}
\footnote{Corresponding to the $l=0$ case of the nodeless harmonic-oscillator basis.}
(i.e. Gaussians with different widths).
For example, in the case of pair-correlation, this leads to the simple expansion
\begin{equation}
C_2(r_{12}) =\sum_k^{k_{max}} c_k {\rm exp}\Big(-\½ \alpha_k r_{12}^2\Big)
\end{equation}
where it is indicated that in practice the sum must be truncated at some finite level.
In general, the expansion of the $n$'th order correlation function becomes
\begin{equation}\label{GaussEx}
C_n(r_{12},r_{13},\dots,r_{(n-1)n}) = \sum_k^{k_{max}} c_k
\widehat{\cal S}\ {\rm exp}\Big(-\½ \sum_{i<j=1}^n \alpha^{(k)}_{ij}r_{ij}^2\Big)
\end{equation}
where the symmetrization operator, $\widehat{\cal S} = \frac{1}{\sqrt{n!}}\sum_P \widehat{P}$,
includes perturbation terms for the first $n$ particles only.
There are many, possibly an infinite number of expansion sets, which approximate a given
function by Gaussians equally well
\footnote{Heuristic discussions on the completeness and fast convergence of Gaussians
can be found in C6.1 of \cite{SVM}, the appendix of
\cite{Bukowski} and in \cite{King}.}
. This makes the Gaussians an appropriate basis for stochastic optimization.
 
\subsection{Correlated Gaussian basis}\label{cgauss}
If expressed using the Gaussian expansion (\ref{GaussEx}), the N-body variational trial function
can be written in the desired form, $\Psi=\sum_k^K c_k \psi_k$, where the basis functions are
given by
\begin{equation}\label{ECG}
\psi_k=\Phi(\bm{r}_1, \bm{r}_2, \dots \bm{r}_N)\widehat{\cal S}\ 
{\rm exp}\Big(-\½ \sum_{i<j=1}^N \alpha^{(k)}_{ij}r_{ij}^2\Big)
\end{equation}
and the set $\{\psi_k\}$ is both non-orthogonal and over-complete
(i.e. satisfying the requirement that results obtained
by a systematic increase of the number of basis functions, will converge to the exact eigenvalues).
These functions correspond to symmetrized sums of the {\it explicitly correlated Gaussians}
originally suggested in 1960 independently by Boys \cite{Boys} and Singer \cite{Singer}.
Over the years, they have been demonstrated to be
an excellent basis for high accuracy variational calculations of few-body problems
\cite{SVM,Kukulin,Chen,Kinghorn1,Bubin}.

To utilize the correlated Gaussian basis in a translationally independent N-body solution
it is necessary to write eq. (\ref{ECG}) in terms of the
independent Jacobi coordinates defined by the matrix (\ref{UJ}).
Inverting the linear transformation (\ref{xtrans}), one has 
\begin{equation}\label{invtrans}
\bm{r}_{i} = (\bm{u}^{(i)})^T \bm{x}, \quad {\rm and} \quad
\bm{r}_{ij} = \bm{r}_i - \bm{r}_j = (\bm{u}^{(ij)})^T \bm{x},
\end{equation}
where $\bm{u}^{(ij)} = \bm{u}^{(i)} - \bm{u}^{(j)}$ and the vectors $\bm{u}^{(i)}$ have
components $u^{(i)}_k=(U^{-1})_{ik}$.
The summation in the exponent of the correlated Gaussians can then be written
\begin{align}\label{xAx}
\nonumber
\sum_{i<j=1}^N \alpha_{ij}(\bm{r}_i - \bm{r}_j)^2&= 
\sum_{i<j=1}^N \alpha_{ij}
(\sum^{N-1}_{k=1} u^{(ij)}_k \bm{x}_k)
(\sum^{N-1}_{l=1} u^{(ij)}_l \bm{x}_l)
\\&=\bm{x}^T \hspace{-2.0pt} \bm{A} \bm{x},
\end{align}
where
\begin{equation}\label{Amatrix}
\bm{A} = \sum_{i<j=1}^N \alpha_{ij} \bm{\Upsilon}^{(ij)},
\end{equation}
\begin{equation*}
\Upsilon^{(ij)}_{kl}= \bm{u}^{(ij)} (\bm{u}^{(ij)})^T
=((U^{-1})_{ik}-(U^{-1})_{jk}) ((U^{-1})_{il}-(U^{-1})_{jl}).
\end{equation*}
The matrices
$\bm{\Upsilon}^{(ij)}$ with $i,j=1, \dots,N$, and hence $\bm{A}$,
are symmetric $(N-1) \times (N-1)$ matrices.
Assuming that the model state, $\Phi$, can also be expressed in Jacobi coordinates as
$\Phi(\bm{r}_1,\dots,\bm{r}_N)\rightarrow
\tilde{\Phi}_{int}(\bm{x}_1,\dots,\bm{x}_{N-1})\tilde{\Phi}_{cm}(\bm{x}_N)$,
the correlated Gaussians in the form (\ref{ECG}) are equivalent to
\begin{equation}\label{ECG2}
\psi_k=\tilde{\Phi}_{int}(\bm{x}_1,\dots,\bm{x}_{N-1})\tilde{\Phi}_{cm}(\bm{x}_N) \widehat{\cal S}\ 
{\rm exp}\Big(-\½ \bm{x}^T \hspace{-2.0pt} \bm{A}^{(k)} \bm{x} \Big)
\end{equation}
where the $\frac{N(N-1)}{2}$ independent entries of $\bm{A}$ are related
to the $\alpha_{ij}$ parameters via (\ref{Amatrix}) and
\begin{equation}
\alpha_{ij}=-(\bm{U}^T \hspace{-2.0pt} \bm{A} \bm{U})_{ij} \ \ \ (i<j)
\end{equation}
One should note, that a necessary condition for the correlated Gaussians in (\ref{ECG}) and
(\ref{ECG2}) to be square integrable, and thus have a finite
norm, is that the parameters $\alpha_{ij}$ are positive and $\bm{A}$ is positive
definite (i.e. $\bm{x}^T \hspace{-2.0pt} \bm{A} \bm{x} > 0$,
for all nonzero vectors $\bm{x} \in  \mathbb{R}^{(N-1)}$) \cite{PosDef}. 
This must be explicitly checked when trying to optimize $\psi_k$ by ``guessing'' with the SVM.

The success of the correlated Gaussian basis is mainly linked to the amazingly simple form
and consequently fast computation of the resulting matrix elements (the expressions are
evaluated in detail in appendix \ref{MatElm}).
Moreover, the form (\ref{ECG}), with explicit correlation
parameters $\alpha_{ij}$, presents a natural physical interpretation.
For a one-dimensional Gaussian function, $e^{-\½ \alpha r^2}$,
the position expectation value is $\langle r \rangle=2 /\sqrt{\alpha \pi}$.
Hence, when using the form (\ref{ECG}) and $\alpha_{ij}$ as variational parameters,
$1/\sqrt{\alpha_{ij}}$ can be viewed as an average ``distance''
between particles \cite{SVM}. This makes (\ref{ECG}) advantageous when
setting up initial values for a variational procedure or in random selection,
since valid intervals for the particle distances can be estimated from physical intuition
(bound or trapped particles are not expected to move far away from each other!).

\subsection{Correlated exponential basis ($N=3$ only)}

The Gaussian expansion is not always economical in describing the asymptotic behavior of the wave
function at large distances. Only for Gaussian-type interaction potentials, harmonic oscillator
potentials and in a few other cases does the exact wave function have a Gaussian asymptotic
dependence. In the case of Coulomb systems and a wide number of similar potentials, e.g. exponential
and Yukawa-type, the wave function has an exponential asymptotic. 
Moreover, the Gaussian expansion does not give a correct value for some
specific short-range quantities such as the Kato cusp condition (see \cite{Krivec}).
It is therefore interesting to consider the SVM with a trial function based on
a correlated exponential basis.

An exponential basis is, however, not amenable to analytical evaluation
of the matrix elements for a system of more than three particles \cite{Frolov5}. Consequently,
only the $N=3$ case will be considered here. The application of the exponential expansion, gives
\begin{equation}
C_3(r_{12},r_{13},r_{23}) = \sum_k^{k_{max}} c_k
\widehat{\cal S}\ {\rm exp}(-\alpha_k r_{12}-\beta_k r_{13}-\gamma_k r_{23})
\end{equation}
for the triplet-correlation function. Written in terms of the interparticle distances,
denoted by $\bm{x}^T=(\bm{r}_{12},\bm{r}_{13},\bm{r}_{23})$ for notational convenience,
the corresponding correlated exponential basis functions becomes
\begin{equation}\label{ECE2}
\psi_k=\tilde{\Phi}_{int}(\bm{x}_1,\bm{x}_2,\bm{x}_3)
\widehat{\cal S}\ {\rm exp}(-\bm{a}^{(k)T} \bm{x} )
\end{equation}
where $\bm{a}^{(k)T}=(\alpha_k,\beta_k,\gamma_k)$.
The details of calculating the matrix elements for this basis can be found in appendix \ref{ExpElm}.
During the SVM optimization procedure, the inverse parameters $\alpha_k^{-1},\beta_k^{-1}$
and $\gamma_k^{-1}$, are advantageously selected from those intervals in which the average distances
between particles is expected to vary
\footnote{Follows from the discussion at the end of section \ref{cgauss}, since in the case of a
normalized exponential function, $n(\alpha)e^{-\½ \alpha r}$, the expectation
value of the distance, $r$, is $\langle r \rangle \sim \alpha^{-1}$, \cite{Dahl}.}.

\section{Symmetrization}\label{symmet}

As outlined in section \ref{sym} the variational wave function should be either totally
symmetric (bosons) or totally antisymmetric (fermions) under the interchange of identical particles.
In the description leading to the correlated basis functions (\ref{ECG2}) and (\ref{ECE2}),
the model state, $\Phi_{int}$, is assigned the role of imposing the proper overall symmetry.
The most convenient way to achieve this with the SVM is by operating on
the basis functions with the proper sum of permutation operators as defined by $\widehat{\cal P}$
in (\ref{P}).  The specific permutations included in the sum of $\widehat{\cal P}$ depends on
which particles are identical.

In practice, it is helpful to represent a permutation $P=(p_1, p_2, \dots, p_N)$ of $N$ particle
indices by a matrix, $\bm{T}_P$, having elements \cite{SVM}
\begin{equation}
(T_P)_{ij}=\delta_{jp_i}, \ \ i,j=1,2,\dots,N
\end{equation}
Then, in the case of $N$ identical particles, there are $N!$ different $\bm{T}_P$ matrices of size
$N\times N$ and a specific permutation of the single-particle coordinates,
$P:\bm{r}_i \rightarrow \bm{r}_{p_i}$, is simply written
\begin{equation}
\widehat{P} \bm{r} = \bm{T}_P \bm{r}
\end{equation}
Applying transformation (\ref{xtrans}), the corresponding permutation will induce a linear
transformation of the relative (e.g. Jacobi) coordinates, given by
\begin{equation}\label{permJ}
\widehat{P} \bm{x} = \tilde{\bm{T}}_P \bm{x},
\ \ {\rm where} \ \tilde{\bm{T}}_P = \bm{U} \bm{T}_P \bm{U}^{-1}
\end{equation}
Since the center-of-mass coordinate is unchanged under coordinate permutation and
hence can be ignored, the size of $\tilde{\bm{T}}_P$ is $(N-1)\times(N-1)$.
This way, the effect of $P$ operating on the explicitly correlated
functions, $\psi_k$, in (\ref{ECG2}) and (\ref{ECE2}) can be reduced to the simple replacements
\begin{equation}
P:\bm{A}^{(k)} \rightarrow \tilde{\bm{T}}_P^T \hspace{-1.0pt} \bm{A}^{(k)} \tilde{\bm{T}}_P
\quad {\rm and} \quad 
P:\bm{a}^{(k)T} \rightarrow \bm{a}^{(k)T} \tilde{\bm{T}}_P
\end{equation}

With this approach, one can operate on $\psi_k$ with the symmetrizer
$\widehat{\cal S} \equiv \frac{1}{\sqrt{N!}}\sum_P \widehat{P}$,
where all $N!$ permutations are included in the sum, in the cases where the basis function
is required to be symmetric,
If the spatial function is to be antisymmetric one should use the antisymmetrizer
$\widehat{\cal A} \equiv \frac{1}{\sqrt{N!}}\sum_P (-1)^p\widehat{P}$,
where p=0,1 is the parity of permutation $P$.
In both cases, however, this will produce $N!^2$ terms in a single matrix element calculation.
This general number is readily reduced to $N!$ since the correlated basis functions are of the
product form $\psi_k = \prod_{i<j}^N \phi_{ij}(r_{ij})$ (with $\phi_{ij}$ being either
a Gaussian of an exponential) so that any symmetric operator $\widehat{O}$ satisfies
$\langle \widehat{P}_{ij} \psi_{k^\prime} | \widehat{O} | \psi_{k} \rangle
= \langle \psi_{k^\prime} | \widehat{O} | \widehat{P}_{ij} \psi_{k} \rangle$.
All matrix elements for $\widehat{\cal P} \psi_k$ functions can then be written 
\begin{equation}\label{symbas}
\langle \widehat{\cal P} \psi_{k^\prime} | \widehat{O} | \widehat{\cal P} \psi_{k} \rangle
= \frac{2}{N!}\langle \psi_{k^\prime} | \widehat{O} | \sum_P \alpha_P \widehat{P} \psi_{k} \rangle
\end{equation}
where $\alpha_P=1$ for bosons and $\alpha_P=(-1)^p$ for fermions and
the single sum is over all $N!$ permutations of identical particles
for both $\widehat{\cal P} = \widehat{\cal S}, \widehat{\cal A}$.
Unfortunately, the matrix element (\ref{symbas}) is still an ${\cal O}(N!)$ computation
which is only tractable for few-body systems ($N <\sim 5$).
Further simplifications of $\psi_k$ has to be assumed to handle many-body problems
(see e.g. section \ref{BECTwo}).
 
\section{Few-body system: The Helium atom} \label{SecHe}

The nonrelativistic ground state energy of the Helium atom has been a benchmark test for three-body
calculations since the pioneering work of E. A. Hylleraas \cite{Hylleraas}, 75 years ago. 
Recently, this subject has attracted much attention
\cite{Frolov1,Drake1,Korobov,Schwartz,Koga,Persson} and significant
progress has been made, with the accuracy of the energy now at 36 decimals \cite{Schwartz}.
As a brief illustrative example, the SVM is now applied to the (1$^1$S) ground state of $^\infty$He.
The numerical results is presented in the next chapter and used to test the implemented
computer program and the rate of convergence.

The Helium three-body system consists of two indistinguishable electrons (labels 1 and 2) and
an $\alpha$-nucleus (3). Neglecting relativistic effects, the two-body interaction is
exclusively Coulombic, with $\sum_{i<j}^N V_{ij}=\frac{1}{r_{12}}-\frac{2}{r_{13}}-
\frac{2}{r_{23}}$.
The Hamiltonian (\ref{Hj}) written in relative coordinates
$\bm{x}^T=(\bm{r}_{12},\bm{r}_{13},\bm{r}_{23})$ is then
\begin{equation}\label{HHe}
\widehat{H} = -\½  \widehat{\bm{\nabla}}_{\bm{x}}^T \bm{\Lambda} \widehat{\bm{\nabla}}_{\bm{x}}
+\frac{1}{x_1}-\frac{2}{x_2}-\frac{2}{x_3},
\ \ \rm{with} \ 
\bm{\Lambda} =
\begin{pmatrix} 
\frac{1}{2} & 1 & -1 \\ 
1 & \frac{1}{\mu_{\alpha}} & \frac{1}{m_{\alpha}} \\ 
-1 & \frac{1}{m_{\alpha}} & \frac{1}{\mu_{\alpha}} \\ 
\end{pmatrix}
,
\end{equation}
where $m_{\alpha}$ is the mass of the $\alpha$-nucleus and
$\mu_{\alpha}=\frac{m_{\alpha}}{(1+m_{\alpha})}$ is the reduced mass. In the present
calculation we use $m_{\alpha}=\infty$ making $\mu_{\alpha}=1$, 
\footnote{Alternatively one might use the exact value $m_{\alpha}=7294.2618241$.}.
Following the approach described in the previous sections the trial function for the
Helium ground state can be constructed from exponential basis functions:
\begin{equation}\label{HeB}
\Psi=\sum_{k=1}^K c_{k} \psi_{k}, \ \ \ \ 
\psi_{k}=\widehat{\cal A}\{\chi_{00}\ {\rm exp}(-\alpha_k x_1-\beta_k x_2-\gamma_k x_3)\}
\end{equation}
where $\widehat{{\cal A}}$ is the antisymmetrizer,
$\chi_{00}=\frac{1}{\sqrt{2}}\{\alpha(1)\beta(2)-\beta(1)\alpha(2)\}$
is the two-electron singlet spin function arising from the coupling of two spin-$\½$ particles
\cite{Dahl} and  $\alpha_k, \beta_k$ and $\gamma_k$ are nonlinear parameters.
An angular part in the trial function is not necessary for the $L=0$ ground state
calculation. Since $\chi_{00}$ is constant although antisymmetric under the interchange of the
identical electrons, it can be omitted if the antisymmetrizer is changed to
$\widehat{\cal A} \rightarrow 1+ \widehat{P}_{12}$ ($\equiv \widehat{\cal S}_{12}$),
where $\widehat{P}_{12}$ denotes a simple exchange of labels 1 and 2.

All the necessary matrix elements are evaluated in appendix \ref{ExpElm}. Since the basis functions
are chosen to be real, both the overlap and Hamiltonian matrices
are symmetric and the secular equation, $\bm{H}\bm{c} = \epsilon\bm{S}\bm{c}$, can be
solved effectively by well-known linear algebra methods \cite{Mar}. The lowest of the eigenvalues
found, $\epsilon_1$, will then be the approximate ground state energy of Helium. The
quality of the result will depend on the specific values chosen for the nonlinear parameters,
$\alpha_k$,$\beta_k$ and $\gamma_k$, and the size of the basis, $K$.

\section{Many-body system: Bose-Einstein Condensate}\label{BEC}

The main goal of this thesis is to discuss correlations in three- and four-boson systems.
In the following
is described how to employ the SVM to a system of $N$ identical bosons trapped by
an isotropic harmonic oscillator and interacting via two-body potentials $V_{ij}$.
Moreover, four different levels of correlation are explicitly allowed in the variational trial
function, $\Psi$, ranging from mean-field to the {\it full} $N$-body correlated treatment (as
derived in chapter \ref{correlations}).

Expressed in terms of the Jacobi coordinates defined in (\ref{Jacobi}) the many-body Hamiltonian
describing the internal motion of a trapped N-boson system can be written 
\begin{equation}\label{Hjac}
\widehat{H}_{int} =\widehat{H} - \widehat{H}_{cm}=
\sum^{N-1}_{i=1} \Big[-\frac{\hbar^2}{2m}\widehat{\bm{\nabla}}_{\bm{x}_i}^2
 + \½ m \omega^2 \bm{x}_i^2\Big] + \sum^N_{i<j}V_{ij}
\end{equation}
where $m$ is the boson mass and $\omega$ is the trapping frequency and
\begin{equation}\label{Hcmjac}
\widehat{H}_{cm} = -\frac{\hbar^2}{2Nm}\widehat{\bm{\nabla}}_{\bm{x}_N}^2 +\½Nm\omega^2 \bm{x}_N^2
\end{equation}
is the center-of-mass Hamiltonian. It is apparent, that $\widehat{H}_{cm}$ represents the standard
form of the three-dimensional harmonic oscillator having ground state energy
$E_{cm,0} = \½ \hbar \omega$, \cite{Dahl}. The total BEC energy is $E_0=E_{int,0}+E_{cm,0}$,
where $E_{int,0}$ is to be calculated by the SVM.

\subsubsection{Numerical computation in harmonic oscillator units}
In numerical calculations with limited precision arithmetic the optimal average order
of magnitude of the numbers handled is $\sim 1$. To meet this demand in the case of BEC problems
it is convenient to abandon the atomic units and do the numerical computation in the harmonic
oscillator units, given by
\begin{equation}
\epsilon_{ho} \equiv \hbar \omega, \quad a_{ho} \equiv \sqrt{\frac{\hbar}{m\omega}}
\end{equation}
where $\epsilon_{ho}$ is the unit energy and $a_{ho}$ the unit length.
With all lengths ($\bm{r}_i, b$, etc.) in units of $a_{ho}$ and all energies
($\widehat{H}, \widehat{H}_{cm}, V_{ij}$, etc.) in units of $\epsilon_{ho}$, the BEC Hamiltonian
(\ref{Hjac}) becomes
\begin{equation}
\widehat{H}_{int} =
-\½ \sum^{N-1}_{i=1} \Big[\widehat{\bm{\nabla}}_{\bm{x}_i}^2
 - \bm{x}_i^2\Big] + \sum^N_{i<j}V_{ij}
\end{equation}
and the ground state energy of a noninteracting trapped gas is just
$E_0 = \frac{3}{2} N \epsilon_{ho}$. This means,
that for a reasonable number of particles $< 10^6$, the magnitude of the results
in the harmonic oscillator units are also reasonable ($\sim N$). However,
care must be taken during evaluation of the matrix elements since they can reach much greater values
and are the main source for loss of accuracy. For very large $N$ and in some other cases
additional rescaling is required
\footnote{A technique for scaling the magnitude of the overlap
$\langle \psi_{k^\prime} | \psi_{k} \rangle$ is demonstrated in appendix \ref{AppScale}}.

\subsection{Selecting the BEC ground state}

In the case where the bosons are interacting attractively, the eigenstate with lowest energy
is not necessarily the BEC state we are looking for. If the scattering length is large
enough and $N>2$, the bosons may form ``molecular-type'' many-body bound
states even when the boson-boson interaction potential is too shallow to support two-body bound
states. These states could as well be characterized as condensed (N-body) states but they exists
only at high densities making them unstable to recombination processes. In addition, such bound states
do not have the distinctive BEC features (e.g. density profile) obtained in experiments
\cite{Pitaevskii2} and it is therefore important to select the correct ``gas-like'' condensate
state as the target of the SVM.

\begin{figure}[bt]
\psfrag{XLABEL}{$\rho/b_t$}
\psfrag{YLABEL}{$U(\rho)\ [\hbar\omega]$}
\includegraphics[width=16cm]{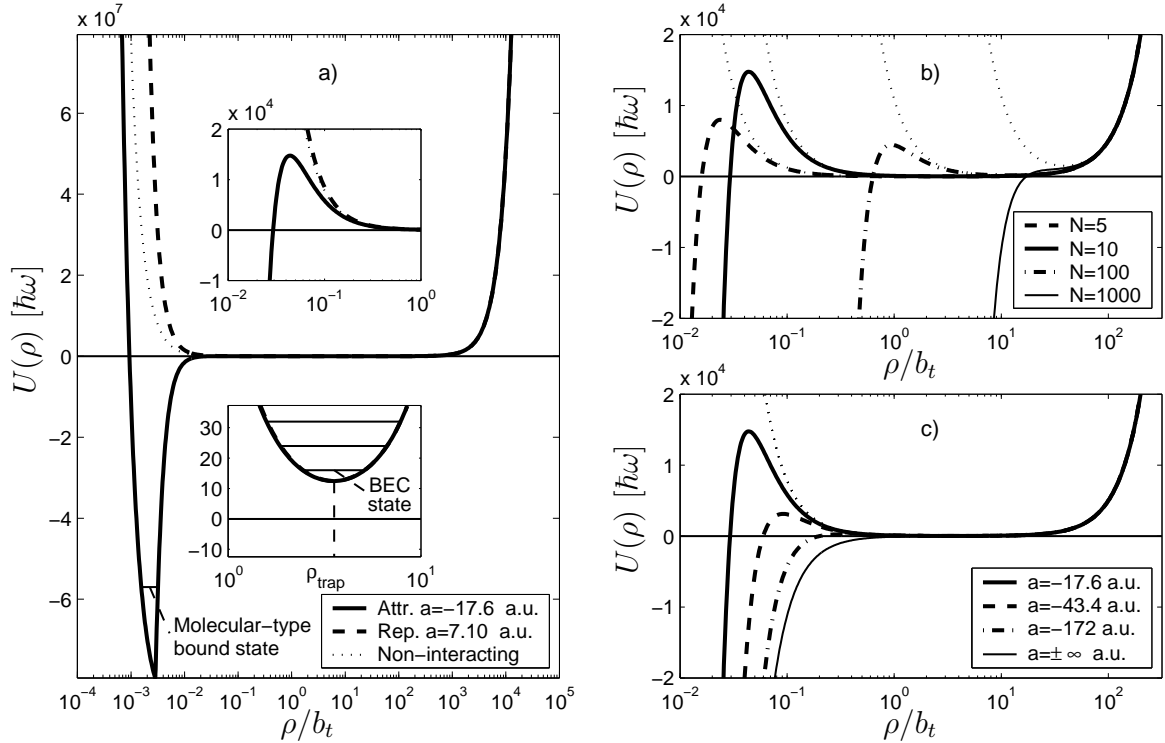}
\caption{The effective boson-boson potential (\ref{effpot}) as a function of the
hyperradii for a Gaussian two-body interaction, $V(x) = V_0 e^{-x^2/b^2}$, with $b=18.9$ a.u. 
a) The case of a weak attractive and a weak repulsive
interaction for $N=10$. The insets show the finer details of the barrier region and the
trap center.
b) The details of the barrier region when the scattering length is fixed at $a=-17.6$ a.u. and
the number of particles is varied and c) when the scattering length is varied.
}
\label{effpotgraph}
\end{figure}

The characteristic difference between the self-bound and the trapped condensate in the attractive
boson system is their spatial extension. To illustrate this, it is convenient to examine the
effective potential experienced by a boson as a function of the hyperradius, $\rho$, i.e.
the average distance between the bosons in the trap (see definition
(\ref{hyperradius})). A rough sketch of the behavior (details can be found in \cite{Ole, Sogo})
for the $N=10$ case is outlined in figure \ref{effpotgraph}
\footnote{The model potential used for this graph is derived in \cite{Ole} by use of the adiabatic
hyperspherical expansion method and composed of terms for the external trap
$(\sim\rho^2)$, the generalized centrifugal barrier $(\sim\rho^{-2})$ and an interaction
part from the angular equation $(\sim\lambda(\rho))$ as
\begin{equation}\label{effpot}
U(\rho)=\frac{\hbar^2}{2m}\Big[\frac{\lambda(\rho)}{\rho^2}
+\frac{(3N-4)(3N-6)}{4\rho^2}+\frac{\rho^2}{b_{t}^4}\Big]
\end{equation}
where $b_t = a_{ho} \equiv \sqrt{\hbar/m\omega}$ is the trap length.
}
a) (solid line) and shows a global minimum at low $\rho$ and a second minimum (with $U(\rho)>0$) at
hyperradii around the center of the trap ($\rho_{trap}\equiv b_t \sqrt{3N/2}$). The effective
potential for the corresponding repulsive interaction (dashed line) has only the second minimum and
is almost indistinguishable from the solid line in the bottom inset. This second minimum supports
(quasi-stationary) states with the characteristic features of Bose-Einstein condensates and the
lowest of the corresponding eigenstates is the BEC ground state of interest here.

Considering the attractive boson system in detail, it becomes apparent that the barrier enclosing the
BEC minimum gradually declines as the number of particles or the scattering
length increases. This is demonstrated in graph b) and c) of fig. \ref{effpotgraph}.
The barrier completely disappears roughly when $|a|N/b_t > \sim0.67$, as derived previously by many
authors \cite{Pethick,Blume2,Sogo}. This is the well-known limit where the Gross-Pitaevskii mean-field
theory breaks down. However, in the current method, the BEC eigenstate does not collapse when $a$ is
increased to where the barrier vanishes, but instead, transcends smoothly down the potential hill,
to become a weakly bound so-called {\it Efimov state}
\footnote{The many-body Efimov states are unavoidable for large scattering lengths and
are located in the barrier-absent plateau region of fig. \ref{effpotgraph} c)
(for $a=\pm \infty$), far outside the range of the two-body interaction but before the confining
wall of the trap \cite{Esben}.}. At the same time, increasing the scattering length also makes
the first potential minimum deeper and, independently of the forming Efimov
state, allows more and more (lower lying) molecular-type bound states to appear.

Taking these conditions into account, the SVM has to be targeted
to calculate the energy of the BEC state in a special way.
From the outset, the trial and error procedure can be designed to minimize the variational
energy, $\epsilon_i$, of any given eigenstate, $\Psi_i$, in accordance
with the general solution (\ref{super}) and the variational theorem (\ref{varthe2}).
However, to be able to specify the number, $i$, of the BEC eigenstate, requires that one knows
the exact number of lower lying self-bound states (including Efimov states) before the calculation.
All though crude estimates for the number of bound state exists (see \cite{Ole}) they are not
nearly precise enough to be applicable.
The best alternative, which has been chosen from the many different schemes tested during this work,
is to have the SVM always minimize the lowest positive eigenvalue. This means, that the algoritm
will determine, at runtime, what eigenstate of the current basis corresponds to the lowest positive
eigenvalue, and add the particular basis candidate that minimizes this eigenvalue.
The target state will then automaticly increase by one each time another bound state appears
in the solution. This procedure is illustrated in practice in section \ref{N4res}.

\subsection{Mean-field}
The single-particle wave functions, $\phi_0(\bm{r}_i)$,
for the ground state of non-interacting bosons trapped by a spherical symmetric external field,
have a Gaussian form \cite{Pethick}, i.e. $\phi_0(\bm{r}_i)\sim {\rm exp}(-r_i^2/2b_t)$,
where $b_t = a_{ho} \equiv \sqrt{\hbar/m\omega}$ is the trap length. Using relation
(\ref{hyperradius}), this leads to a Hartree mean-field wave function that can be expressed in
terms of the hyperradius $\rho$:
\begin{equation}
\Phi_{HF}=\prod_{i=1}^N \phi_0(\bm{r}_i)\sim {\rm exp}\Big(-\sum_{i=1}^N r_i^2/2b_t\Big)
={\rm exp}\Big(-NR^2/2b_t\Big){\rm exp}\Big(-\rho^2/2b_t\Big)
\end{equation}
where $\bm{R}$ is the center-of-mass coordinate. Since the hyperradius is defined by
$\rho^2 \equiv \frac{1}{N} \sum_{i<j}^{N} \bm{r}_{ij}^2$,
any explicit dependence of $\rho$ corresponds to a mean-field influence, that is, the same
correlation between any pair of particles. As Boson-boson interactions modify the
Gaussian shape of the non-interacting system, an appropriate mean-field trial function
would be an expansion in Gaussians depending solely on $\rho$, e.g.
\begin{equation}
\Psi_{\rm MF}=\sum_{k=1}^K c_{k} \psi_{k}, \quad \quad
\psi_k={\rm exp}\Big(-\½ N\alpha^{(k)} \rho^2\Big)
\end{equation}
where the center-of-mass dependence is explicitly removed. Transforming to Jacobi coordinates
with $\rho^2=\sum_{i=1}^{N-1} \bm{x}_i^2$, these basis functions become simple editions of the
explicitly correlated Gaussians (i.e. (\ref{ECG2}) with $\bm{A}^{(k)}=\alpha^{(k)}\bm{I}$ and
$\tilde{\Phi}_{int}=\tilde{\Phi}_{cm}=1$). It is straight forward to insert
$\bm{B}^{-1}=(\alpha^{(k^\prime)}+\alpha^{(k)})^{-1}\bm{I}$ together with identities (\ref{JacId})
in the matrix element expressions for the correlated Gaussians given in appendix \ref{GaussElm} and
find \footnote{Expressions for $v(1/2)$ can be derived from (\ref{vCoulomb})-(\ref{vDelta}).}
\begin{align}
S_{k^\prime k} &= \langle \psi_{k^\prime} | \psi_{k} \rangle =
\bigg( \frac{2\pi}{\alpha^{(k^\prime)}+\alpha^{(k)}} \bigg)^{3(N-1)/2}\\
H_{k^\prime k} &= \langle \psi_{k^\prime} | \widehat{H}_{int} |\psi_{k} \rangle=\bigg(\frac{3(N-1)}{2}
\frac{m\omega^2 - \alpha^{(k^\prime)}\alpha^{(k)}}{N\alpha^{(k^\prime)}+N\alpha^{(k)}}
+ \frac{N(N-1)}{2}v(1/2) \bigg) S_{k^\prime k}
\end{align}
for the overlap and Hamiltonian mean-field matrix elements.

\subsection{Two-body correlations}\label{BECTwo}
In a dilute system of interacting particles one can often assume that, at any given time,
only two particles are close enough to each other to interact \cite{Ole1}.
The rest of the particles are only ``feeling'' the mean-field.
In such a situation, most often defined by  $n|a|^3 \ll 1$, where $n$ is the density \cite{Pitaevskii2},
one should expect two-body correlations to be the dominant interparticle
relationship and an appropriate trial function to describe this would be
\begin{equation}\label{dilute}
\Psi_{\rm 2B}=\sum_{k=1}^K c_{k} \psi_{k}, \quad \quad \psi_k=
\widehat{{\cal S}}\ {\rm exp}\Big(-\½ (\beta^{(k)}-\alpha^{(k)}) \bm{r}_{12}^2\Big)
{\rm exp}\Big(-\½ N\alpha^{(k)} \rho^2\Big)
\end{equation}
where all pairs have the same mean-field correlation parameter, $\alpha^{(k)}$, except one pair
(particle 1 and 2 in the first term of $\widehat{{\cal S}}$) that are correlated by $\beta^{(k)}$.
The symmetrization makes sure that all separate pairs are taken into account in the same fashion.
This $\psi_{k}$ is already in the favorable form of a sum of the explicitly correlated Gaussian
basis functions (\ref{ECG}). In addition, due to the fact that they are so simple, it is
possible to derive analytical expressions for the matrix elements of $\widehat{H}_{int}$,
that are independent of the number of particles in the system (i.e.
with computational complexities ${\cal O}(1)$). The formulas and further details can be found in
appendix \ref{TwoElm}.

\subsection{Higher-order correlations}\label{BECHigher}
While the dominant effect of interactions in dilute gases when $n|a|^3\ll 1$ is due to two-body
encounters, three- and higher-order correlations should become more and more
important with increasing density, $n$, or scattering length, $a$.
The specific SVM trial function
that includes the up to m-body correlations ($m \le N$) of these cases, can be written
\begin{equation}\label{higher}
\Psi_{\rm mB}=\sum_{k=1}^K c_{k} \psi_{k}, \quad \quad \psi_k=
\widehat{{\cal S}}\ 
{\rm exp}\Big(-\½ \sum_{i<j}^m \alpha^{(k)}_{ij}\bm{r}_{ij}^2\Big)
{\rm exp}\Big(-\½ N\alpha^{(k)} \rho^2\Big)
\end{equation}
where $\psi_k$ is a symmetrized sum of explicitly correlated Gaussians.
In this expression, the individual correlations of the $(m+1)m/2$ particle pairs is represented by an
equal number of nonlinear parameters $\alpha^{(k)}_{ij}$. Unfortunately, a {\it full} correlated
treatment with $m=N$ requires $N!$ different term in $\widehat{{\cal S}}$ and limits it's
usability to BEC's of only a few atoms (maximum of $N=5$ in this work). 

\chapter{Numerical results}\label{Results}

This chapter presents the numerical results obtained by applying the SVM to the
N-body systems introduced in sections \ref{SecHe} and \ref{BEC}. Starting with the well explored
$^\infty$He atom, the implemented computer program is first thoroughly tested and
bench marked. It is illustrated that the convergence rate is fastest when the basis functions
represent the asymptotic behavior of the exact wave function well. The main calculations will
subsequently treat BEC systems of sizes $N=3, 4$ and $10$, where the bosons are interacting
attractively over a wide range of scattering lengths (e.g. $-\infty < a < 0$).
Detailed graphs of the lowest energy levels are presented and it is shown how to distinguish between
``gas-type'' and ``molecular-type'' eigenstates. In addition, each individual calculation is
repeated three times with the SVM trial function including different degrees of correlation
({\it mean-field}, {\it two-body} and {\it full} correlations) giving a clear indication of
the importance and effect of correlations in such systems.

\section{Method test: $^\infty$He}\label{ResHe}
\begin{figure}[bt]
\includegraphics[width=16cm]{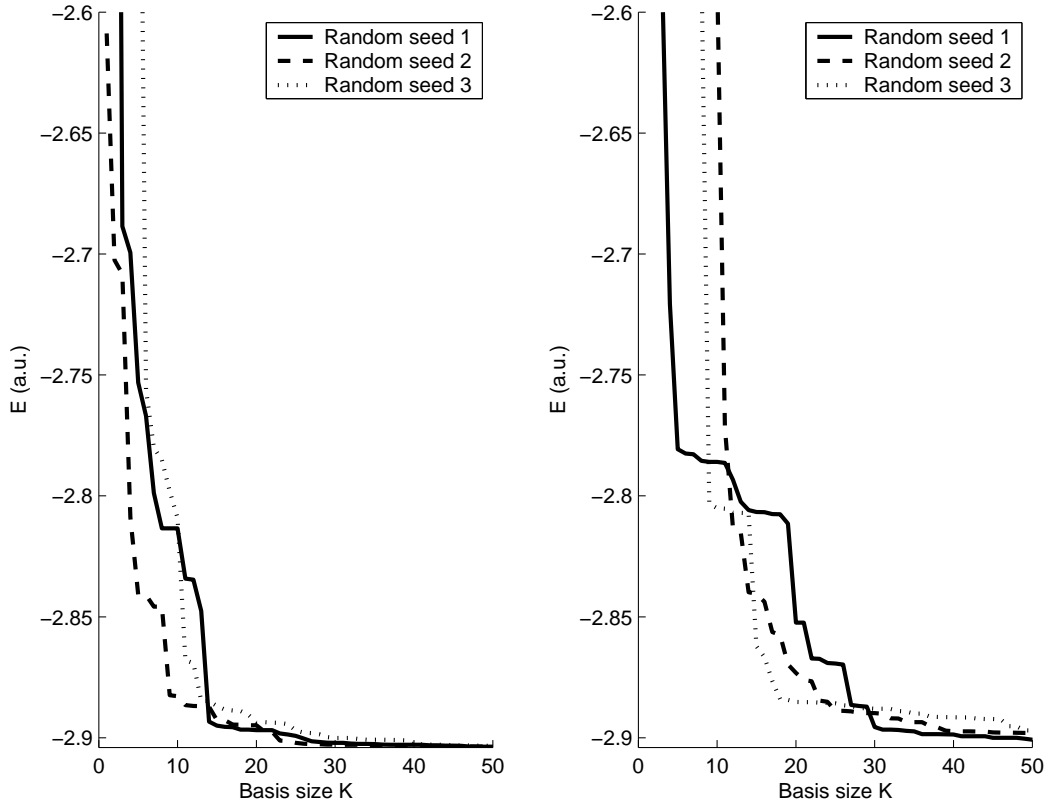}
\caption{The energy convergence of the $^\infty$He system as a function of the
basis size with (left) random exponential basis functions and (right) random Gaussian basis
functions.}
\label{ConvRandom}
\end{figure}

The strengths and weaknesses of the SVM can be rigorously investigated by considering the
$^\infty$He three-body system. For a definite basis size there is a total of $3K$ nonlinear parameters
(see (\ref{HeB})) which, ideally, have to be optimized. 
Postponing optimization for a moment by considering completely random parameters
reveals the result of expanding the function space ${\cal V}_K$ by functions that are
far from optimal. The graph on the left of figure \ref{ConvRandom}
shows the energy convergence of Helium as the basis size is increased from $1$ to $50$ by
adding exponential basic functions given in (\ref{HeB}), where the inverse
parameters, $\alpha_k^{-1}$,$\beta_k^{-1}$ and $\gamma_k^{-1}$, are selected randomly in
the intervals $[0,4]$,$[0,2]$ and $[0,2]$ respectively
\footnote{Because the average distance between the electrons is expected to be twice the
average distance between an electron and the nucleus, the interval for $\alpha_k^{-1}$,
corresponding to $r_{12}$, is set to twice that of $\beta_k^{-1}$ and $\gamma_k^{-1}$.
}.
The three curves correspond to three different random seeds.
The graph on the right shows the same convergence in the case of a Gaussian-type-basis
\footnote{i.e. $\Phi_{k}=(1+ \widehat{P}_{12}){\rm exp}(-\½\alpha_k x_1^2-\½\beta_k x_2^2
-\½\gamma_k x_3^2)$. See section \ref{cgauss} for further details.}.

\begin{figure}[bt]
\includegraphics[width=16cm]{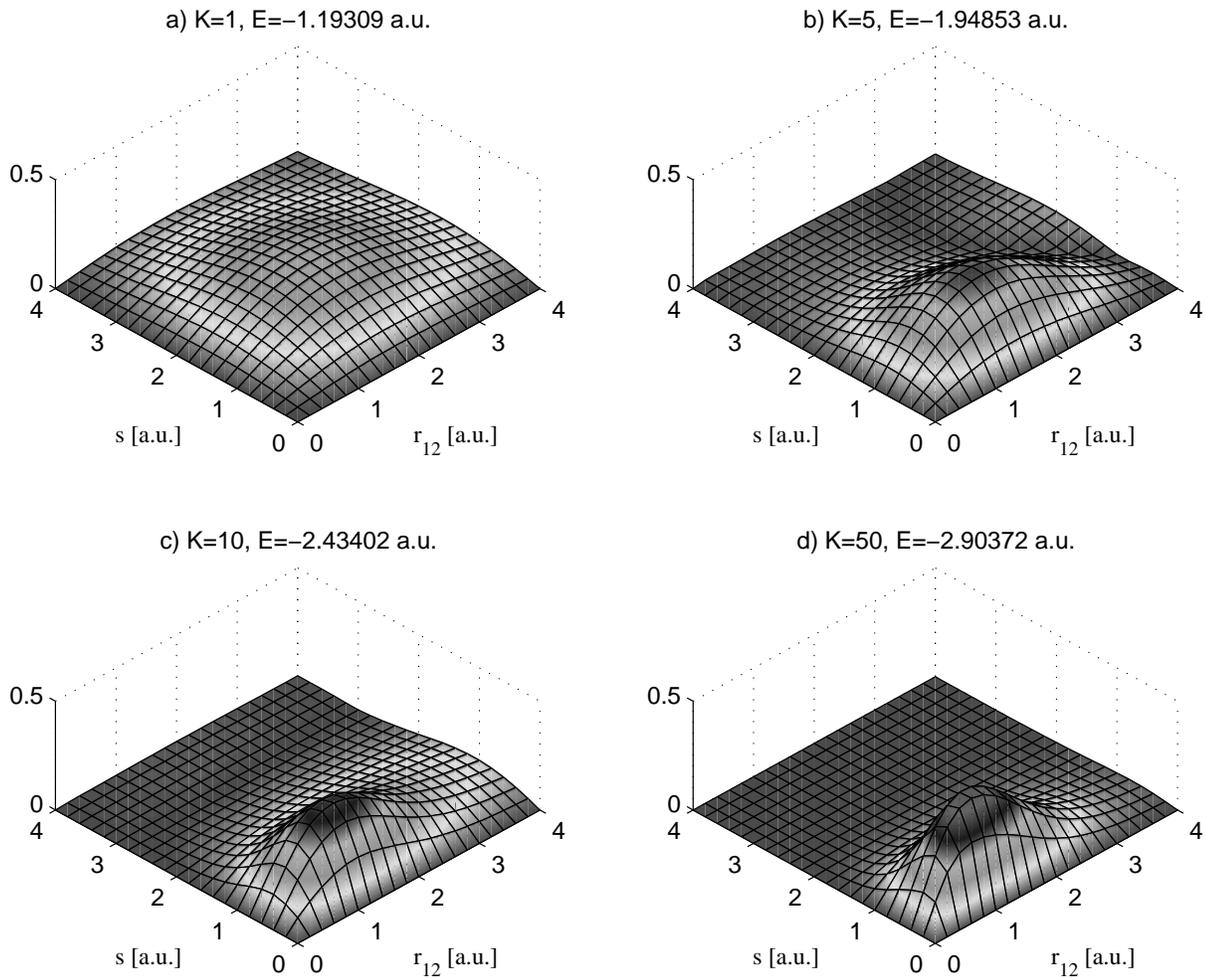}
\caption{Graphic illustration of the normalized wave function, $r_{12}s\Psi(r_{12},s)$, of
$^\infty$He, where $s^2=r_{13}^2+(r_{12}/2)^2=r_{23}^2+(r_{12}/2)^2$, as calculated by the SVM
with a Gaussian basis of increasing size reaching better and better energies. The three cases
(a),(b) and (c) with basis sizes $K=1,5$ and $10$, corresponds to a completely random selection,
while case (d) is the result of an optimized trial and error selection producing the exact
ground state energy, $E=-2.90372$.}
\label{HeWF}
\end{figure}

It is apparent from the convergence shown in fig. \ref{ConvRandom} that the crude adding
of linearly independent basis functions is actually an effective way to reach the accurate ground
state energy. This is the case for both types of basis functions. The corresponding
stepwise construction of the appropriate wave function is shown in figure \ref{HeWF},
where the coordinate $s$ is the length of the vector from the center of mass of the two electrons
to the $\alpha$ nucleous and it is assumed that $\bm{r}_{12}\perp\bm{s}$ for the sake of
illustration. Clearly, if the SVM trial function is flexible enough, the variational principle
ensures slow but sure convergence to the optimal representation as the basis size increases. 

\begin{figure}[bt]
\includegraphics[width=16cm]{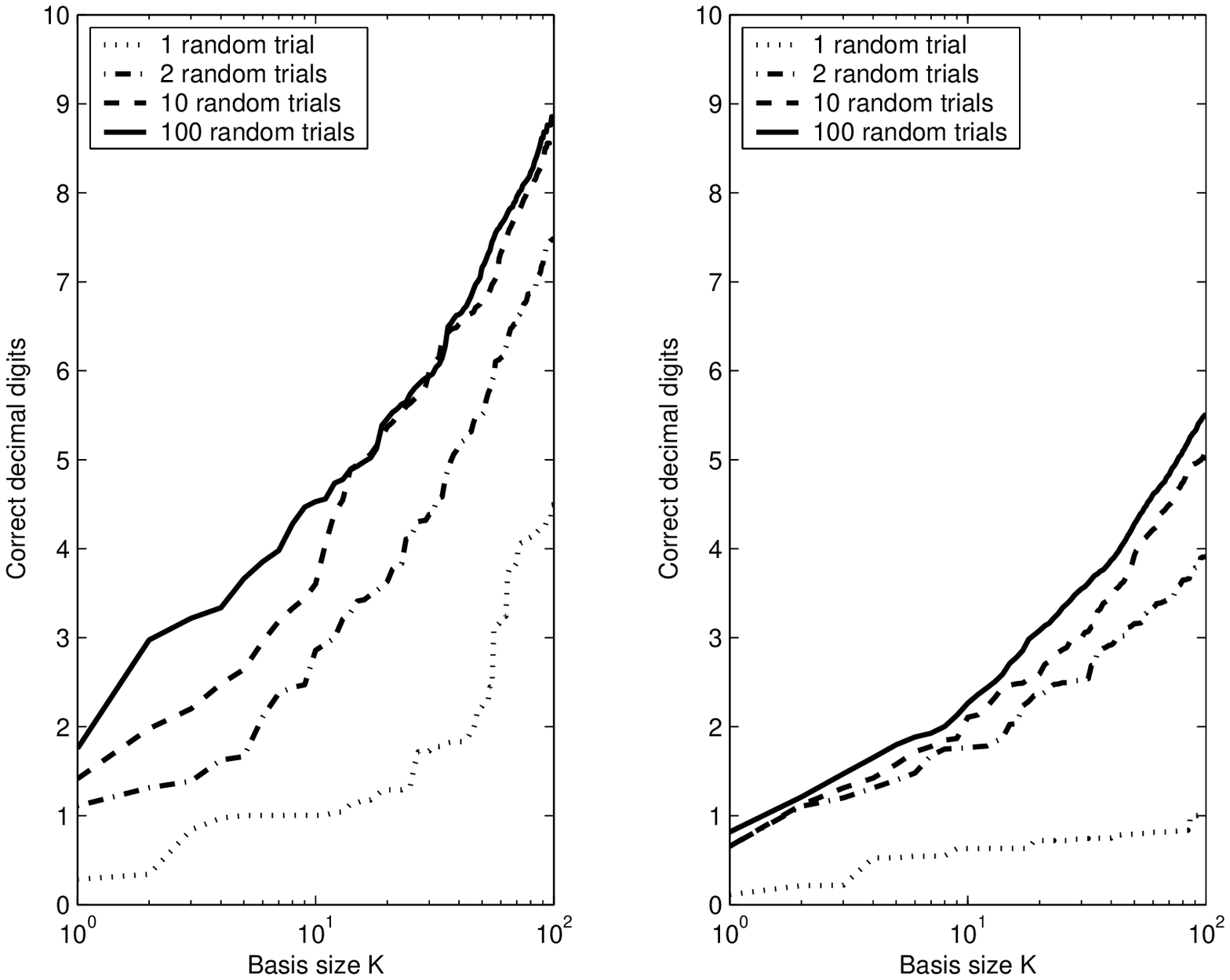}
\caption{The accuracy in correct decimal digits
(i.e. log$_{10}(\frac{E_{\rm exact}}{E_{\rm exact}-E})$)
of the nonrelativistic ground state energy of $^\infty$He as a function of the basis size
when the basis functions are optimized by testing 1, 2, 10 and 100 random values for each
nonlinear parameter. The left figure corresponds to exponential basis functions and the
right figure to Gaussian basis functions.}
\label{OptRandom}
\end{figure}

Further optimization of the nonlinear parameters will improve the rate of convergence
and limit a possibly excessive use of basis functions. In figure \ref{OptRandom}
the results of the SVM trial and error optimization is illustrated.
The accuracy in correct decimal digits is displayed as a function of the basis size
in the cases where 1, 2, 10 and 100 random values are tested for each nonlinear parameter
before adding the best trial. Again, the graph on the left corresponds to the exponential basis
(\ref{HeB}) where the inverse parameters, $\alpha_k^{-1}$,$\beta_k^{-1}$ and $\gamma_k^{-1}$,
are selected randomly, and the right graph corresponds to a Gaussian basis where the squared inverse
parameters, $\alpha_k^{-2}$,$\beta_k^{-2}$ and $\gamma_k^{-2}$, are selected randomly.
To avoid frequent linear dependency in the basis the random number intervals are doubled, i.e.
$[0,8]$,$[0,4]$ and $[0,4]$ respectively, for all calculations. 
Since the exact wave function of Coulombic systems will have an exponential asymptotic behavior
at large distances \cite{Frolov5}, the exponential basis produces much better results than
the Gaussian basis. More importantly, however, this means that in the case of BEC calculations,
as treated in the following section, the asymptotic is expected to have a Gaussian form,
and therefore the Gaussian basis would be the best choice.

From the curves it is obvious that optimization of the nonlinear parameters improve the accuracy
of the results to a certain extend. The positive effect saturates, however,
when the flexibility of the basis functions, which is very limited in this case, is completely
exploited by testing many different random values for the parameters.
Still, the variational theorem guarantees better results with every increase in the basis size.
This illustrates the generic trade-off between high optimization and large basis size. 
On the one hand, focusing on flexible basis functions with many nonlinear parameters
and high optimization costs, gives good results even for very low basis sizes
\footnote{Thakkar and Koga \cite{Koga} reach an impressive 15 decimal accuracy
in the $^\infty$He ground state energy with a basis size of only $K=100$.}.
On the other hand, keeping the optimization cost to a minimum by using simple basis
functions, allows a huge basis size and correspondingly precise results.
One of the best values (24 decimal accuracy) of the nonrelativistic $^\infty$He energy
has been achieved by V. I. Korobov \cite{Korobov} using $K=5200$ simple exponential basis
functions like (\ref{HeB}) with complex parameters selected pseudo-randomly
from optimized intervals. However, as explained in the next section, this is not an effective approach
for more complicated systems, like the case of trapped bosons, because of the much wider random
intervals needed.

One may say, that achieving the high accuracy obtained in this section is redundant
and has no physical meaning.
However, the extraordinary precision is a consequence of the variational stability
of the energy eigenvalue and does not necessarily reflect that the correct analytical structure
of the wave function has been found
\footnote{The exact wave function must satisfy the Kato cusp conditions \cite{Drake,Frolov1}
and includes e.g. logarithmic terms, which have negligible effect on the value of the variational
energy \cite{Varga3}.}
. The method above gives much poorer accuracy for the calculation of observables
other than the energy, e.g. relativistic or QED corrections \cite{Varga3}.
Obviously though, the results show the power of the SVM and of modern computers
\footnote{All the results presented in this section were computed on an (old) Athlon-650 CPU
using 32-bit floating point arithmetic in less than one hour.}
and their ability to solve at least quantum three-body problems to any number of digits.

\section{Bose-Einstein Condensate}\label{ResBEC}

This section presents the numerical results obtained from applying the SVM to specific systems of
attractively interacting bosons in the case of a spherical harmonic trap.
The default physical parameters used in all the
calculations are:
\begin{align}\label{parm}
\nonumber
{\rm Mass\ of\ }^{87}{\rm Rb}\quad\quad&m=86.91\ {\rm amu}\ (1.443\times 10^{-25}\ {\rm Kg})\\
\nonumber
{\rm Frequency\ of\ trap} \quad\quad&\nu=77.87\ {\rm Hz}\\
\nonumber
{\rm Lenght\ of\ trap}\ (\equiv a_{ho}) \quad\quad&b_t=23024\ {\rm a.u.}\\
\nonumber
{\rm Har.\ Osc.\ energy} \quad\quad&\epsilon_{ho}=1.183\times 10^{-14}\ {\rm a.u.}\ 
(5.160\times 10^{-32}\ {\rm J})\\
\nonumber
{\rm Two-body\ interaction} \quad\quad&V(r)=V_0 e^{-r^2/b^2}\ {\rm or}
\ V(r) = 4\pi\hbar^2 a\delta(r)/m \\
\nonumber
{\rm Range\ of\ potential} \quad\quad&b=11.65\ {\rm a.u.}\\
\nonumber
{\rm Depth\ of\ potential} \quad\quad&V_0=[-1.248261\times 10^{-7};0]\ {\rm a.u.}\\
\nonumber
{\rm S-wave\ scattering\ length} \quad\quad&a=[-10^{7};0]\ {\rm a.u.}\\
\nonumber
{\rm Two-body\ bound\ states} \quad\quad&N_b=0
\end{align}
Note, that in the case of the Gaussian two-body interaction, it is the potential depth which is
changed in order to vary the scattering length, while the potential range is fixed. Moreover,
the potential depth is limited to values where no two-body bound states are supported.

Four different combinations of trial wave functions and two-body interaction potentials
have been considered and is denoted by the following names:
\begin{itemize}
\item{{\it Hartree}: Denotes a SVM calculation with the Hartree single-particle product in
(\ref{hartree}) as the trial wave function and the Gaussian potential as the two-body interaction
\footnote{This does not correspond to a genuine self-consistent Hartree-Fock calculation since
the range and depth of the interaction potential are not variational parameters in the current
approach.}.}
\item{{\it Mean-field}:  This corresponds to {\it Hartree} case but
with the zero-range pseudopotential as the two-body interaction.}
\item{{\it Two-body}: A SVM calculation with a trial wave function given by (\ref{dilute})
that explicitly includes pair correlation and a Gaussian two-body interaction.}
\item{{\it Full}: This name designates a SVM calculation that explicitly allows up to
N-body correlations in the variational trial function (see (\ref{higher})), and has the Gaussian
potential as the two-body interaction.}
\end{itemize}
In all of the above cases, the random value interval from which the nonlinear variational
parameters are selected, is given by
\begin{equation}\label{randint}
(\alpha^{(k)})^{-2},(\alpha^{(k)}_{ij})^{-2} \in [0.0001;10]\ b_t
\end{equation}
However, since this interval spans an impressive 5 orders of magnitude in the attempt to
account for both molecular-type bound states and gas-type BEC states, the random value generator
has to be specially designed to output an equal number of values at each order (e.g. as many
parameters selected in the range $[0.0001;0.001]$ as in the range $[1; 10]$)
\footnote{The simplest way to do this, is by choosing a random number, $v$, from the interval
$[-4;1]$ and then assign the nonlinear parameters as $(\alpha^{(k)})^{-2}=log_{10}\ v$.}.
 
Many previous calculations on BECs suggest that reasonably dilute condensates are well described
by the s-wave scattering length alone \cite{Blume, Blume2}. In addition, the validity of the widely
used Gross-Pitaevskii mean-field theory, which has been recently examined
depends on the factor $a^3$. Therefore it is convenient, also in the current context,
to describe the properties of trapped N-boson systems as a function of the s-wave
scattering length, $a$.

\subsection{System with $N=3$, and $-\infty < a < 0$}\label{N3res}

The translationally invariant three-body problem for identical particles has only two degrees
of freedom in coordinate space. This means, that the application of the SVM with basis functions
having two nonlinear parameters is sufficient to include all correlations of three trapped
particles. Consequently, the restriction to $N=3$ leads to a computationally simple problem
and presents an opportunity to shed some initial light on the characteristics of BECs.

\subsubsection{Energy levels}

\begin{figure}[tb]
\psfrag{XLABEL}{$a/b_t$}
\psfrag{YLABEL}{$E_{total}\ [\hbar\omega]$}
\includegraphics[width=16cm]{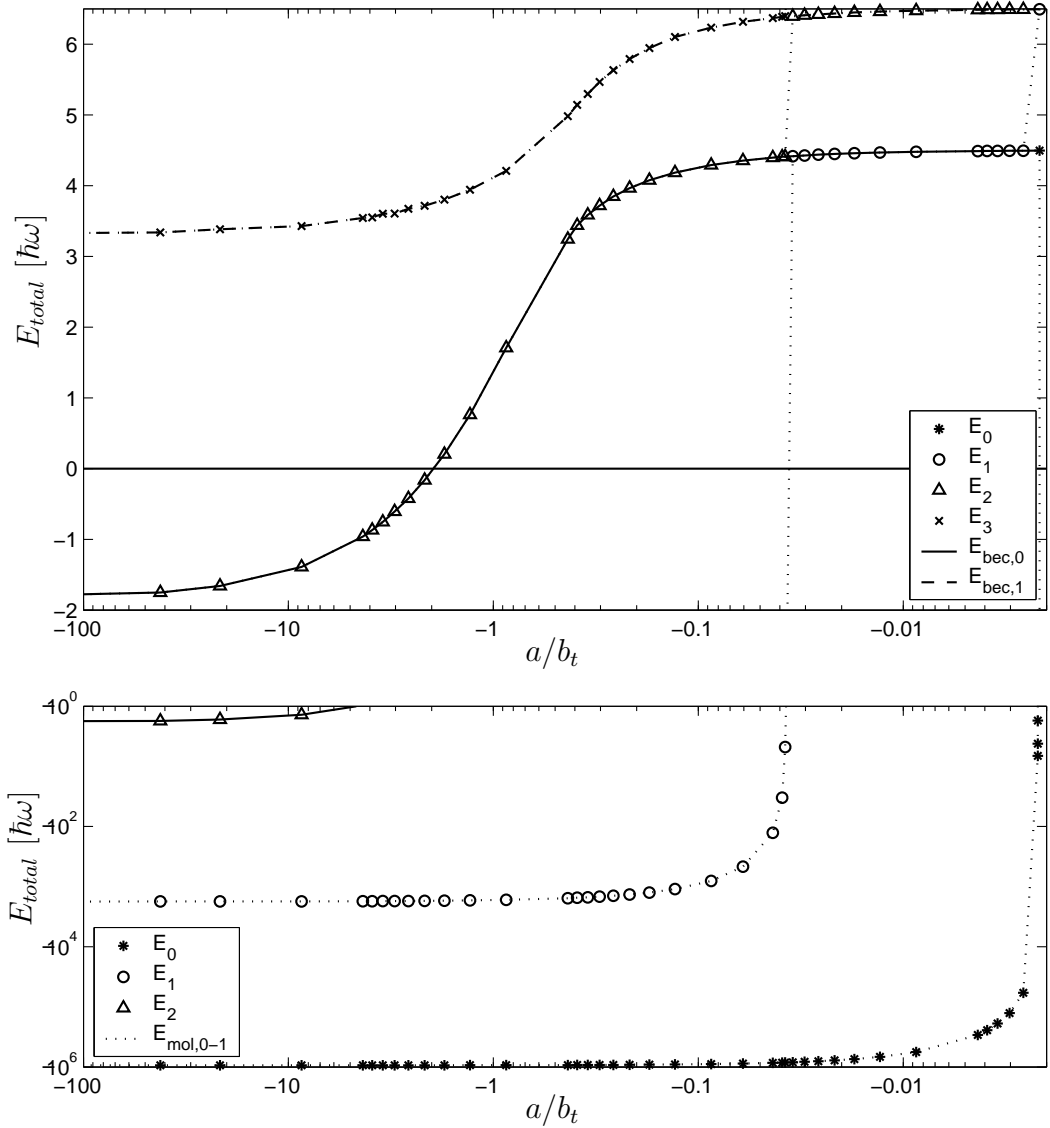}
\caption{Energy levels for the $N=3$ system of trapped bosons as a function of the scattering length.
The data points show the energies of the ground state, $\Psi_0$, and the three lowest
excited states, $\Psi_1$-$\Psi_3$, as indicated. The upper figure details the
energy levels of the the gas-type BEC states and the lower figure illustrates
the ``plunging'' molecular-type energy levels.}
\label{N3energies}
\end{figure}

The overall behavior of the internal energy levels as a function of the s-wave scattering length,
$a$, for three $^{87}$Rb atoms in a spherical trap is shown in figure \ref{N3energies}.
Data points for the energies, $E_0$,$E_1$,$E_2$ and $E_3$, corresponding to the four lowest
eigenstates, $\Psi_0$,$\Psi_1$,$\Psi_2$ and $\Psi_3$, are plotted. The upper graph shows
the energies that fall in the vicinity of zero and the lower graph displays the energies
which are large and negative on a logarithmic axis.
This rather complicate energy level structure is interpreted as follows.
In the case where $a$ is very close to zero, the energy levels correspond to the non-interaction
boson gas result, given by $E_n=(3N/2+2n)\hbar\omega$. 
Increasing the attraction slightly to where $a\sim -0.002\ b_t$, creates the condition for a
molecular-type three-body bound state, and the lowest energy, $E_0$, ``plunges'' downwards.
The second lowest energy, $E_1$, then takes the
place of the lowest gas-like energy which is not noticeably affected by the forming molecular
state. This sequence of events is repeated when the scattering length is further decreased
to roughly $a\sim -0.04\ b_t$. From here on, down to where $a\sim -100\ b_t$, it is
only $\Psi_2$ and higher eigenstates that give energies above or close to zero.
Moreover, these energies change dramatically in the region around $a\sim -1\ b_t$, where $E_2$
becomes negative and $E_3$ drops by almost $3\hbar\omega$.

As a first conclusion, it is apparent that in the specific range $-100\ b_t < a < 0$ (with
$N_b=0$), the $N=3$ system has at most two strongly bound states ($E_0$ and $E_1$ for large
negative $a$), which are interpreted as true molecular-type states.
Secondly, there are additional unbound eigenstates, which seem independent of the
molecular states, and, in the limit of weak interaction, are
equidistantly spaced by $2\hbar\omega$, corresponding to an ideal trapped gas.
These states are interpreted as gas-type BEC states. The significant change that the gaseous
energy levels undergo around $a\sim -1\ b_t$ is linked to the disappearance of the barrier in
the effective boson-boson potential (see figure \ref{effpotgraph}). As the barrier vanishes the
energy level ``located'' in the trap minimum of the potential is free to descend down in the global
minimum. At some point, when the scattering length is much larger than the size of the trap,
the energy level stabilizes. The corresponding bound state is considered to be a so-called
Efimov state since it is weakly bound and has a large spatial extension (see below). 

Before ending this subsection it is worth noting that the gross behavior
of the $N=3$ energy levels presented here, agree both quantitatively and qualitatively with the
results obtained by D. Blume and Chris H. Greene in \cite{Blume2} with the adiabatic
hyperspherical expansion method. Consequently, one can confidently assume that the SVM also produces
correct results when continuing into the (unchartered) many-body regime in the following.

\subsubsection{Characteristics of the BEC state}

\begin{figure}[h!]
\begin{minipage}[c]{\textwidth}
\includegraphics[width=16cm]{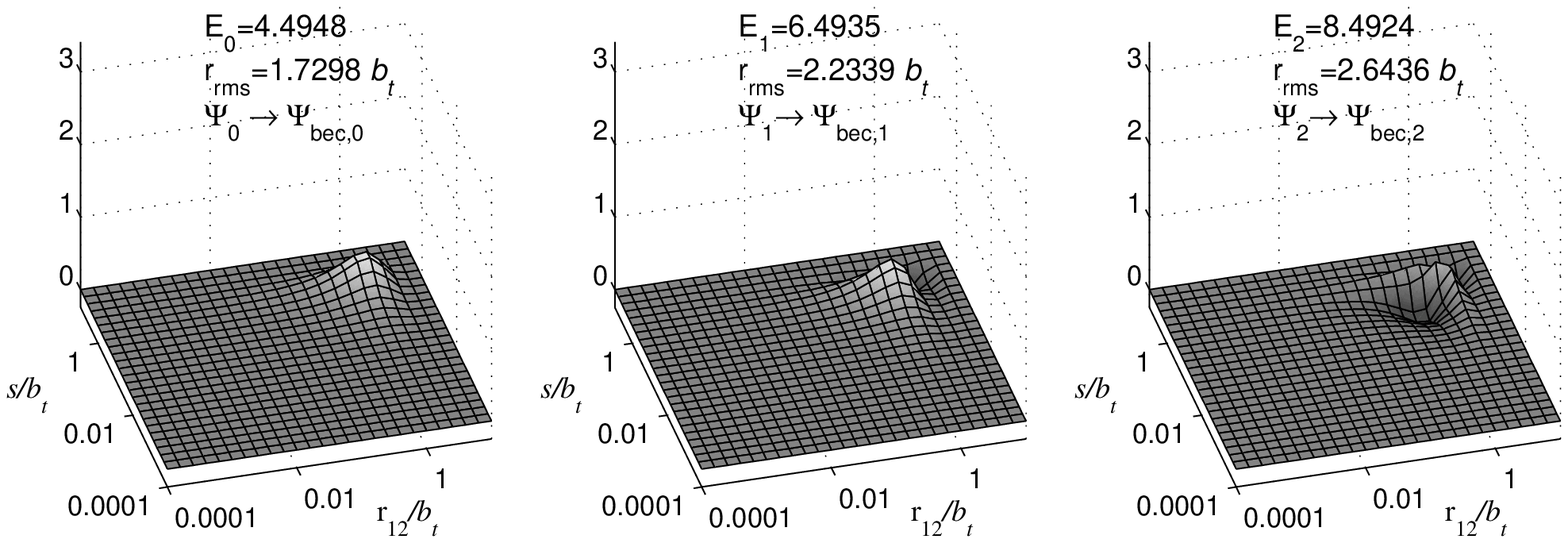}
\caption{Graphic illustration of $r_{12}s\Psi_0,r_{12}s\Psi_1$ and $r_{12}s\Psi_2$,
as a function of $r_{12}$ and $s$ (see caption to fig. \ref{HeWF}), where $\Psi_0$, $\Psi_1$ and
$\Psi_2$ are normalized and correspond to the three lowest eigenstates determined by the SVM for
$N=3$ and $a=-50$ a.u. ($=-0.0022\ b_t$). These eigenstates have nonzero regions at
$r_{12}\sim s\sim \sqrt{N}b_t$ only, and are consequently interpreted as the ground, first and
second excited BEC states. Notice the logarithmic scale on the $r_{12}$ and $s$ axes.}
\label{N3wave1}
\end{minipage}
\begin{minipage}[c]{\textwidth}
\includegraphics[width=16cm]{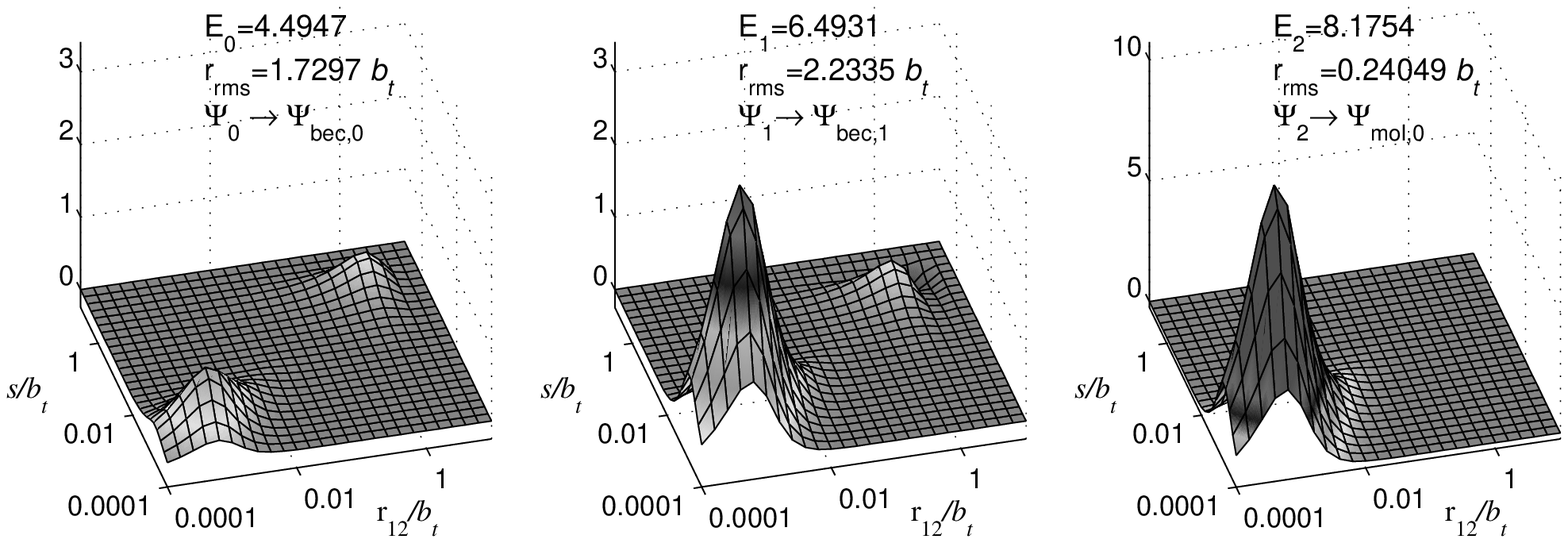}
\caption{As above, but for scattering length $a=-57$ a.u. ($=-0.0025\ b_t$), where the first
molecular-type bound state is close to appearing. At this scattering length, $\Psi_0$ and $\Psi_1$
still correspond to the ground and first excited BEC states since almost all the amplitude is
concentrated in the $r_{12}\sim s\sim \sqrt{N}b_t$ region (N.B. the logarithmic axes).
The $\Psi_2$ eigenstate, however, now resembles an unbound $(E>0)$ molecular-type state with the
amplitude distributed at very small interparticle distances, $r_{12}\sim s\sim<10^{-1}\ b_t$.}
\label{N3wave2}
\end{minipage}
\begin{minipage}[c]{\textwidth}
\includegraphics[width=16cm]{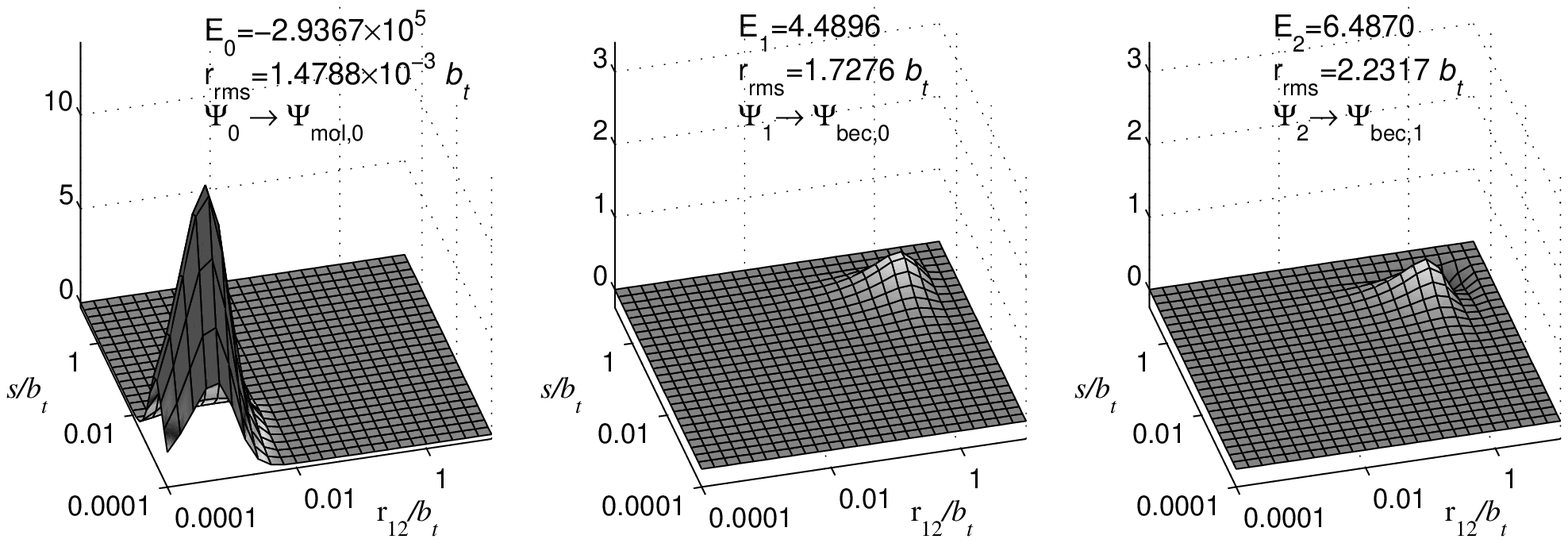}
\caption{As above, but for scattering length $a=-100$ a.u. ($=-0.0043\ b_t$). In this case, the
lowest eigenstate, $\Psi_0$, is a true molecular-type bound state almost unaffected by the
external trap. The first and second excited eigenstates, $\Psi_1$ and $\Psi_2$, correspond to
the ground and first excited BEC states respectively.}
\label{N3wave3}
\end{minipage}
\end{figure}
It is of interest, also in the current context, to examine the defining features of the BEC
eigenstates in more detail. In the simple $N=3$ case, one can illustrate the spatial
dependence of the calculated wave function graphically as a function of the two degrees of freedom
(e.g. $r_{12}$ and $s$, see fig. \ref{HeWF}). This is done in figures
\ref{N3wave1}-\ref{N3wave3} for three different scattering lengths.
\clearpage
In addition, the corresponding energy and the result from calculating the root-mean-square radius,
given by
\begin{equation}
<r_{rms}^2>
=\langle \Psi| \sum_{i=0}^N (\bm{r}_i-\bm{R})^2/N |\Psi\rangle
=\langle \Psi| \rho^2/N |\Psi\rangle
=\sum_{k=0}^{K}\sum_{k^\prime=0}^{K}\sum_{i=0}^{N-1}
\langle \psi_k| \bm{x}_i^2|\psi_{k^\prime}\rangle
\end{equation}
is indicated above each image. The expected and apparent conclusion from these graphs is that
there are distinctive differences in the spatial distribution of the particles in the bound
molecular-like eigenstates and the gas-like eigenstates. This significant difference in the
spatial extension, or equally in the density, is then an effective measure to determine the type
of a given eigenstate calculated by the SVM. In the following, it is therefore assumed that 
an eigenstate having $r_{rms} > \sim 10^{-1}\ b_t$ is a BEC type state.
Look in the captions for further details of the interpretation. 

\subsubsection{Correlations}

In reference to the above assumption, the reader might wonder about the wave functions in
figure \ref{N3wave2}, where the amplitude
is separated in two distinct peaks corresponding to different densities.
Whether this mixture of wave function amplitude characteristic to both gas-like and molecular-like
states is an actual physical fact or just a consequence of the stochastic method employed,
is not clear from these calculations.
It is quite possible that the peak at low interparticle
distances in the first two illustrations of fig. \ref{N3wave2} is a remnant of previously
optimal basis configurations in the stepwise trial and error procedure. If this is the case,
further rigorous optimization should slowly remove the high density peaks. In the discussion
of correlations below, however, this is not really relevant as long as the root-mean-square
distance gives a clear indication of the type of a given eigenstate.

With the preliminary treatment completed it is now possible to turn the attention to the
interparticle correlations of the three-boson system and concentrate on the effects they
produce in a particular eigenstate. The most convenient way to illustrate these effects
is by isolating the interaction energy contribution to the total energy, since,
in accordance with the definition adopted in section \ref{defcor}, this equals
the correlation energy, $E_{corr}$, apart from a sign. 
In general, for the ground state of N interacting bosons in a spherical symmetric trap, one has a
simple relationship, given by $E_{corr}=3N\hbar\omega/2 - E_{total}$,
where the term $3N\hbar\omega/2$ represents the energy of the non-interacting gas.

\begin{figure}[tb]
\psfrag{XLABEL}{$a/b_t$}
\psfrag{YLABEL}{$E_{corr}\ [\hbar\omega]$}
\includegraphics[width=16cm]{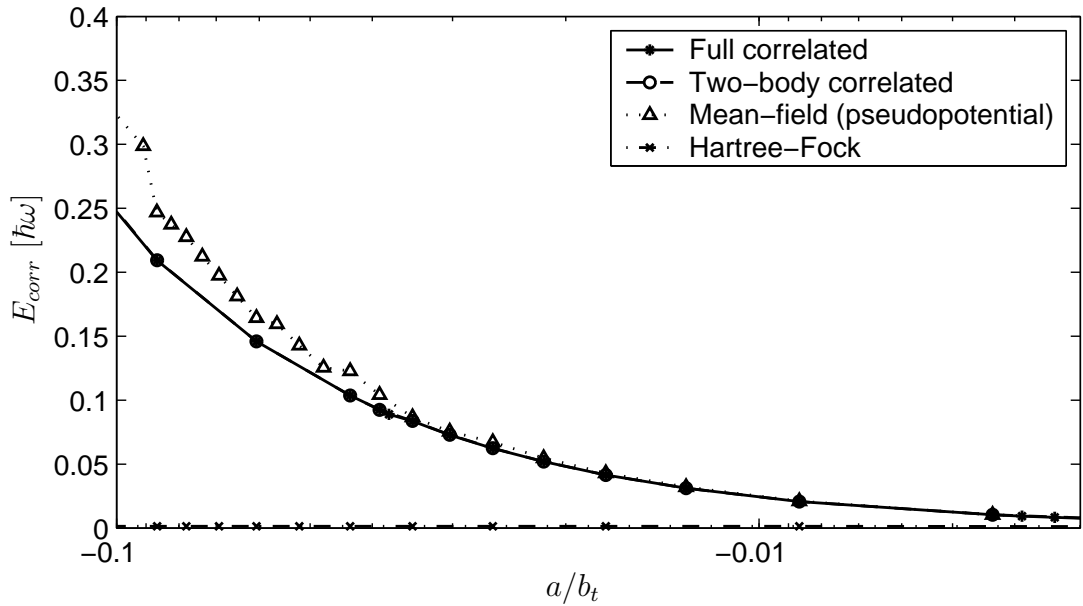}
\caption{Correlation energy of the lowest BEC state, $E_{bec,0}$, in the $N=3$ boson system,
as a function of the scattering length for SVM calculations including different levels of correlation.
Note that the data points produced by the {\it full} treatment and the {\it two-body} treatment are
almost identical up to the precision considered here, and consequently indistinguishable
in the graph.
}
\label{N3correlation}
\end{figure}
Figure \ref{N3correlation} shows the correlation energy as a function of the scattering length
for the lowest BEC state of the $N=3$ boson system. Four different sets of data points
are plotted, corresponding to the correlation levels included in the {\it Hartree, Mean-field,
two-body} and {\it full} SVM calculations. The solid line, representing the full
correlated treatment, is here regarded as the closest candidate to the ``exact'' correlation curve.
As noted above, there is only
two independent spatial coordinates in the $N=3$ case, and consequently, the two-body and the
full correlated treatments should be equivalent. Evidently, this is also the case, in that
the dashed line of the two-body correlated calculation is indistinguishable from the solid curve.

The most apparent feature in fig. \ref{N3correlation} is the obvious failure of the
{\it mean-field} treatment to account correctly for the correlations in the system when the
scattering length becomes large. This can be directly related to the break down of 
pseudopotential approximation as was predicted in section \ref{valpseudo}.
In the recent article \cite{DuBois}, DuBois {\it et al} establishes the condition,
$(N-1)|a/b_t|^3 <\sim 10^{-3}$, for the validity of the GP mean-field calculation of
bosons in a trap. With $N=3$ this amounts to $a >\sim -0.08\ b_t$ in the current case.
Looking at the curve for the {\it mean-field} SVM calculation, this rough limit
agrees very well with the observed validity range.

One last lesson may be learned from comparing the {\it Hartree} and the {\it mean-field}
curves. Both of these treatments are based on the Hartree product trial function which is
explicitly uncorrelated and the only difference is in the adopted two-body interaction
potential. However, the results are very distinct and the {\it Hartree} calculations
using the finite Gaussian potential as the two-body interaction is clearly an unlucky choice
which leads to a terrible treatment of correlations. The {\it mean-field} calculation, on the
other hand, reveals that the application of an effective interaction as opposed to a realistic one,
can be quite powerful in the attempt to include correlation effects. This might explain why
the mean-field approaches have been exceptionally successful in treating dilute systems of bosons.

\subsection{System with $N=4$, and $-\infty < a < 0$}\label{N4res}

The system of four trapped bosons, in particular, has the features to be a great source of
knowledge in a discussion of many-body correlations in BECs. With $N=4$ particles it is expected,
that the interparticle correlations governing the dynamical motion will include both three- and
four-body effects. At the same time, such a system is simple enough to facilitate
numerous and accurate calculations of the supported eigenstates and corresponding energies.
In the present application of the SVM, these features makes the $N=4$ system unique, since a
subsequent increase in the number of particles makes the {\it full} calculations intractable.
\subsubsection{Energy levels}

\begin{figure}[tb]
\psfrag{XLABEL}{$a/b_t$}
\psfrag{YLABEL}{$E_{total}\ [\hbar\omega]$}
\includegraphics[width=16cm]{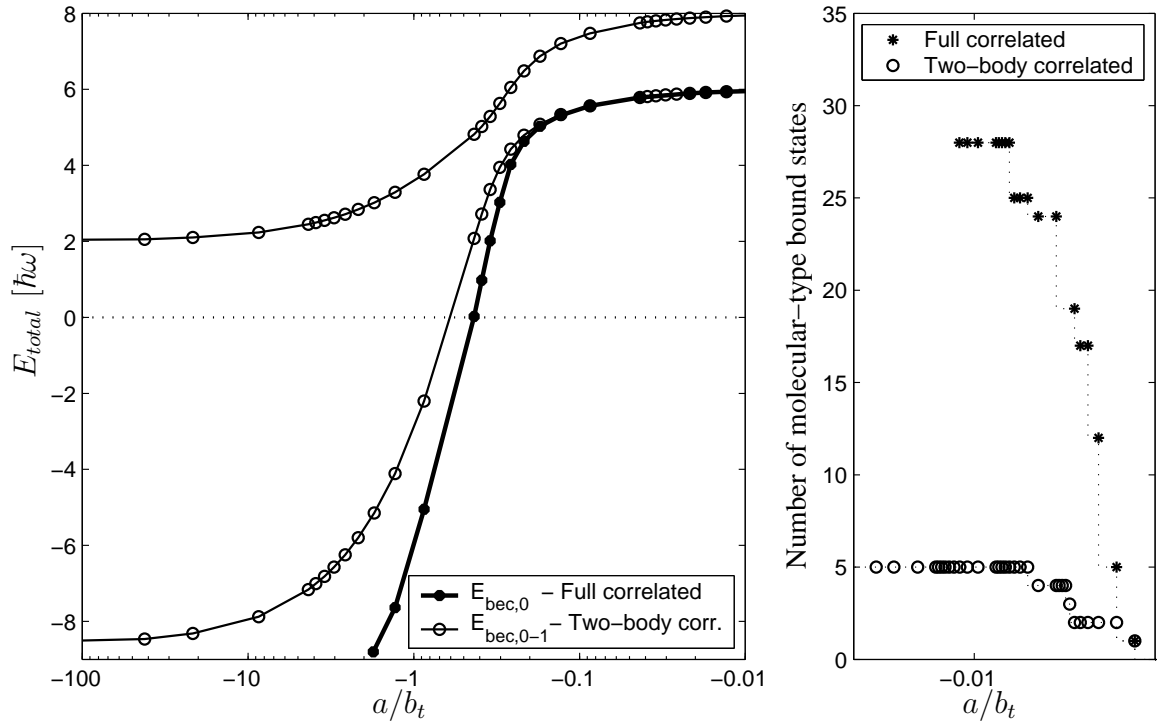}
\caption{Left: Energy levels of the two lowest BEC states for the $N=4$ system of trapped bosons
as a function of the scattering length. Right: The corresponding number of lower lying
(with respect to the BEC state) molecular-type states, both in the case of the {\it full}
SVM calculation and the {\it two-body} SVM calculation.
}
\label{N4energies}
\end{figure}
Energies corresponding to the two lowest BEC states, $\Psi_{bec,0}$ and $\Psi_{bec,1}$,
of the $N=4$ system, have been calculated with the SVM in the {\it two-body} description.
The resulting levels are displayed in left part figure \ref{N4energies}. In the same graph,
the energy curve obtained with the {\it full} SVM calculation is also plotted, however,
only for the lowest BEC state and for a limited range of the scattering length ($-2\ b_t \le a<0$).
Apparently, the characteristic behavior of the energy levels is very similar to the $N=3$ case
in figure \ref{N3energies}. As is the related interpretation.
One slight change can be observed by taking a closer look at the range where $E_{bec,0}$
falls off significantly. This now happens around $a\sim -0.3\ b_t$, and indicates that the barrier
in the effective boson-boson potential vanishes at a smaller scattering length than in the three-boson
case. This comes as no surprise since the disappearance of the barrier is known occur when
$|a|\approx0.67b_t/N$, \cite{Sogo}.

On the right hand side of figure \ref{N4energies}, a step-graph shows the number of
molecular-type bound states determined by SVM calculation as a function of the scattering length
and corresponding to the data points in the left figure.
Contrary to the $N=3$ system, there are now a large number of lower lying bound
states to take into account. However, this depends heavily on the correlations allowed in the
trial wave function. While the {\it two-body} treatment finds at most five ``plunging''
energy levels, the {\it full} calculation determines up to 28 in the reduced scattering
length range considered, and even more appear for larger attraction.
As is pointed out later, this is the reason why the {\it full} calculation has been limited to
scattering lengths above $-2\ b_t$. Moreover, the significant difference in the number
molecular-type states supported at particular levels of correlations was very much expected,
since the density for these states is so high, that higher-order effects should play an
important role in the interparticle dynamics. The data presented here confirms this prediction.

In light of the main topic of this thesis, the most interesting feature visible in the $N=4$ energy
level figure, is the revealing separation between the {\it two-body} curve and the
{\it full} curve in the region where they are both calculated. Clearly, the {\it full} treatment
results in a significantly lower total energy than the {\it two-body} treatment when the scattering
length becomes lower than $\sim -0.2\ b_t$. In other words, a description including
all higher-order correlations, produces a distinctively better variational upper bound to the
BEC ground state energy, than one including only two-body correlations. 
This discovery is very important and will be further investigated below.

\subsubsection{Density profile}

One of the defining properties of a BEC is the characteristic density profile of the condensed
gas. To further support the understanding of the lowest gas-like state from the SVM
calculations as a true BEC state, it is therefore convenient to consider the one-body density
function, which is given by \cite{Bishop}
\begin{equation}
n(r)=\langle \Psi| \sum_{i=0}^N \delta(\bm{r}_i-\bm{R}-\bm{r}) |\Psi\rangle
=\sum_{k=0}^{K}\sum_{k^\prime=0}^{K}\sum_{i=0}^{N-1}
\langle \psi_k|\delta\big((\bm{u}^{(i)})^T\bm{x}_i-\bm{r}\big) |\psi_{k^\prime}\rangle
\label{density}
\end{equation}
where $\bm{r}_{i} = (\bm{u}^{(i)})^T \bm{x}$ and $\bm{R} = \bm{x}_N$ was used
\footnote{The resulting matrix elements are easily derived from formula (\ref{I2}).}
. 
\begin{figure}[tb]
\includegraphics[width=16cm]{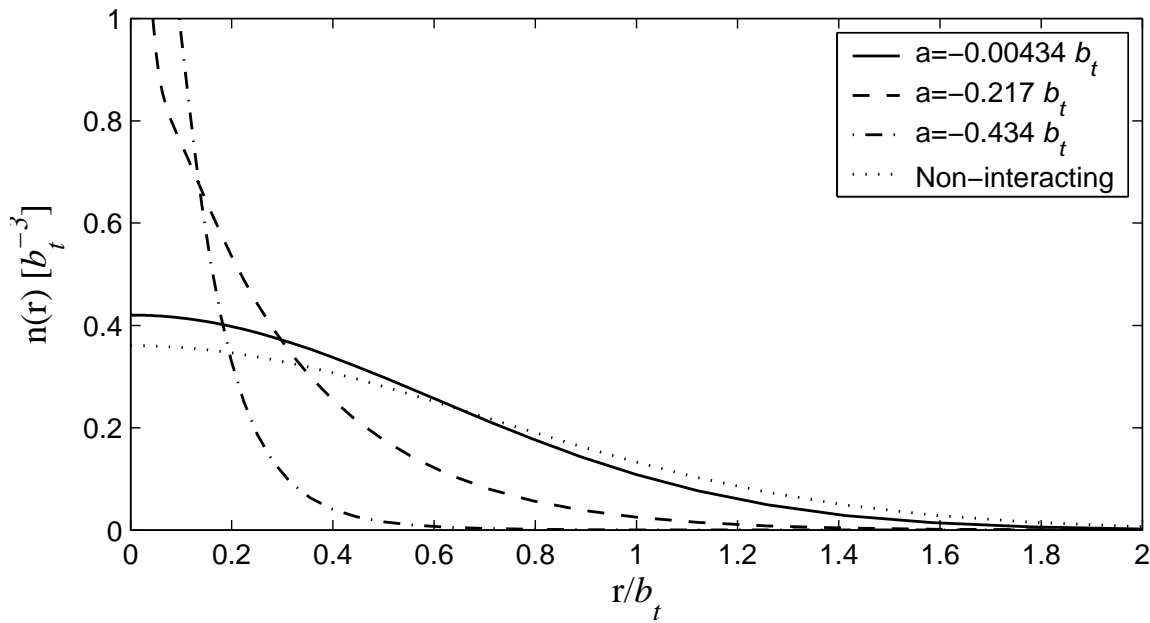}
\caption{The radial one-body density function defined in (\ref{density}) for the BEC ground state of
a system of four trapped bosons, as a function of the distance from the center-of-mass. The
different curves correspond to SVM calculations for increasingly negative scattering lengths as
indicated, and have been normalized by $\int n(r) d\bm{r}=N$.}
\label{N4density}
\end{figure}
Figure \ref{N4density} illustrates this function for the lowest gas-type eigenstate determined
by the SVM in three specific cases of increasingly negative scattering length.
Obviously, the solid line, calculated for very low $a=-0.00434\ b_t=-100$ a.u., lies very close to
the form of the analytically available Gaussian shape ($n(r)\propto e^{-r^2}$)
of the non-interacting ideal gas. And as expected, when the interaction becomes more and more
attractive the density profile becomes narrower since the bosons are forced closer together.
The overall behavior coincides very well with other calculations on condensed bosons
\cite{Ole, Blume} and with experimental determined profiles \cite{Soding}.
This is convincing proof that the correct interpretation of the SVM results has been made.

\subsubsection{Correlations}

\begin{figure}[tb]
\psfrag{XLABEL}{$a/b_t$}
\psfrag{YLABEL}{$E_{corr}\ [\hbar\omega]$}
\includegraphics[width=16cm]{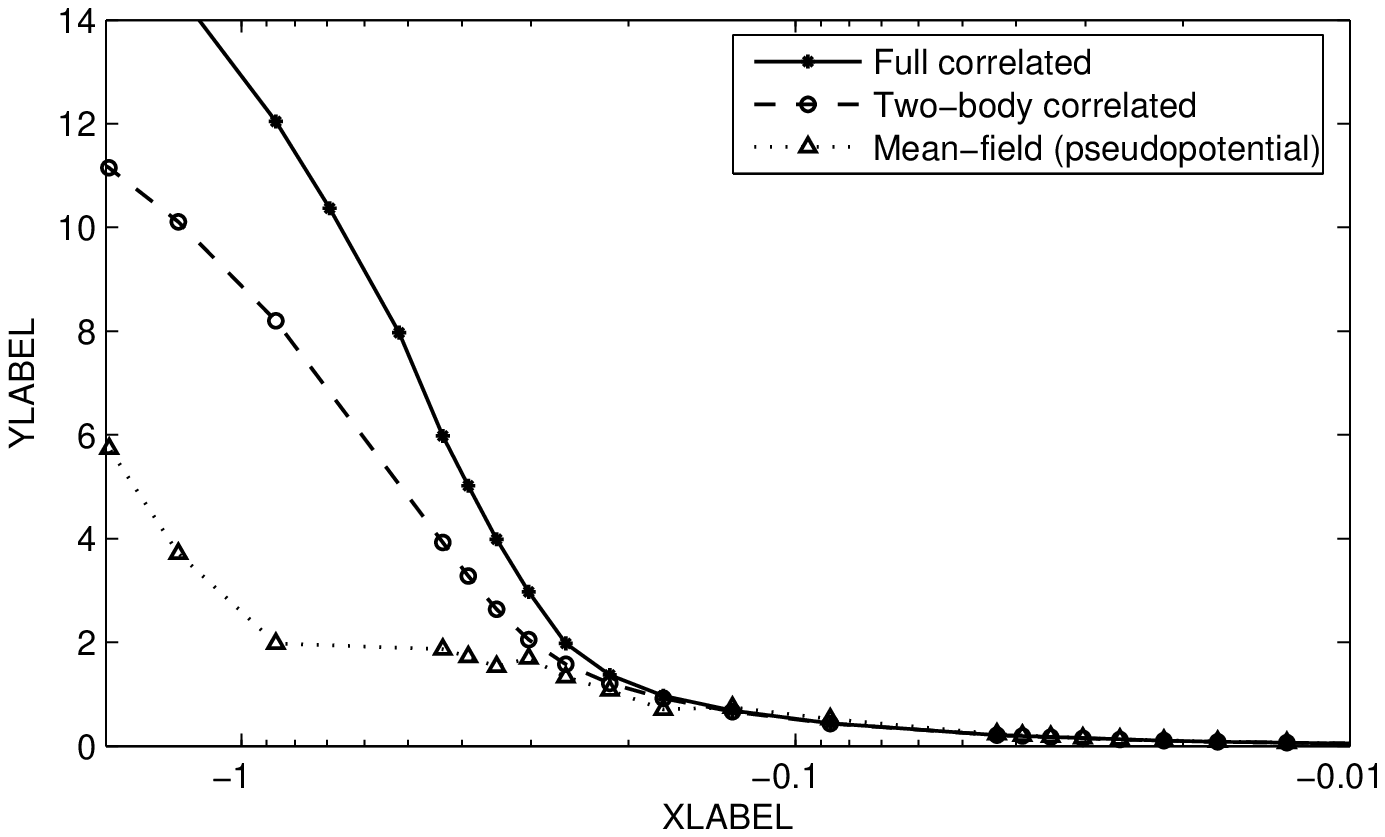}
\caption{Correlation energy of the lowest BEC state, $E_{bec,0}$, in the $N=4$ boson system,
as a function of the scattering length for SVM calculations including different levels of correlation.
}
\label{N4correlation}
\end{figure}
In accordance with the analysis of the three-boson system, the correlation energy of the lowest
BEC state in the $N=4$ case, $E_{corr}=6\hbar\omega - E_{total}$, is plotted in figure
\ref{N4correlation} for the {\it full}, the {\it two-body} and the {\it mean-field} calculations
(the {\it Hartree} combination is not considered here, since it was proven above to be
insufficient for treating correlations). The curves from these three data sets
indicate a clear discrepancy for large negative scattering length, resulting in significantly
different correlation energies. Focusing first on the {\it mean-field} treatment, it seems
that this description can account for many of the higher-order correlation effects as far as
$a\sim -0.1\ b_t$. This is roughly the same validity region as for the $N=3$ case and also
roughly the same as the GP mean-field validity range established in \cite{DuBois} (see $N=3$
discussion above). However, the current {\it mean-field} calculations are not
extensive enough to be conclusive in terms of validity estimates.

Before leaving the {\it mean-field} description in this presentation, it is worth noting
that in the $N=4$ calculations, the {\it mean-field} correlation energy is lower than the
{\it full} correlation energy for $a<-0.1\ b_t$, while the opposite is the case for $N=3$. 
This might indicate, that the pseudopotential approximation adopted in most Hartree-Fock type
mean-field theories, are destined to include too much correlation energy in the situations
where two-body effects are dominant, and subsequently too little in the case where higher-order
correlations are important. However, since almost all GP mean-field theories consider either repulsive
interactions or only very small negative scattering lengths (in order to satisfy the barrier
condition $|a|N/b_t > \sim0.67$), this conclusion cannot be confirmed elsewhere.

\begin{figure}[tb]
\psfrag{XLABEL}{$a/b_t$}
\psfrag{LLABELLLABELLLABEL}{$(E_{coor}^{full}-E_{coor}^{2B})/E_{corr}^{full}$}
\includegraphics[width=16cm]{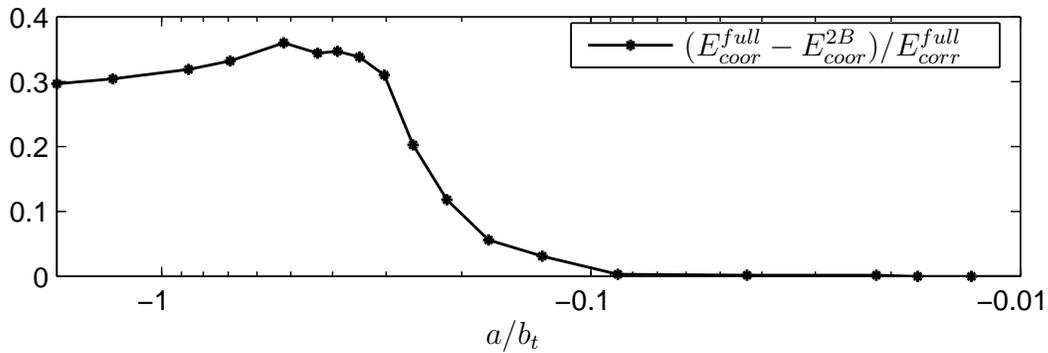}
\caption{The correlation energy difference, $E_{coor}^{full}-E_{coor}^{2B}$ normalized
by $E_{corr}^{full}$, as a function of the scattering length for the $N=4$ system of trapped bosons.
}
\label{N4correlation2}
\end{figure}
Turning the attention now to the results of the {\it two-body} and {\it full} SVM calculations,
the answers to some of the interesting questions asked in the introduction of this thesis
are revealed. First of all, it is evident from the curves in figure
\ref{N4correlation}, that three-body and higher-order correlations are an integrated part of
the dynamics of the four-boson system. In addition, these correlations seem to be increasingly
important, at least up to a certain point, as the interaction between the bosons become more
attractive.

During the further analysis it is convenient to display the difference between the
{\it full} correlation energy, $E_{coor}^{full}$, regarded as the ``exact'' correlation energy,
and the {\it two-body} correlation energy, in terms of the former, since
this will illustrate the relative importance of the higher-order correlations.
Such a graph is available in figure \ref{N4correlation2} and indicates that the relative effect of
higher-order correlations in the four-boson system, saturates when the scattering length reaches
$\sim -0.5\ b_t$, i.e. a length similar to half the size of the trap. At this point, however,
the relative deviation between the {\it two-body} and the {\it full} treatments correspond
to almost $35\%$. In conclusion one may then state, that in the range where
$-\infty < a < -0.5\ b_t$, three- and four-body correlations contribute approximately one third of the
correlation energy in the $N=4$ system while two-body correlations contribute
the remaining two thirds. For $a>-0.1\ b_t$, however, the three- and four-body correlations
are negligible.

\subsection{System with $N=10$, and $-0.2\ b_t < a < 0$}\label{N10}

In order to investigate correlations in many-boson systems, it is off course of great interest
to consider system with more than four particles. The problem is, that the application
of the SVM for the {\it full} correlated description, is only feasible in practice for systems with
$N < \sim 5$ (because of the symmetrization requirement, see section \ref{symmet}).
Fortunately, this is not the case for the {\it two-body} description where the corresponding
computations are independent of $N$.
The next best option in this work is therefore to focus on possible similarities between the
two-body correlated treatment of the $N=4$ system and, for example, the $N=10$ system.

\subsubsection{Energy levels}

\begin{figure}[tb]
\psfrag{XLABEL}{$a/b_t$}
\psfrag{YLABEL}{$E_{total}\ [\hbar\omega]$}
\includegraphics[width=16cm]{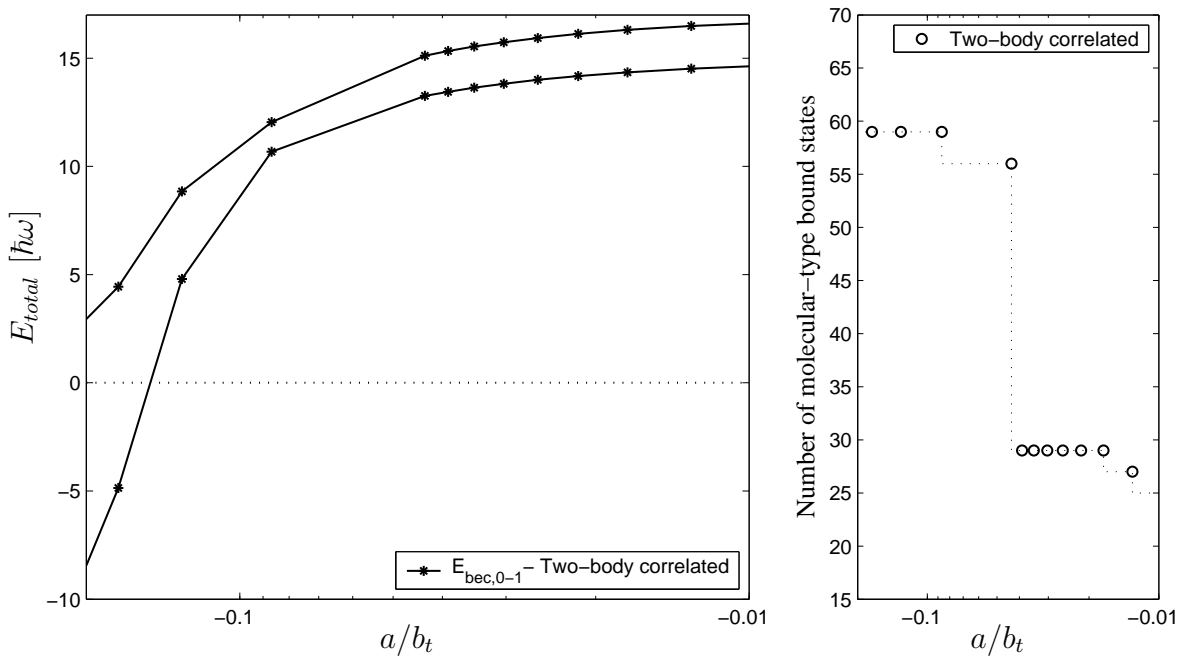}
\caption{Left: Energy levels of the two lowest BEC states for the $N=10$ system of trapped bosons
as a function of the scattering length. Right: The corresponding number of lower lying
(with respect to the BEC state) molecular-type states. All data points refer to the
{\it two-body} correlated SVM calculation.
}
\label{N10energies}
\end{figure}

The energy levels for the lowest two BEC states of the $N=10$ system has been calculated
in the {\it two-body} description for scattering lengths in the range $-0.2\ b_t < a < 0$
(\footnote{The SVM convergence is severely complicated by the fast growing number of lower
lying states (as clarified in section \ref{discussion}), which means that the $N=10$
calculations have to be limited to $a > -0.2\ b_t$.}),
as shown in figure \ref{N10energies}. When looking at the resulting curves,
one immediately recognizes the same overall behavior as in the three- and four-boson
systems. The only clear, but expected, discrepancy is in the number of lower lying molecular-type
bound states, as can be observed on the right of the figure

\section{Additional remarks about the results}\label{discussion}

The study of boson systems presented in this chapter raises several discussion points,
where the most vital are concerning the accuracy of the data obtained and, assuming
this is acceptable, the validity of the interpretation, especially in terms of
generalizing the conclusions to many-boson systems with $N>4$. These questions are
addressed in this section.

\subsubsection{Accuracy of the SVM results}

\begin{figure}[tb]
\psfrag{YLABEL}{$E_{total}\ [\hbar\omega]$}
\includegraphics[width=16cm]{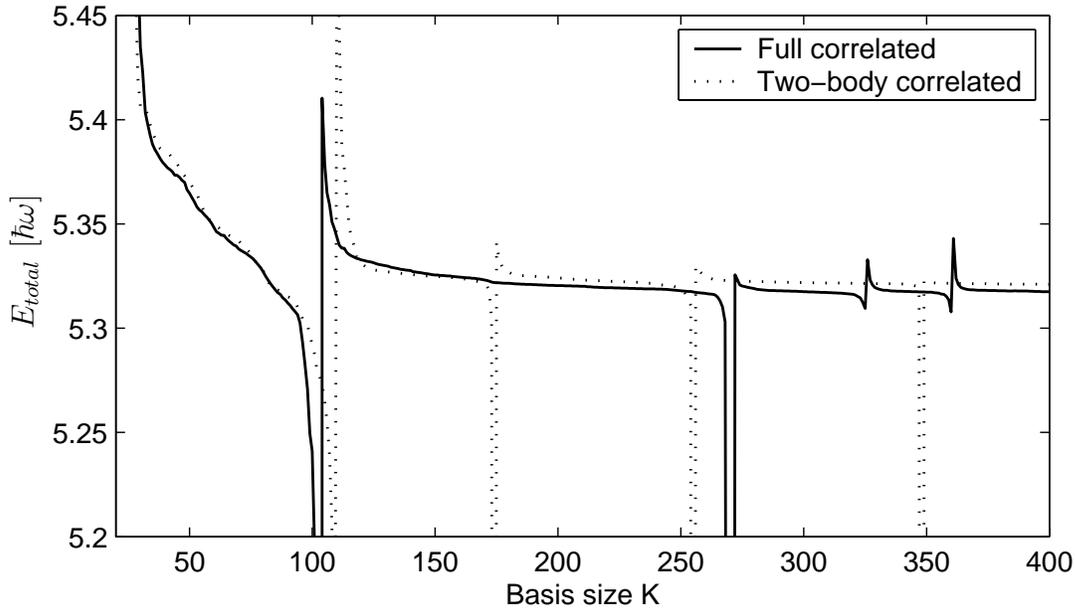}
\caption{The energy convergence of the lowest BEC state for the $N=4$ system of trapped bosons
in the case where $a=-0.13\ b_t$. Both the {\it full} SVM calculation and the {\it two-body}
SVM calculation is shown.
}
\label{N4convergence}
\end{figure}
As the observant reader might have noticed, that no comments or indications have been made so
far about the accuracy of the energies calculated in this work.
The reason for this is, that the stochastic nature of the SVM makes it inherently difficult to
estimate error bars on the results since these are all variational upper bounds.
In other words, there is no definite way to determine how
close the stochastic optimization procedure has come to reproducing the ``exact'' wave
function. However, one can get a general idea of the accuracy achieved in a given
calculation from studying the amount of energy gained when adding an additional basis function
to the basis.

To exemplify, consider the BEC energy convergence graph displayed in figure
\ref{N4convergence}, where the iterative trial and error strategy is illustrated for both the
{\it two-body} and {\it full} calculations in the specific $N=4$ case where $a=-0.13\ b_t$.
As a first impression, one notices the several dramatic falls followed by abrupt jumps
that will occur each time a new lower lying molecular bound-state has been found.
This behavior is a consequence the specific algoritm employed since it automaticly selects a higher
target state if the current target state becomes bound.
Overlooking such ``resonances'', the overall
tendency in the curves is an initially fast and subsequently slow decline towards the optimal
variational energy that can be achieved with the respective trial functions.
The key assumption is then, that focusing on the energy gain from the last basis increase,
and after making sure that this is not within a resonance, one can roughly estimate the accuracy
of the result from its size.

Following this example, the accuracy was accordingly checked for the calculations
presented here. In conclusion to this process, it seems convincing to the writer, that the results
are accurate to at least three significant digits, which should be enough in the current context.

\subsubsection{The molecular-type states}

Although it is the high density molecular-type states
that are expected to lead to the strongest interparticle relationships, these states are
not feasible in a discussion of many-body correlations in the current context. The reason is, that
the particles in such states are so close that they cannot be characterized as weakly
interacting and independent of the exact shape of the interaction potential, and this means,
that the adoption of the Gaussian model potential for the two-body interaction, as opposed to
one with a hard-core dependence, is inappropriate.
Furthermore, one can be inclined to think that the influence of correlations in the
molecular-type systems is so pronounced, that is makes no sense to try and distinguish between
two-body, three-body etc. effects. In other words, an accurate treatment of the molecular-type
many-boson systems is inherently tied with the {\it full} correlation description.

\subsubsection{Generalization to many-body systems}

In the numerical calculations presented in this chapter, most attention has been given
to the system of four bosons in a trap. The data obtained for this system is both extensive and
accurate, and reveal new important information about correlations in a four-body system.
However, in reference to the introduction of this thesis, it still remains to determine whether
the conclusions drawn for the four-boson system can be generalized to systems with $N>4$.

As is apparent from the analysis of the results, the SVM has limited usability
in the investigation of large $N$ systems. During the restricted time period of this study, the only
attempt to compare the four-boson results with the features of larger systems, has
been with the {\it two-body} treatment of the $N=10$ case presented in section \ref{N10}.
From this effort it is clear, that there are definite similarities between the $N=4$
and the $N=10$ energy levels, at least in the limited range considered. However, the lacking
ability to do the same comparison in the case of the {\it full} correlated treatment, is a severe
drawback.

In conclusion, the stand point taken by this writer is, that the available comparison does
not constitute enough evidence to allow the interpretations, regarding the effects of many-body
correlations, to be generalized to $N>4$ systems as they are. In particular, the estimate of the
contribution from two-body correlations to the total correlation energy (of two thirds) seems
inappropriate. It is more reasonable to expect this fraction to fall off as the number of particles
increase because the number of possible many-body correlations is higher.
However, the question whether higher-order correlations play an important part in the dynamics of
many-body systems, has, for all $N$, been clearly answered (yes!) from the four-body results.

\chapter{Conclusion and outlook}

In this thesis the intricate nature of correlations in many-body systems has been 
studied theoretically. The numerical results have been obtained with the Stochastic
Variational Method using several different forms for the variational trial function.

First the {\it ab initio} theoretical framework of the SVM was derived
from the foundation of the time-independent Schrödinger equation and an N-body Hamiltonian.
It was shown how to include different levels of correlation in a variational description by
incorporating interparticle dependencies explicitly in the functional form of the trial function. 
The particular descriptions treated were the uncorrelated Hartree-Fock description,
the pseudopotential mean-field description and the explicitly correlated description. While
the first two correspond to common mean-field theories, the latter includes correlations beyond
the mean-field and can be designed to introduce only pair-, triplet- or any
given higher-order correlation. This makes such a description ideal for investigating
the significance of correlation effects.

The SVM was subsequently developed to work with the different correlation descriptions
by expanding the trial functions in a basis of contracted Gaussians or, for the three-body case, in
a basis of exponential functions. All the necessary matrix elements were derived in
the cases where the two-body interaction is given by either a finite Gaussian potential,
a zero-range pseudopotential or a Coulomb potential. A particular elegant expression was obtained
for the Jastrow-type two-body correlated trial function, resulting in
matrix elements that where independent of the number of particles. The trial and error
optimization procedure, which is the heart of the SVM, was implemented in C++ and
thoroughly testet. The original few-body algoritm \cite{SVM} was upgraded to produce fast
convergence for systems of trapped bosons which require a random value interval over several orders of
magnitude.

After some initial bench marking by extensive calculations on the $^\infty$He atom,
the main part of the numerical work was concentrated on the systems of three and four
weakly interaction bosons in a spherical symmetric trap. Attractive interactions
corresponding to s-wave scattering lengths of $-\infty < a < 0$ were considered, in the attempt to
simulate the conditions of experiments on $^{87}$Rb near a Feshbach resonance.
The results obtained reveal the detailed behavoir of the energy levels as a function of $a$,
and confirms the gross features found by others for the $N=3$ system \cite{Blume2}.
In addition, both molecular-like and gas-like states are included, whereas
standard mean-field methods only treat the latter states.
The plotted energy curves describe a smooth non-diverging behavior and provides a detailed
picture of the corresponding physics in combination with an analysis of the effective
boson-boson potential.

Several interesting conclusions were drawn from the numerical results. 
First of all, it was apparent that the mean-field pseudopotential treatment
is insufficient for large negative scattering lengths even for few-boson systems.
However, the current study was not extensive enough to consider the validity criteria in detail.
Secondly, the next best treatment corresponds to assuming that one pair in the system is close
in space, but even this two-body correlated case eventually fails to
reproduce the correct correlation energy, although the error is at most 35\%
in the four-boson system. Subsequently, one would expect this error percentidge to climb
for system of more particles. In other words, it is not possible to study many-body systems
accurately over a wide range of scattering lengths, if only two-body correlations
are taken into account.

Because all the lower lying molecular-type bound states
have to be determined in the SVM, the running time of some calculations
came close to 12 hours. Therefore, and due to the lack of time, it was not possible to do widespread
investigations of the $N=4$ system for the full correlated treament. 
Hence a more systematic investigation of the large negative $a$ region could be carried out
in a future study, as this scattering range represents a ``blind'' spot in figure
\ref{N4correlation2}, which might include additional physics.

Another immediate extension of the present work is to investigate the five-boson system for
the two-body and three-body correlated treatments, at least in a limited range of
scattering lengths. Such calculations would represent feasable computations, and
the produced results can further clarify the relative importance between two-body
and higher-order correlations in an N-body system. Other direct improvements
include the adoption of more realistic two-body potential models, for example, the
simple sum of two Gaussians, where one corresponds to a repulsive hard core.
For people devoted to computer science, the new implemented code also opens the opportunity
of rigorous experimentation with the trial and error algoritm. This might lead to the
discovery of new and better stochastic optimization techniques which are becomming
increasingly important in computational physics.

In conclusion, the present investigation of correlations yielded insight into the
dynamics of many-body systems by investigating the systems of three and four bosons in a trap.
The results build conclusive evidence of the assumption that higher-order
correlations play an important role in many-body systems, and that such interparticle mechanisms
should not be neglected in future studies of this field.

\appendix
\chapter{Angular momentum functions}\label{AppBasis}
\label{angular}
In this appendix, it is shown how to treat systems having non-zero definite angular momentum
($L\ne0,S\ne0$) in a variational approach with the general variational trial function (\ref{BF}).
The goal is to have a the trial wave function that described a system of particles
having individual orbital angular momenta, $l_i$, spins, $s_i$, and isospins, $t_i$. 
The problem is to find a set of angular momentum operators that commute with the Hamiltonian and with
each other, so that common eigenfunctions exist. The corresponding good quantum numbers
are adapted by the orbital, spin and isospin parts of the basis function.

\section{Orbital angular momentum: $\theta_{LM_L}(\hat{\bm{x}})$}
Considering orbital angular momentum first, the single-particle
operators, $\widehat{\bm{l}}_j=-i \bm{r}_j \times \bm{\nabla}$, do not commute with
the kinetic energy term of the many-body Hamiltonian (\ref{Hr}). However, since this Hamiltonian
has no relativistic terms, the sum of orbital angular momentum operators,
$\widehat{\bm{L}}=\sum_{i=1}^N \widehat{\bm{l}}_i$, does commute with $\widehat{H}$\cite{Dahl}, i.e. 
\begin{equation}
[\widehat{H}, \widehat{\bm{L}}]=0, \quad
\Bigg\{{
\begin{aligned}
\widehat{L}^2 \Psi &= L(L+1)\Psi, \quad L=0,1,2,\dots \\
\widehat{L}_z \Psi &= M_L\Psi, \quad \quad \quad M_L=L, L-1, \dots, -L,
\end{aligned}
}
\end{equation}
making the definite orbital angular momentum $L$ and projection $M_L$ good quantum numbers.
The common eigenfunctions of the single-particle operators $\widehat{l}^2_i$ and
$\widehat{l}_{iz}$ is the surface spherical harmonics, $Y_{l_im_i}(\hat{\bm{r}}_i)$.
Generalizing, the angular part of the basis function, $\theta_{LM_L}(\hat{\bm{x}})$,
is a vector-coupled product
\footnote{Theory for addition of angular momentum can be found in \cite{Sak}.}
of spherical harmonics
\begin{align}
\nonumber
\theta_{LM_L}(\hat{\bm{x}})&=\Big[ \big[ [Y_{l_1m_1}(\hat{\bm{x}}_1)
\otimes Y_{l_2m_2} (\hat{\bm{x}}_2)]_{L_{12}M_{12}}
\otimes Y_{l_3m_3} (\hat{\bm{x}}_3)\big]_{L_{123}M_{123}}
\\&\quad\otimes\cdots \otimes Y_{l_Nm_N} (\bm{x}_N) \Big]_{LM_L}
\label{angpart1}
\\&= \sum_{k=\{ m_1,m_2,\cdots,m_N \} } C_k \prod_{i=1}^N Y_{l_im_i}(\hat{\bm{x}}_i),
\end{align}
where $C_k$ is a product of Clebsch-Gordan coefficients
\begin{align}
\nonumber
C_k &= \langle l_1m_1l_2m_2 | l_1l_2L_{12}m_1+m_2 \rangle
\langle L_{12}m_1+m_2l_3m_3 | L_{12}l_3L_{123}m_1+m_2+m_3 \rangle
\\ & \quad \cdots \langle L_{12\cdots N-1}m_1+m_2
\cdots m_{N-1}l_Nm_N|L_{12\cdots N-1}l_N LM_L \rangle
\end{align}
In this way, each relative orbital motion is being assigned a definite angular momentum
and $\theta_{LM_L}(\hat{\bm{x}})$ is dependent on the specific set of angular momenta,
$\{l_1,l_2,\dots,l_N;L_{12}, L_{123}, \dots\}$, chosen. The intermediate momenta,
$L_{12}$, $L_{123}$, etc., do not in general constitute good quantum numbers.
Thus, for a realistic description, it is often necessary to include several different
channels, i.e. sets of angular momenta, which is then referred to as the method of
partial wave expansion \cite{SVM}.

The various possible partial wave channels obviously increase the size of the basis.
In addition, this form of $\theta_{LM_L}(\bm{x})$, demanding the coupling of $(N-1)$
angular momenta, becomes increasingly complicated as the number of particles goes up.
With a variational approach all this can be avoided by adopting
another, closely related
\footnote{Any function of the form (\ref{angpart1}) can be written as a linear combination
of terms like (\ref{angpart2}) and vice versa. See the detailed proof in \cite{SVM}, sec. 6.2.},
choice for $\theta_{LM_L}(\bm{x})$, proposed by Varga and Suzuki in \cite{Suzuki};
\begin{equation}\label{angpart2}
\theta_{LM_L}(\bm{x})= |\bm{v}|^{2K}Y_{LM_L}(\bm{\hat{v}}),
\ \ \ {\rm with}\ \bm{v} = \sum_{i=1}^N u_i\bm{x}_i=\bm{u}^T \hspace{-2.0pt} \bm{x}
\end{equation}
Only the total orbital angular momentum enters in this expression. The real vector
$\bm{u}^T=\{u_1, \dots, u_N\}$, defining a linear combination of the relative coordinates,
may be considered a variational parameter. $K$ is a positive integer, most often small
\footnote{Using $K>0$ in (\ref{angpart2}) corresponds to including many
higher partial waves in the expression (\ref{angpart1}) for $\theta_{LM_L}(\bm{x})$,
\cite{Suzuki}.}.
Like before, several terms like (\ref{angpart2}) in the angular part of the basis function
will improve the description.
The continuity of the parameters $u_i$ yield continuous changes in the evaluated
energy expectation value, $\cal{E}$. This can be more advantageous in a variational calculation
than assigning sets of discrete angular momenta.

\section{Spin angular momentum: $\chi_{SM_S}$}
The spin and isospin angular momentum is treated in the same way as the orbital angular momentum.
If $\widehat{H}$ has no spin terms, the single-particle spin operator,
$\widehat{\bm{s}}_i= \½\widehat{\bm{\sigma}}_i$,
commutes with $\widehat{H}$. However, only symmetric operators commute with
every permutation operator, $\widehat{P}$, used below to ensure the proper symmetry.
The symmetric operators $\widehat{S}^2$ and $\widehat{S}_z$ corresponding to the
total spin, $\widehat{\bm{S}}=\sum_{i=1}^N \widehat{\bm{s}}_i$, are thus convenient.
The spin part of the basic functions then depend on the good quantum numbers $S$ and $M_S$, and is
given by successively coupled single-particle spin functions;
\begin{equation}
\chi_{SM_S}=\Big[ \big[ [\chi_{s_1m_1} \otimes \chi_{s_1m_1}]_{S_{12}M_{12}}
\otimes \chi_{s_3m_3} \big]_{S_{123}M_{123}}\otimes\cdots \otimes \chi_{s_Nm_N} \Big]_{SM_S}
\label{spinpart}
\end{equation}
The set  of spin quantum numbers, $\{s_1,s_2,\dots,s_N;S_{12}, S_{123}, \dots\}$, specifies the
particular coupling. Again, several sets may be needed to obtain a good wave function.
Alternatively, one can use a spin function based on continues parameters (see \cite{SVM}, sec.
6.4). The isospin function, $\eta_{TM_T}$, can be constructed in exactly the same manner as
the spin function, $\chi_{SM_S}$.

\section{L-S coupling} 

As already mentioned, the above angular momentum description is valid under the assumption
that the Hamiltonian does not contain any relativistic terms.
For the systems under consideration in this thesis (see chapter \ref{Results}) a non-relativistic
treatment is sufficient. However, this is far from always the case. Accurate calculations
of atomic energy levels have to account of atomic fine structure, generated by the prominent
spin-orbit term, $\sum_{i=1}^N \xi(r_i) \widehat{\bm{s}}_i \cdot \widehat{\bm{l}}_i$, \cite{BJ}.
The nucleon-nucleon interaction also has a strong spin-isopin dependence
\footnote{Modern two-body NN potentials are Argonne $v_{18}$, Nijmegen II, Reid93,
CN-Bonn \cite{Fabrocini}.}.
Obviously, neither $\widehat{\bm{L}}$ nor $\widehat{\bm{S}}$ commutes with the spin-orbit term,
while the vector sum, $\widehat{\bm{J}}=\widehat{\bm{L}}+\widehat{\bm{S}}$, does.
Consequently, $J$ and $M_J$ will serve as good angular momentum quantum numbers in
the presence of relativistic terms. The orbital and spin angular momenta
are coupled
\footnote{Known as {\it Russell-Saunders} or {\it L-S} coupling, \cite{Dahl}.}
, by applying the Clebsch-Gordan series to the $L$ and $S$ quantum numbers
of the separate parts, $\theta_{LM_L}(\bm{x})$ and $\chi_{SM_S}$, as indicated in the form
of the basis functions (\ref{BF}).

\chapter{Hartree-Fock ground state of identical fermions}\label{AppHFferm}

As Wolfgang Pauli's famous exclusion principle states, identical fermions cannot occupy the same
quantum state at the same time. This means that in an idealized case at $T=0$
the Hartree-Fock many-fermion ground state, $\Psi_{HF}^{(0)}$, should be formed by occupying N
single-particle energy levels from the lowest up.
If all interactions where neglected the many-fermion state would then represent a filled Fermi sphere
where all stationary orbitals corresponding to an energy less than the Fermi energy
$(E_F=k_B T_F)$ is occupied by exactly one particle. However, interactions pertube this picture
and modify the energy levels and single-particle states.

Within the Hartree-Fock method this modification is well described by the variational single-particle
wave functions. Assuming that the Pauli principle is still forced on the product wave function
(\ref{hartree}) by making it explicitly antisymmetric
\footnote{E.g. by adopting the Slater determinant description,
$\Psi_{HF} = {\cal A} \Psi_{H} = {\rm det}\ | \phi_1 \phi_2 \dots \phi_N|$, \cite{Dahl}.}
and that the N single-particle states $\phi_1 \phi_2 \dots \phi_N$ with lowest energy
are given by the spin-orbitals
\begin{equation}
\phi_i(\bm{r})=\psi_i(\bm{r})\chi_{1/2,s_i}, \quad i=1,2,\dots,N
\end{equation}
where $s_i$ is the spin of the $i$th fermion with $<\chi_{1/2,s_i}|\chi_{1/2,s_j}>=\delta_{s_is_j}$,
the interpretation of the Hartree-Fock equations
as eigenvalue problems \cite{BJ} with the Lagrangian multipliers, ${\cal E}_i$, as the one-particle
eigenvalues, yields solutions 
\begin{equation}\label{HFsol}
{\cal E}_i = <\psi_i|\widehat{h}_i|\psi_i>
+ \sum^N_{j=1} <\psi_i \psi_j|V_{ij}|\psi_i \psi_j - \delta_{s_is_j}\psi_j \psi_i>
\end{equation}
These eigenvalues obey Koopman's theorem \cite{Koopman}, $E_{HF}(N)-E_{HF}(N-1) = {\cal E}_N$.
Then ${\cal E}_i$ is the energy needed to remove the $i$th fermion from the system provided
the change in the wave function for the other particles can be neglected (e.g. when $N\gg1$).
Summing over ${\cal E}_i$, and comparing with eqs. (\ref{HFH})-(\ref{HFV}), 
the total energy of the fermion ground state in the Hartree-Fock approximation is
not $\sum_{i=1}^N {\cal E}_i$ as might be expected but rather
\begin{equation}
E_{HF}^{(0)} = <\Psi_{HF}^{(0)}|\widehat{H}|\Psi_{HF}^{(0)}> =\sum_{i=1}^N {\cal E}_i
- \sum^N_{i<j} <\psi_i \psi_j|V_{ij}|\psi_i \psi_j - \delta_{s_is_j}\psi_j \psi_i>
\end{equation}
The ``additional'' term can be understood as eliminating the double counting of pairs of particles
since $\sum_{i=1}^N {\cal E}_i$ includes the energy for each particle interacting with every other
particle and so counts the contribution from a given pair twice.
 
Equation (\ref{HFsol}) illustrates the influence of fermion-fermion interactions
on the single-particle energy levels. This effect is well described in the Hartree-Fock approach.
However, another effect of interactions is that the fermions might be scattered in and out of the
single-fermion levels, which are no longer stationary. This is not supported within the independent
particle approximation.
Fortunately, a fermionic particle can only change energy in a collision if the
final energy state is unoccupied. In the ultra-cold quantum regime, it is highly likely
that low-energy states are already occupied. Thus the validity of the Hartree-Fock method for
the fermionic ground-state relies on Pauli blocking since it tends to suppress any process
in which fermions change energy states.

\chapter{Matrix elements}\label{MatElm}

\section{Explicitly correlated Gaussian basis}\label{GaussElm}
All matrix elements of the correlated Gaussian basis functions (\ref{ECG})
are given by integrals that can be evaluated to simple analytical results.
Since the systems considered in this thesis are limited to central interactions and
zero angular momentum, all that is needed to calculate the matrix elements are the
three basic integral formulas \cite{SVM}
\begin{align}
&{\cal I}_0 \equiv 
\int_{-\infty}^{\infty} \cdots \int_{-\infty}^{\infty} d\bm{x}_1 d\bm{x}_2 \cdots d\bm{x}_{N-1}
e^{ -\½ \bm{x}^T \hspace{-2.0pt} \bm{A} \bm{x} }
= \bigg( \frac{(2\pi)^{N-1}}{{\rm det} \bm{A}} \bigg)^\frac{3}{2}
\\
&{\cal I}_1 \equiv
\int_{-\infty}^{\infty} \cdots \int_{-\infty}^{\infty} d\bm{x}_1 d\bm{x}_2 \cdots d\bm{x}_{N-1}
\bm{x}^T \hspace{-2.0pt} \bm{C} \bm{x} e^{ -\½ \bm{x}^T \hspace{-2.0pt} \bm{A} \bm{x} }
= 3 {\rm Tr} (\bm{A}^{-1} \bm{C} ){\cal I}_0 
\\
&{\cal I}_2 \equiv 
\int_{-\infty}^{\infty} \cdots \int_{-\infty}^{\infty} d\bm{x}_1 d\bm{x}_2 \cdots d\bm{x}_{N-1}
\delta(\bm{c}^T \hspace{-2.0pt}\bm{x}-\bm{r}) e^{ -\½ \bm{x}^T \hspace{-2.0pt} \bm{A} \bm{x} }
= \Big( \frac{c}{2\pi} \Big)^\frac{3}{2} e^{-\½ c \bm{r}^2} {\cal I}_0 
\label{I2}
\end{align}
where $\bm{x}^T= (\bm{x}_1,\bm{x}_2,\dots,\bm{x}_{N-1})$ are independent coordinates,
$\bm{A}$ and $\bm{C}$ are symmetric positive definite matrices and 
$c^{-1} = \bm{c}^T \bm{A}^{-1} \bm{c}$.

Introducing $\bm{B} = \bm{A}^{(k^\prime)} + \bm{A}^{(k)}$ for notational convenience and
using transformations $\bm{r}_{i} = (\bm{u}^{(i)})^T \bm{x}$ and
$\bm{r}_{ij} = (\bm{u}^{(ij)})^T \bm{x}$, defined in (\ref{invtrans}), the above integrals give
the following analytical expressions for the overlap, kinetic, trap and interaction matrix
elements respectively
\begin{equation}\label{overlap}
\langle \psi_{k^\prime} | \psi_{k} \rangle =
\bigg( \frac{(2\pi)^{N-1}}{{\rm det} \bm{B}} \bigg)^\frac{3}{2}
\end{equation}
\begin{align}
\nonumber
\langle \psi_{k^\prime}|
-\½ \widehat{\bm{\nabla}}_{\bm{x}}^T \bm{\Lambda} \widehat{\bm{\nabla}}_{\bm{x}}
 |\psi_{k}  \rangle
&=-\½ \langle \widehat{\bm{\nabla}}_{\bm{x}}^T \psi_{k^\prime}| \bm{\Lambda} |
\widehat{\bm{\nabla}}_{\bm{x}} \psi_{k}  \rangle
\\
\nonumber
&=-\½\langle \psi_{k^\prime} | \bm{x}^T \hspace{-2.0pt}
\bm{A}^{(k^\prime)} \bm{\Lambda} \bm{A}^{(k)} \bm{x} | \psi_{k} \rangle
\\&=-\frac{3}{2}{\rm Tr}( \bm{B}^{-1} \bm{A}^{(k^\prime)} \bm{\Lambda} \bm{A}^{(k)})
\langle \psi_{k^\prime}|\psi_{k}  \rangle
\label{kinetic}
\end{align}
\begin{align}
\nonumber
\langle \psi_{k^\prime} | \sum_{i=1}^N \½ m_i \omega^2 \bm{r}_i^2|\psi_{k} \rangle &=
\frac{3}{2} \omega^2
\langle \psi_{k^\prime} | \bm{x}^T \hspace{-2.0pt} \sum_{i=1}^N m_i  \bm{u}^{(i)} (\bm{u}^{(i)})^T
\bm{x} |\psi_{k} \rangle
\label{trap}
\\&=
\frac{3}{2} \omega^2 {\rm Tr} \Big(\bm{B}^{-1} \sum_{i=1}^N m_i \bm{u}^{(i)} (\bm{u}^{(i)})^T \Big)
\langle \psi_{k^\prime}|\psi_{k}  \rangle
\end{align}
\begin{align}
\nonumber
\langle \psi_{k^\prime} | \sum_{i<j}^N V_{ij} |\psi_{k} \rangle
&=\sum_{i<j}^N\int_{-\infty}^{\infty}d\bm{r} V(\bm{r})
\langle \psi_{k^\prime}| \delta(\bm{r}_{ij}-\bm{r}) |\psi_{k}  \rangle 
\\&=\sum_{i<j}^N
v\Big(\frac{1}{\bm{u}^{(ij)T} \bm{B}^{-1} \bm{u}^{(ij)}}\Big)
 \langle \psi_{k^\prime}|\psi_{k} \rangle
\label{potential}
\end{align}
where in the last expression the dependency on the specific form of the central 
potential describing the interaction, $V(\bm{r})$, is isolated in the single integral
\begin{equation}\label{vij}
v(c_{ij})=\Big( \frac{ c_{ij}}{2\pi} \Big)^\frac{3}{2} 
\int_{-\infty}^{\infty}d\bm{r} V(\bm{r}) e^{-\½ c_{ij} \bm{r}^2}.
\end{equation}
In the case of equal masses, $m_i\equiv m$, one may simplify the trap expression (\ref{trap})
with the identities
\begin{equation}\label{JacId}
\sum_{i=1}^N \bm{u}^{(i)} (\bm{u}^{(i)})^T = \bm{I}\quad {\rm and} \quad 
\bm{u}^{(ij)T} \bm{u}^{(ij)}=2
\end{equation}
that are satisfied by the Jacobi transformation.
To complete the evaluation of the matrix elements the $v(c_{ij})$ expressions for the three simple
central potentials used in this work are listed:
\begin{itemize}
\item For few body atomic systems the Coulomb interaction, $V(\bm{r})=\frac{q_i q_j}{r}$, is used
where $q_1$ and $q_2$ are the charges. In this case the integral (\ref{vij}) gives \cite{Schaum}
\begin{equation}\label{vCoulomb}
v_{Coulomb}(c_{ij})=4\pi q_i q_j \Big( \frac{c_{ij}}{2\pi} \Big)^\frac{3}{2}
\int_{0}^{\infty}dr \ r e^{-\½c_{ij} r^2} =2q_i q_j\sqrt{\frac{c_{ij}}{2\pi}}
\end{equation}

\item In the calculations of N-body boson systems the interaction is described by a
Gaussian potential, $V(\bm{r})=V_0 e^{-\bm{r}^2/b^2}$, giving
\begin{equation}\label{vGauss}
v_{Gauss}(c_{ij})=V_0\Big( \frac{c_{ij}}{2\pi} \Big)^\frac{3}{2}  
\int_{-\infty}^{\infty}d\bm{r} e^{-\½(c_{ij} + \frac{2}{b^2}) \bm{r}^2}
= V_0\Big( \frac{1}{1+2/b^2 c_{ij}} \Big)^\frac{3}{2}
\end{equation}

\item The commonly used ``mean-field'' description of Bose-Einstein Condensates has a two-body 
interaction given by a zero-range delta function potential \cite{Pitaevskii2},
$V(\bm{r})=\frac{4 \pi \hbar^2 a}{m}\delta (\bm{r})$, in which case
\begin{equation}\label{vDelta}
v_{Delta}(c_{ij})=\frac{4\pi \hbar^2 a}{m} \Big( \frac{ c_{ij}}{2\pi} \Big)^\frac{3}{2}
\end{equation}
where $a$ is the s-wave scattering length.
\end{itemize}

\section{Two-body correlated Gaussian basis}\label{TwoElm}

In this section, the matrix elements
$\langle \psi_{k^\prime} |\widehat{H}_{int}| \psi_{k}\rangle$
and $\langle \psi_{k^\prime} | \psi_{k}\rangle$ for the two-body correlated version
of the explicitly correlated Gaussian basis functions defined in (\ref{dilute}), are
evaluated. Introducing
$\gamma^{(k)}=\beta^{(k)}-\alpha^{(k)}$ for notational convenience, the basis functions
written in Jacobi coordinates are given by
\begin{equation}
\psi_k=\widehat{{\cal S}}\ \psi_k^{(12)}=\frac{1}{\sqrt{N!}}\sum_{i<j}^N \psi_k^{(ij)}, \quad \quad
\psi_k^{(ij)}={\rm exp}\Big(-\½ \bm{x}^T \hspace{-2.0pt} \bm{A}^{(ij,k)} \bm{x} \Big)
\end{equation}
where
\begin{equation}\label{defAij}
\bm{A}^{(ij,k)}=\gamma^{(k)}\bm{u}^{(ij)}(\bm{u}^{(ij)})^T+ N\alpha^{(k)}\bm{I}
\end{equation}
Since we are only considering systems of identical particles the Hamiltonian, $\widehat{H}_{int}$, is
symmetric, and one can initially apply (\ref{symbas}) to obtain
\begin{equation}\label{2mat}
\langle \psi_{k^\prime} |\widehat{H}_{int}| \psi_{k}\rangle=
\langle \widehat{\cal S} \psi_{k^\prime}^{(12)}|\widehat{H}_{int}|\widehat{\cal S}\psi_{k}^{(12)}
\rangle
= \frac{2}{N!}\langle \psi_{k^\prime}^{(12)} |\widehat{H}_{int}| \sum_{i<j}^N \psi_k^{(ij)} \rangle
\end{equation}
and likewise for the overlap matrix element ($\widehat{H}_{int}\rightarrow 1$).
Using the previously derived matrix element formulas (\ref{overlap})-(\ref{potential}) with
$m_i\equiv m$, the expressions for the overlap, kinetic, trap and interaction matrix elements between
$\psi_{k^\prime}^{(12)}$ and $\psi_{k}^{(ij)}$ are simply
\begin{align}
\langle \psi_{k^\prime}^{(12)}|\psi_{k}^{(ij)} \rangle &=
\bigg( \frac{(2\pi)^{N-1}}{{\rm det} \bm{B}} \bigg)^\frac{3}{2}
\\
\langle \psi_{k^\prime}^{(12)}|
\sum^{N-1}_{i=1} -\frac{\hbar^2}{2m}\widehat{\bm{\nabla}}_{\bm{x}_i}^2
 |\psi_{k}^{(ij)}  \rangle
&=-\frac{3\hbar^2}{2m}{\rm Tr}( \bm{B}^{-1} \bm{A}^{(12,k^\prime)}\bm{A}^{(ij,k)})
\langle \psi_{k^\prime}^{(12)}|\psi_{k}^{(ij)} \rangle
\\
\langle \psi_{k^\prime}^{(12)} | \sum_{i=1}^N \½ m \omega^2 \bm{r}_i^2|\psi_{k}^{(ij)} \rangle &=
\frac{3}{2} m\omega^2 {\rm Tr} (\bm{B}^{-1})
\langle \psi_{k^\prime}^{(12)}|\psi_{k}^{(ij)} \rangle
\\
\langle \psi_{k^\prime}^{(12)} | \sum_{m<n}^N V_{mn} |\psi_{k}^{(ij)} \rangle &=
\sum_{m<n}^N v\Big(\frac{1}{\bm{u}^{(mn)T} \bm{B}^{-1} \bm{u}^{(mn)}}\Big)
\langle \psi_{k^\prime}^{(12)}|\psi_{k}^{(ij)} \rangle
\end{align}
where $\bm{B}=\bm{A}^{(12,k^\prime)}+\bm{A}^{(ij,k)}$. Such (naive) adoption of the
general formulas for the evaluation of the two-body matrix elements apparently leads
to $N$ different contributions to (\ref{2mat}) for the overlap, trap and kinetic terms and even
$N^2(N-1)/2$ for the interaction term. This is however not the case since most of the
terms are identical. Consider the explicit Jacobi coordinate representation of $\psi_{k}^{(12)}$,
$\psi_{k}^{(13)}$, $\psi_{k}^{(23)}$ and $\psi_{k}^{(34)}$, defined by the matrices
\begin{equation}
\bm{A}^{(12,k)}=
\begin{pmatrix}
\sqrt{2}\gamma^{(k)} & 0 & \cdots & 0 \\ 
0 & N\alpha^{(k)} & \cdots & 0 \\ 
\vdots & \vdots & \ddots & \vdots \\ 
0 & 0 & \cdots & N\alpha^{(k)} \\ 
\end{pmatrix}
\end{equation}
\begin{equation}
\bm{A}^{(13,k)}
\begin{pmatrix}
\frac{1}{2}\gamma^{(k)}+N\alpha^{(k)} &
\sqrt{3}\frac{1}{2}\gamma^{(k)}+N\alpha^{(k)} & 0 & \cdots & 0 \\ 
\sqrt{3}\frac{1}{2}\gamma^{(k)}+N\alpha^{(k)} &
\frac{3}{2}\gamma^{(k)}+N\alpha^{(k)} & 0 & \cdots & 0 \\ 
0 & 0 & N\alpha^{(k)} & \cdots & 0 \\ 
\vdots & \vdots & \vdots & \ddots & \vdots \\ 
0 & 0 & 0 & \cdots & N\alpha^{(k)} \\ 
\end{pmatrix}
\end{equation}
\begin{equation}
\bm{A}^{(23,k)}
\begin{pmatrix}
\frac{1}{2}\gamma^{(k)}+N\alpha^{(k)} &
-\sqrt{3}\frac{1}{2}\gamma^{(k)}+N\alpha^{(k)} & 0 & \cdots & 0 \\ 
-\sqrt{3}\frac{1}{2}\gamma^{(k)}+N\alpha^{(k)} &
\frac{3}{2}\gamma^{(k)}+N\alpha^{(k)} & 0 & \cdots & 0 \\ 
0 & 0 & N\alpha^{(k)} & \cdots & 0 \\ 
\vdots & \vdots & \vdots & \ddots & \vdots \\ 
0 & 0 & 0 & \cdots & N\alpha^{(k)} \\ 
\end{pmatrix}
\end{equation}
and
\begin{equation}
\bm{A}^{(34,k)}=
\begin{pmatrix}
N\alpha^{(k)} & 0 & 0 & 0 & \cdots & 0 \\ 
0 & \frac{2}{3}\gamma^{(k)}+N\alpha^{(k)} &
\sqrt{2}\frac{2}{3}\gamma^{(k)}+N\alpha^{(k)} & 0 & \cdots & 0 \\ 
0 & \sqrt{2}\frac{2}{3}\gamma^{(k)}+N\alpha^{(k)} &
\frac{4}{3}\gamma^{(k)}+N\alpha^{(k)} & 0 & \cdots & 0 \\ 
0 & 0 & 0 & N\alpha^{(k)} & \cdots & 0 \\ 
\vdots & \vdots & \vdots & \vdots & \ddots & \vdots \\ 
0 & 0 & 0 & 0 & \cdots & N\alpha^{(k)} \\ 
\end{pmatrix}
\end{equation}
Using these matrices and the general definition (\ref{defAij}) one may convince oneself
\footnote{E.g. using \cite{LA}:
$\begin{vmatrix}
a & b \\ 
b & c \\ 
\end{vmatrix}
=ac-bb
$,
and
$\begin{bmatrix} a & b \\ b & c \\ \end{bmatrix}^{-1}
=\frac{1}{ac-bb}\begin{bmatrix} d & -b \\ -b & a \\ \end{bmatrix}
$}
that both ${\rm det} \bm{B}$,${\rm Tr}(\bm{B}^{-1})$,
${\rm Tr}(\bm{B}^{-1} \bm{A}^{(12,k^\prime)}\bm{A}^{(ij,k)})$ and even
$\bm{u}^{(mn)T} \bm{B}^{-1} \bm{u}^{(mn)}$, will evaluate to only three different values
and that this, in turn, allows the very important simplification
\begin{equation}
\langle \psi_{k^\prime}^{(12)}|\widehat{H}_{int}|\psi_{k}^{(ij)} \rangle =
\Bigg\{{
\begin{aligned}
\langle \psi_{k^\prime}^{(12)}|\widehat{H}_{int}|\psi_{k}^{(12)} \rangle&, \quad \quad i=1,\ j=2\\
\langle \psi_{k^\prime}^{(12)}|\widehat{H}_{int}|\psi_{k}^{(13)} \rangle&,
\quad \quad i=1,2,\ j=3, \dots,N  \\
\langle \psi_{k^\prime}^{(12)}|\widehat{H}_{int}|\psi_{k}^{(34)} \rangle&,
\quad \quad i=3,\dots,N,\ j=4, \dots,N
\end{aligned}
}
\end{equation}
and correspondingly for the overlap matrix elements. Moreover, the $N(N-1)/2$ terms in the
sum of the interaction matrix element are also limited to a constant number of different
values. The evaluation of these terms is strait forward but rather extensive
and here only the end results will be listed:
\begin{align}
\nonumber
\langle \psi_{k^\prime}^{(12)}|\sum_{m<n}^N V_{mn}|\psi_{k}^{(12)} \rangle &=
\Big[v(c_{12})
+2(N-2)v(c_{13})
+\½(N-2)(N-3)v(c_{34})\Big]
\\&\quad\times\langle \psi_{k^\prime}^{(12)}|\psi_{k}^{(12)} \rangle
\end{align}
\begin{align}
\nonumber
\langle \psi_{k^\prime}^{(12)}|\sum_{m<n}^N V_{mn}|\psi_{k}^{(13)} \rangle &=
\Big[v(c_{12})
+v(c_{13})
+v(c_{23})
+(N-3)\big(v(c_{14})
+v(c_{24})
+v(c_{34})\big)
\\&\quad+\½(N-3)(N-4)v(c_{45})\Big]
\times\langle \psi_{k^\prime}^{(12)}|\psi_{k}^{(13)} \rangle
\end{align}
\begin{align}
\nonumber
\langle \psi_{k^\prime}^{(12)}|\sum_{m<n}^N V_{mn}|\psi_{k}^{(34)} \rangle &=
\Big[v(c_{12})
+4v(c_{13})
+v(c_{24})
+v(c_{34})
+2(N-4)\big(v(c_{15})
+v(c_{35})\big)
\\&\quad+\½(N-4)(N-5)v(c_{56})\Big]
\times\langle \psi_{k^\prime}^{(12)}|\psi_{k}^{(34)} \rangle
\end{align}
where it is assumed that $N>4$ (or that only the appropriate terms that would
be available if $N\le 4$ are taken into account) and
$c_{nm}^{-1} = \bm{u}^{(mn)T} \bm{B}^{-1} \bm{u}^{(nm)}$. The interaction potential function,
$v(c_{nm})$, has been derived in (\ref{vCoulomb})-(\ref{vDelta}) for the central interactions of
interest in this thesis.

Collecting the above results, the matrix elements of $\widehat{H}_{int}$
for the two-body correlated basis functions
(and similarly for the overlap elements using $\widehat{H}_{int} \rightarrow 1$),
can be written as a sum of three terms, i.e.
\begin{equation}
\langle \psi_{k^\prime} | \widehat{H}_{int} |\psi_{k} \rangle
=\frac{2}{N!}\Big[\langle \psi_{k^\prime}^{(12)} | \widehat{H}_{int} |\psi_{k}^{(12)} \rangle
+N_{13}\langle \psi_{k^\prime}^{(12)} | \widehat{H}_{int} |\psi_{k}^{(13)} \rangle
+N_{34}\langle \psi_{k^\prime}^{(12)} | \widehat{H}_{int} |\psi_{k}^{(34)} \rangle\Big]
\end{equation}
where $N_{13}=2(N-2)$ and $N_{34}=(N(N-1)/2-1-2(N-2))$ and, more importantly, each of the individual
terms is given by a combination of a few expressions with computational complexity ${\cal O}(1)$.

\section{Correlated exponential basis ($N=3$ only)}\label{ExpElm}

In this section, the explicit analytical expressions for matrix elements needed in the 
exponential basis variational solution of a nonrelativistic Coulombic three-body system with
$L=0$, are presented. Additional formulas for arbitrary values of total angular momentum can
be found in \cite{Frolov0}.

Evaluation of the elements of  the overlap matrix (\ref{Smatrix}) and the Hamiltonian matrix
(\ref{Hmatrix}) corresponding to the Coulombic three-body Hamiltonian (in a.u. units)
\begin{equation}
\widehat{H}_{int} = -\½\widehat{\bm{\nabla}}_{\bm{x}}^T \bm{\Lambda} \widehat{\bm{\nabla}}_{\bm{x}}
+\frac{q_1q_2}{x_1}+\frac{q_1q_3}{x_2}+\frac{q_2q_3}{x_3}
\end{equation}
where $\bm{x}^T=(\bm{r}_{12},\bm{r}_{13},\bm{r}_{23})$ and $\bm{\Lambda}$ is defined by
(\ref{Lambda}), requires only the elements
\begin{equation}\label{IntHe1}
\langle \psi_{k^\prime} | \widehat{\cal S}\psi_{k} \rangle,
\ \langle \psi_{k^\prime} |\frac{1}{x_i}| \widehat{\cal S}\psi_{k} \rangle,
\ \langle \psi_{k^\prime} |\widehat{\bm{\nabla}}_{\bm{x}_i} \hspace{-2.0pt}
\cdot \widehat{\bm{\nabla}}_{\bm{x}_j}| \widehat{\cal S}\psi_{k} \rangle,
\end{equation}
to be calculated for all $i,j=1,2,3$ and
$\widehat{\cal S} = \frac{1}{\sqrt{6}}(1+\widehat{P}_{12}+\widehat{P}_{13}+\widehat{P}_{23}
+\widehat{P}_{12}\widehat{P}_{13}+\widehat{P}_{12}\widehat{P}_{23})$.
With the simple exponential basis function,
$\psi_{k}={\rm exp}(-\alpha_k x_1-\beta_k x_2-\gamma_k x_3)$, the scalar integral is
advantageously defined by \cite{Frolov5}
\begin{equation}
\langle \ \rangle
= \int\int\int r_{12} r_{13} r_{23} dr_{12}dr_{13}dr_{23}
= \int\int\int x_1 x_2 x_3 dx_1 dx_2 dx_3
\end{equation}
even though the interparticle distances are not independent variables. In this case,
however, the elements in (\ref{IntHe1}) are conveniently expressible in terms of the
{\it basic three-body integral}
\begin{equation}\label{BasicInt}
F(n_1,n_2,n_3)=\int\int\int x_1^{n_1} x_2^{n_2} x_3^{n_3} dx_1 dx_2 dx_3
\ {\rm exp}(-\alpha x_1-\beta x_2-\gamma x_3)
\end{equation}
where $\alpha=\alpha_{k^\prime}+\alpha_k$,$\beta=\beta_{k^\prime}+\beta_k$ and
$\gamma=\gamma_{k^\prime}+\gamma_k$. Simple analytical formulas for all
necessary $F(n_1,n_2,n_3)$ integrals are derived at the end of this section.

The overlap integral in the above notation is then simply
\begin{align}
\nonumber
\langle \psi_{k^\prime} | \psi_{k} \rangle
&=\int\int\int x_1 x_2 x_3 dx_1 dx_2 dx_3 \ {\rm exp}(-\alpha x_1-\beta x_2-\gamma x_3)
\\&= F(1,1,1)
\end{align}
Correspondingly, the matrix elements of the potential energy terms become
\begin{align}
\langle \psi_{k^\prime} |\frac{1}{x_1}| \psi_{k} \rangle &= F(0,1,1)
\\
\langle \psi_{k^\prime} |\frac{1}{x_2}| \psi_{k} \rangle &= F(1,0,1)
\\
\langle \psi_{k^\prime} |\frac{1}{x_3}| \psi_{k} \rangle &= F(1,1,0)
\end{align}
To determine the matrix elements of the kinetic energy terms the gradient
operator, $\widehat{\bm{\nabla}}$, is needed.
In the relative coordinates it has the form \cite{Frolov0}
\begin{equation}
\widehat{\bm{\nabla}}_{\bm{x}_i}=\hat{\bm{x}}_i \frac{\partial}{\partial x_i}
+\frac{1}{x_i}\widehat{\bm{\nabla}}_{\Omega_{\bm{x}_i}}
\end{equation}
The angular part in the gradient does not contribute when working on
purely radial basis functions, hence 
\begin{equation}
\widehat{\bm{\nabla}}_{\bm{x}_i} \hspace{-2.0pt} \cdot \widehat{\bm{\nabla}}_{\bm{x}_j}
=\Bigg\{{
\begin{aligned}
&\frac{x_i^2+x_j^2-x_k^2}{2x_i x_j} \frac{\partial^2}{\partial x_i \partial x_j}
, \quad i \neq j \neq k \\
&\frac{\partial^2}{\partial x_i^2}+\frac{2}{x_i}\frac{\partial}{\partial x_i},
\quad \quad \quad \quad i=j
\end{aligned}
}
\end{equation}
where the law of cosines \cite{Schaum}, $\hat{\bm{x}}_i \cdot \hat{\bm{x}}_j=
\frac{x_i^2+x_j^2-x_{ij}^2}{2x_i x_j}$, was used. The kinetic energy elements are then
\begin{align}
\langle \psi_{k^\prime} |\widehat{\bm{\nabla}}_{\bm{x}_1}^2 | \psi_{k} \rangle &= 
\langle \psi_{k^\prime} |\frac{\partial^2}{\partial x_1^2}+\frac{2}{x_1}
\frac{\partial}{\partial x_1}| \psi_{k} \rangle = -\alpha_k^2 F(1,1,1)+2\alpha_k F(0,1,1)
\\\langle \psi_{k^\prime} |\widehat{\bm{\nabla}}_{\bm{x}_2}^2 | \psi_{k} \rangle &= 
\langle \psi_{k^\prime} |\frac{\partial^2}{\partial x_2^2}+\frac{2}{x_2}
\frac{\partial}{\partial x_2}| \psi_{k} \rangle = -\beta_k^2 F(1,1,1)+2\beta_k F(1,0,1)
\\\langle \psi_{k^\prime} |\widehat{\bm{\nabla}}_{\bm{x}_3}^2 | \psi_{k} \rangle &= 
\langle \psi_{k^\prime} |\frac{\partial^2}{\partial x_3^2}+\frac{2}{x_3}
\frac{\partial}{\partial x_3}| \psi_{k} \rangle = -\gamma_k^2 F(1,1,1)+2\gamma_k F(1,1,0)
\end{align}
\begin{align}
\nonumber
\langle \psi_{k^\prime} |\widehat{\bm{\nabla}}_{\bm{x}_1} \hspace{-2.0pt}
\cdot \widehat{\bm{\nabla}}_{\bm{x}_2}| \psi_{k} \rangle &= 
\langle \psi_{k^\prime} |
\frac{x_1^2+x_2^2-x_3^2}{2x_1 x_2} \frac{\partial^2}{\partial x_1 \partial x_2}
| \psi_{k} \rangle
\\&= -\½\alpha_k\beta_k\big[F(2,0,1)+F(0,2,1)-F(0,0,3)\big]
\end{align}
\begin{align}
\nonumber
\langle \psi_{k^\prime} |\widehat{\bm{\nabla}}_{\bm{x}_1} \hspace{-2.0pt}
\cdot \widehat{\bm{\nabla}}_{\bm{x}_3}| \psi_{k} \rangle &= 
\langle \psi_{k^\prime} |
\frac{x_1^2+x_3^2-x_2^2}{2x_1 x_3} \frac{\partial^2}{\partial x_1 \partial x_3}
| \psi_{k} \rangle
\\&= -\½\alpha_k\gamma_k\big[F(2,1,0)+F(0,1,2)-F(0,3,0)\big]
\end{align}
\begin{align}
\nonumber
\langle \psi_{k^\prime} |\widehat{\bm{\nabla}}_{\bm{x}_2} \hspace{-2.0pt}
\cdot \widehat{\bm{\nabla}}_{\bm{x}_3}| \psi_{k} \rangle &= 
\langle \psi_{k^\prime} |
\frac{x_2^2+x_3^2-x_1^2}{2x_2 x_3} \frac{\partial^2}{\partial x_2 \partial x_3}
| \psi_{k} \rangle
\\&= -\½\beta_k\gamma_k\big[F(1,2,0)+F(1,0,2)-F(3,0,0)\big]
\end{align}
Since the permutation of particles does not change the explicit form of $\psi_k$
(e.g. $\widehat{P}_{12}\psi_{k}={\rm exp}(-\beta_k x_1-\alpha_k x_2-\gamma_k x_3)$), one can
determine all terms in the matrix element expressions in (\ref{IntHe1}) from the above results.

\subsubsection{Analytical formulas for the basic three-body integral}
\label{basicint}
To complete the derivation of the three-body matrix elements a few cases of
the basic three-body integral $F(n_1,n_2,n_3)$ in (\ref{BasicInt}) has to be calculated.
Transfor\-ming to truly independent perimetric coordinates given by
\begin{equation}
u_i=\½(r_{ik}+r_{ij}-r_{jk}),\ \ i\ne j\ne k=(1,2,3),
\end{equation}
the integral $F(0,0,0)$ is trivial \cite{Frolov5}
\begin{equation}
F(0,0,0)
=\frac{1}{(\alpha+\beta)(\beta+\gamma)(\alpha+\gamma)}
\end{equation}
Other cases of $F(n_1,n_2,n_3)$ can be derived by differentiating or integrating this
expression with respect to $\alpha$,$\beta$ and $\gamma$. Introducing the function
\begin{equation}
D(a,b,c)=\frac{1}{(\alpha+\beta)^a(\beta+\gamma)^b(\alpha+\gamma)^c}
\end{equation}
the necessary $F(n_1,n_2,n_3)$ integrals are given by
\begin{align}
F(1,1,0)&=D(2,2,1)+D(2,1,2)+D(1,2,2)+2*D(3,1,1)
\\F(3,0,0)&=2*3*\big[D(3,1,2)+D(2,1,3)+D(4,1,1)+D(1,1,4)\big]
\\F(1,1,1)&=2*\big[D(3,2,1)+D(2,3,1)+D(1,3,2)+D(1,2,3)
\nonumber
\\&\quad \quad \quad +D(2,1,3)+D(3,1,2)+D(2,2,2) \big]
\\F(2,1,0)&=2*\big[D(3,2,1)+D(2,1,3)+D(1,2,3)+D(2,2,2)
\nonumber
\\&\quad \quad \quad +2*D(3,1,2)+3*D(4,1,1)\big]
\end{align}
Permutation of the coordinates $x_1 \leftrightarrow x_2\leftrightarrow x_3$ in the
basic integral (\ref{BasicInt}) corresponds to permutation of
$(n_1,\alpha)\leftrightarrow(n_2,\beta)\leftrightarrow(n_2,\gamma)$
and allows easy construction of the remaining cases, e.g.
$F(0,1,2)=F(2,1,0;\alpha\leftrightarrow\gamma)$.

\chapter{C++ implementation of the Stochastic Variational Method}
As with most numerical calculations in physics the main effort during implementation
is on precision and speed. With this particular method, heavy duty demands during
matrix element calculations and large eigenvalue problems, makes an efficient routine
essential. In this section, considerations on key aspects
of implementing SVM are described.

\section{Arbitrary precision arithmetic}
When working with a very large basis the standard 64-bit precision arithmetic might not
be sufficient to maintain numerical stability in the computations. To this end, the free
multiprecision package {\it doubledouble} \cite{GMP} is applied for calculations with floating
point numbers of an 128-bit length. A C++ class was then wrapped aroud this type to allow
for 64-bit exponents. However, employing this class makes all computations aproximately
7 times slower.

\section{Scaling overlap values to minimize loss of accuracy}\label{AppScale}

The magnitude of the overlaps is the dominant scale of the matrix elements
corresponding to a given trial function. Unfortunately, scaling the overlap to
$\langle \psi_{k^\prime} |\psi_{k} \rangle \sim 1$ is not possible
without breaking up the (binary) power function calculation and scaling concurrently. However, one can
use the overlap magnitude to estimate the maximum and minimum values handled in the
eigenvalue solution and hence set up a validity check for a calculation on a
specific computer.
Neglecting all constant factors
\footnote{For a running calculation $N$ can be viewed as a constant factor, although, when no
scaling is applied it must be kept in the power expression to avoid overflows when computing the
$(N-2)$th power for very large $N$ (i.e. $N\alpha \sim 1$). }
in the overlap expression (\ref{overlap}) while introducing a scale factor $S$ gives 
\begin{align}
\langle \psi_{k^\prime} |\psi_{k} \rangle_{max} &\sim 
\Big( \frac{S}{\alpha_{min} }\Big)^{\frac{3N}{2}-1}
\\
\langle \psi_{k^\prime} |\psi_{k} \rangle_{min} &\sim
\Big( \frac{S}{\alpha_{max} }\Big)^{\frac{3N}{2}-1}
\end{align}
where $\alpha_{max}$ and $\alpha_{min}$ is the maximum and minimum value possible for $\alpha$ 
and correspondingly for $\gamma$. One can center the overall magnitudes
around $10^0$ by multiplying all overlaps with the factor
$S=\sqrt{\alpha_{max} \alpha_{min}}$.

\section{Avoiding linear independence}

At the core of SVM is the random trial and error selection of the basis functions
$\Psi(\alpha_i), i=1..K$. The result is a state-space spanned by
a finite number of dense nonorthogonal functions. Because of the random origin
of the basis functions they cannot be expected to be linearly independent. Although
linear dependence is seldom with the fully random basis, in contrast to
geometric progression and random tempering \cite{SVM}, care must be taking
to avoid it.
In practical problems exact linear dependence, like degeneracy in the basis, is unlikely,
but still close to exact linear dependence between basis functions will lead to poor
precision in the calculation. This is because one or several eigenvalues of $\bm{S}$ gets
very small when the linear dependence is distinct, producing large expansion coefficients in the
trial function. Then a small error in calculation of the matrix elements of $\bm{H}$ and
$\bm{S}$ can result in a large error in the variational energy.

\section{A symmetric-definite generalized eigenvalue problem}

Adding a new trial function to the basis demands solving a symmetric-definite
generalized eigenvalue problem with good precision and efficiency.
Solving eigenvalue problems has been an
intense area of research since the dawn of computers in the 1950's, resulting in numerous
elegant methods, specially designed for different conditions of the eigenproblem (see summary
of research in \cite{Gol}). Fortunately, for real symmetric-definite matrices, the eigenproblem is
relatively simple. The eigenvalues are always real and there exists a complete orthogonal
eigensystem that is exploited in very efficient numerical methods.

For real symmetric matices, the eigenproblem is relatively simple, due to the existence
of a complete orthogonal eigensystem, and the fact that all eigenvalues are real.
These properties are exploited in the most efficient numerical methods, and the symmetric
eigenproblem may be considered as solved: for small matrices $n<=25$ we have the QR
method, one of the most elegant numerical techniques produced in the field of numerical
analysis; for large matrices $25 < n < 1000$, we have a combination of divide and conquer
with QR techniques. For asymmetric matrices the picture is less rosy.

\section{Root finding}\label{apRoot}
The Gram-Schmidt orthogonalization formula (\ref{GramPhi}) implies that
\begin{equation}
\sum_{i=1}^K | \phi_i \rangle \langle \phi_i | 
+\frac{|\phi_{K+1} \rangle \langle \phi_{K+1} |}{\norm{\phi_{K+1}}^2}
=1
\end{equation}
with
\begin{equation}
\norm{\phi_{K+1}}^2 =
\langle \Psi(\alpha_{K+1}) | \Psi(\alpha_{K+1}) \rangle
-\sum_{i=1}^K |\langle \Psi(\alpha_{K+1}) | \phi_i \rangle|^2
=\langle \Psi(\alpha_{K+1}) | \phi_{K+1} \rangle
\end{equation}
and
\begin{equation*}
\begin{split}
\langle \phi_{K+1} | H | f \rangle
&=\sum_{i=1}^{K+1} c^{(K+1)}_i \langle \Psi(\alpha_i) | H | f \rangle
\\&=
\langle \Psi(\alpha_{K+1}) | H | f \rangle
+ \sum_{i=1}^{K} c^{(K+1)}_i \langle \Psi(\alpha_i) | H | f \rangle
\\&=
\langle \Psi(\alpha_{K+1}) | H | f \rangle
- \sum_{i=1}^{K} \left( \sum_{j=1}^{K} c^{(j)}_i \langle \Psi(\alpha_{K+1}) | \phi_j \rangle
\right) \langle \Psi(\alpha_i) | H | f \rangle
\\&=
\langle \Psi(\alpha_{K+1}) | H | f \rangle
- \sum_{j=1}^{K} \langle \Psi(\alpha_{K+1}) | \phi_j \rangle \left( \sum_{i=1}^{K} c^{(j)}_i 
\langle \Psi(\alpha_i) | H | f \rangle \right) 
\\&=
\langle \Psi(\alpha_{K+1}) | H | f \rangle
- \sum_{i=1}^K \langle \Psi(\alpha_{K+1}) | \phi_i \rangle
\langle \phi_i | H | f \rangle
\end{split}
\end{equation*}
Hence $h_j\norm{\phi_{K+1}}^{\½}=\langle \phi_j | H | \phi_{K+1} \rangle$ for $j = 1..K$
and the most efficient way to implement the root finding is by using the expression 
\begin{equation*}
\begin{split}
D(\lambda) &= \sum_{i=1}^K \frac{\abs{h_i}^2}{(\epsilon_i-\lambda)} + h_{K+1} - \lambda 
\\&= \sum_{i=1}^K \frac{\abs{\langle \phi_i | H | \phi_{K+1} \rangle}^2}{(\epsilon_i-\lambda)}
+ \langle \phi_{K+1} | H | \phi_{K+1} \rangle - \lambda \norm{\phi_{K+1}}
\end{split}
\end{equation*}
Notice that this equation also proves that the energy will be lower when the dimension of
the basis increases.

\section{Making sure $\bm{A}$ is positive definite}
An $N \times N$ real matrix $\bm{A}$ is called positive definite if
\begin{equation}
\bm{x}^T \hspace{-2.0pt} \bm{A} \bm{x} > 0,
\nonumber
\end{equation}
for all nonzero vectors $\bm{x} \in  \mathbb{R}^N$. There are various ways
to test if a matrix $\bm{A}$ is positive definite based on the following
observations \cite{PosDef}: (a) all the eigenvalues of a positive definite matrix are positive,
(b) all upper left (i.e. principal) submatrix determinants are
positive and (c) a real {\it symmetric} matrix is positive definite iff there
exists a real non-singular lower triangular matrix $\bm{L}$ such that
\begin{equation}
\bm{A}=\bm{L}\bm{L}^T.
\end{equation}
The latter approach, called Cholesky factorization, is the most efficient
in the case of a large size symmetric matrix. 
This method is implemented based on the ALGOL procedure {\it Choldet1} from \cite{Cholesky}. 
The factorization algoritm fails if the matrix is not positive definite. 
In addition, observation (b) above is used to include
simple and fast special cases for $N = 1,2,3$. For such low $N$, only
a few simple determinants have to be evaluated, making the factorization routine
cumbersome in comparison. This will speed up calculations involving four or less particles.

\section{Inversion of positive definite symmetric matrices}
Inversion of positive definite symmetric matrices is effectively done
by the Gauss-Jordan Method following the lines of the ALGOL procedure
{\it gjdef2} from \cite{Bauer}.
The lower triangular array representation of symmetric matrices allow fast element access.

\section{Symmetry: all possible permutations}
To implement the symmetrization procedure described in section \ref{symmet} one
need to find all possible permutations of the set of identical parameters.
The SEPA algorithm \cite{SEPA} is used to create all permutations from which the corresponding
linear transformations of the Jacobi coordinates for each trial encountered is evaluated.
This is done only once for each trial and then stored for later uses.

\chapter{Program usage information}\label{prog}
\small
\begin{verbatim}
Usage: scatlen [OPTIONS]
Calculates the scattering length for a two-body interaction of identical bosons

  -h --help      Prints this usage message.
  -mass <float>  Mass of the particle in atomic mass units (default 86.9091835)
  -pot <int>     Type of potential ='gauss' or 'square' (default is 'gauss')
  -V0 <float>    The potential amplitude in a.u. (default is -5.986e-8)
  -b <float>     The potential width a.u. (default is 18.9 = 1 nm)
  -rmax <float>  Integrate to this radius in a.u. (default 4b)
  -steps <int>   Force a specific number of integration steps (default is 10000)
  -digits <int>  Force a specific number of correct digits (default 4)
  -printpot      Prints the potential points as 'r1 p1 | r2 p2 | ...'
  -printwave     Prints the radial wave function as 'r1 w1 | r2 w2 | ...'
  -compare       Compare result with analytical square box value or Born approx
  -notxt         Demand scattering length as only output
\end{verbatim}
\newpage
\begin{verbatim}
Usage: bec [OPTIONS]
Calculates the energy for a given state of an N-body system using the Stochastic Variat
ional Method.

  -h --help      Prints this usage message.
  -N <int>       Number of particles (default is 3)
  -state <int>   Specify <int>th,'pos','neg' eigenstate as target (default is 0)
  -basis <type>  Basis type = 'full','twobody','hartree' (default is 'twobody')
  -par <int>     Override the number of nonlinear parameters in the full basis
  -sym [int]     Symmetrize trials [only first <int> particles] (default is N)
  -antisym [int] Antisymmetrize trials [only first <int> particles] (no default)
  -size <int>    Size of the basis to be calculated (default is 10)
  -trials <int>  Number of trials pr. nonlinear parameter (default is 100)
  -times <int>   Number of times to restart trials loop at par. one (default is 10)
  -reps <int>    Number of times to repeat the trial&error procedure (default is 1)
  -recycle <int> Recycle intermission every <int>'th new basis (default is size)
  -rfine <float> Recycle type: 0 = random, <float> = finetune (default is 0)
  -rtimes <int>  Number of times to repeat recycle procedure (default is 0)
  -rbegin <int>  Recycle procedure should begin at basis <int> (default is 1)
  -rend <int>    Recycle procedure should end at basis <int> (default is K)
  -units <type>  Calculation units = 'hou' or 'au' (default is h.o.u.)
  -notrap        Remove trap from system (to calculate bound states)
  -dotrap        Add trap to the system (to cancel previous -notrap)
  -int <type>    Interaction = 'non','zero','gauss','coulomb' (default is non)
  -b <float>     Set the potential range in a.u. (default is 11.65)
  -V0 <float>    Set the potential amplitude in a.u. (default is 1.103130e-7)
  -as <float>    Specify the scattering length in a.u. (used only for output)
  -aB <float>    Override the calculated Born scattering length in a.u.
  -seed <int>    Seed for the random number generator (default is 1)
  -rint <float>  Random interval range for alpha coefficient (default is 10.0)
  -rmin <float>  Override the estimated alpha random interval minimum
  -rmax <float>  Override the estimated alpha random interval maximum
  -rbint <float> Random interval range for beta coefficient (default is 10.0)
  -rbmin <float> Override the estimated beta random interval minimum
  -rbmax <float> Override the estimated beta random interval maximum
  -rlog [<int>]  Use logarithmic random interval [with base <int>]
  -fin <name>    Filename for basis input (default none)
  -fout <name>   Filename for basis output (default none)
  -digits <int>  Number of digits used in rootfinding and output (default is 8)
  -noimps <int>  Succeeding 'no improving trial's allowed (default is 5)
  -ldep <float>  Specify the lowest linear dependency allowed (default is 1e-6)
  -threads <int> Maximum number of cpu threads used (default is 2)
  -save <int>    Save basis every <int>'th minute (default is 10)
  -endtime <int> Time limit for the calculation in minutes (default is no limit)
  -check         Check explicitly for numerical instabilities (default is off)
  -warn          All warnings are displayed (default is off)
  -stat          Post-calculation statistics are displayed (default is off)
  -noinfo        Output only calculation results (default is with info)
  -notxt         Output only: [basisnumber energy] (for use with e.g. MATLAB)
  -eigenvalues   Output all eigenvalues at the end (i.e. show excited states)
  -result        Output only the final energy result (for use with e.g. MATLAB)
  -resultall     Output for MATLAB by writing information and end results like:
                 [N energy as b V0 aB aho eho K state Rrms mean(alpha's) mean(beta's) dE seconds]
 ---------------
  -bec           For Bose-Einstein Condensate calculation (default is Rb87)
  -mass <float>  Mass of BEC boson in a.m.u. (default is m(Rb87)=86.9091835)
  -freq <float>  Specify the trap frequency in Hz (default is 77.87)
 ---------------
  -bound         For N-body bound state calculation (default is Helium atom)
  -masses <list> Set particle masses m1,m2,..mN in a.u. (default is 1,1,1e300)
  -charges <list>Set particle charges q1,q2,..qN in a.u. (default is -1,-1,2)
\end{verbatim}

\end{document}